\documentclass[12pt]{article}
\pdfoutput=1
\usepackage[nosort]{cite}

\usepackage[dvipsnames]{xcolor}

\usepackage{epsfig}
\usepackage{amsfonts}
\usepackage{amscd}
\usepackage{latexsym}
\usepackage{dsfont}
\usepackage{amsmath,amssymb}
\usepackage{bm}
\usepackage{verbatim}
\usepackage{setspace}
\usepackage{color}
\usepackage{fancyhdr}
\usepackage{cite}
\usepackage{hyperref}
\usepackage{tikz}
\usepackage{slashed}
\usepackage{multirow}
\usepackage{amsthm}
\newtheorem*{theorem*}{Theorem}
\newtheorem{lemma}{Lemma}[section]

\usetikzlibrary{topaths}
\usetikzlibrary{decorations}
\usetikzlibrary{decorations.pathmorphing}
\usetikzlibrary{arrows, decorations.markings, cd}
\usetikzlibrary{calc, cd, decorations.pathmorphing}
\tikzset{
->-/.style args={#1rotate#2}{decoration={markings, mark=at position #1 with {\arrow[scale=1.5,rotate = #2 ]{stealth}}}, postaction={decorate}}
}
\usetikzlibrary{shapes.geometric}
\usetikzlibrary{knots}
\usepackage{tikz-3dplot}
\usetikzlibrary{positioning}

\usepackage{draft}
\usepackage{hyperref}
\usepackage{graphicx,subfig}
\usepackage{cite}
\usepackage{mciteplus}
\usepackage{bbm}
\usepackage{braket}
\usepackage[most]{tcolorbox}

\numberwithin{equation}{section}

\def\f{\varphi}
\def\dd{\mathcal{D}}
\def\defect{\mathbb{D}}
\def\op{\mathsf{D}}

\def\cA{\mathcal{A}}
\def\bZ{\mathbb{Z}}
\def\bC{\mathbb{C}}

\def\cO{\mathcal{O}}
\def\cH{\mathcal{H}}

\def\({\left(}
\def\){\right)}

\begin{document}

\begin{titlepage}
\hfill MIT-CTP/6154

\title{\Large\bfseries
  Non-Invertible Quantum Cellular Automata I:\\
  {\large Microscopic Definition of Non-Invertible Symmetries}
}

\author{Sahand Seifnashri$^{1}$ and Shu-Heng Shao$^{2}$}
\address{${}^{1}$School of Natural Sciences, Institute for Advanced Study, Princeton, NJ}
\address{${}^{2}$Center for Theoretical Physics --- A Leinweber Institute, Massachusetts Institute of Technology, Cambridge, MA}
\abstract{
We introduce non-invertible quantum cellular automata (QCAs), defined as locality-preserving quantum channels that are dualizable and pure. This provides an operator-algebraic definition of non-invertible symmetries of quantum lattice systems, formulated in terms of their action on local operators. Focusing on quantum spin chains, we motivate these defining conditions and conjecture that every such quantum channel admits a realization by dualizable topological defects or, equivalently, by dualizable sequential quantum circuits. We construct a broad class of examples, including on-site non-invertible symmetries that are anomalous, as well as higher-dimensional examples. In Part II of this work, we reconstruct topological defects whose boundary algebras yield an index theory that classifies non-invertible QCAs up to blend equivalence.
}
\end{titlepage}

\setcounter{tocdepth}{2}
\tableofcontents

\section{Introduction}

Generalized global symmetries~\cite{Gaiotto:2014kfa}, and in particular non-invertible symmetries, have been shown to be as powerful as ordinary invertible symmetries in quantum field theory and quantum lattice systems. See, for example, \cite{McGreevy:2022oyu, Cordova:2022ruw, Schafer-Nameki:2023jdn, Brennan:2023mmt, Shao:2023gho, shao2025noninvertible} for reviews. However, up to now, there has been no satisfactory microscopic definition that is applicable to quantum lattice systems, as we will now explain.

The key challenge in defining non-invertible symmetries on the lattice is giving a precise formulation of their \emph{locality}. For relativistic quantum field theory in Euclidean signature, non-invertible symmetries are defined as topological defects, whose locality is implicit in the word ``defect''. This is because defects (i.e., extended objects) in quantum field theory respect locality by construction. For instance, order-type defects are constructed from a \emph{local} action along their world-volume, such as Wilson lines in gauge theories. Another construction corresponds to disorder-type defects that remove a neighborhood of their world-volume from spacetime and impose \emph{local} boundary conditions on the deleted region. A canonical example of the latter is the 't Hooft line in four-dimensional gauge theories (see, for instance, \cite{Kapustin:2005py}).

The continuum field theory motivates the definition of generalized symmetries as topological defects. However, in non-relativistic Hamiltonian systems, there is a difference between extended operators that act on the Hilbert space and defects that correspond to modifications of the Hamiltonian and possibly the Hilbert space. Surprisingly, for topological defects there is a one-to-one correspondence between the two, as emphasized in \cite{Seifnashri:2023dpa} for invertible symmetries. The current work concerns the generalization of this correspondence to non-invertible symmetry operators and defects in quantum Hamiltonian systems. 

\subsection*{Locality of non-invertible symmetries}
Concretely, consider a lattice system on a finite lattice with Hamiltonian $H$. We are interested in knowing what kind of non-invertible operator $\mathsf{D}: \mathcal{H} \to \mathcal{H}$ qualifies as a legitimate global symmetry. Of course, we must impose that $\mathsf{D}$ commutes with the lattice Hamiltonian $H$. It remains to formulate a notion of locality for $\mathsf{D}$. 
Otherwise, some of the powerful consequences of symmetry, such as anomaly matching and organizing phases based on patterns of symmetry breaking, are absent even for invertible unitary symmetries. 

An invertible symmetry is generated by a unitary operator $U$ that commutes with the Hamiltonian and acts locally. Specifically, we say $U$ is locality-preserving with range $\ell \in \mathbb{R}_{\geq 0}$, if given a local operator $\mathcal{O}$ supported on a region $R$,\footnote{For the most general spatial/crystalline symmetries, we should also allow translating and rotating the region $R$ in space in addition to thickening it and considering its finite neighborhood.}
\ie \label{locality.c}
    U \mathcal{O} U^\dagger \text{ is supported on a finite neighborhood $R^{+\ell}$ of region }R \,.
\fe
The region $R^{+\ell}$ consists of all the points within distance $\ell$ from $R$. Importantly, $\ell$ is a constant independent of the region $R$ and the operator $\mathcal{O}$. Unitary operators of finite range are called quantum cellular automata (QCAs) \cite{Schumacher:2004hix, Farrelly:2019zds}.

Extending the locality criterion \eqref{locality.c} to non-invertible symmetries is challenging since for a non-invertible symmetry operator $\mathsf{D}$, the action $\mathcal{O} \mapsto \mathsf{D} \mathcal{O} \mathsf{D}^\dagger$ is always non-local. For example, by setting $\mathcal{O}=\mathbbm{1}$ we get $\mathsf{D} \mathsf{D}^\dagger$, which is a sum of other symmetry operators that are supported on the whole space and are not local. Thus, it is not clear how to impose locality directly on the symmetry operator $\mathsf{D}$.

One idea is to assume $\mathsf{D}$ is a Matrix Product Operator (MPO), which encompasses a large class of lattice non-invertible symmetries respecting a certain notion of locality \cite{Bultinck:2015bot, Vanhove:2018wlb, Inamura:2021szw, Lootens:2021tet, Molnar:2022nmh, Lootens:2022avn, Garre-Rubio:2022uum, Lootens:2023wnl, Fechisin:2023odt, Seiberg:2024gek, Chatterjee:2024ych, Lootens:2024gfp, Inamura:2024jke, Pace:2024acq, Zhang:2020pco, Jones:2024lws, Pace:2024oys, Meng:2024nxx, Vancraeynest-DeCuiper:2025fpn, Lu:2026rhb, Inamura:2026hif, Wen:2026ncw, Vernier:2026kxf, Bosgoed:2026ayl}. This turns out to be a necessary condition, since any symmetry operator that admits a topological defect has an MPO realization~\cite{Seiberg:2024gek, Tantivasadakarn:2025txn}. However, this is not sufficient. Indeed, there are many examples of MPOs that do not correspond to any (dualizable) topological defects (see Section \ref{sec:kw.deformed}). (For translation-invariant symmetry operators, a sufficient condition is that $\mathsf{D}$ admits a topological injective MPO representation, as introduced in~\cite{Inamura:2026hif}.)

\subsection*{Non-invertible symmetries as locality-preserving quantum channels}

The idea we follow here is to define non-invertible symmetries directly by their action on local operators and impose that this action is locality-preserving.

For invertible symmetries, the action on local operators $\alpha(a)=UaU^\dagger$ is a $*$-automorphism since it satisfies $\alpha(ab)=\alpha(a)\alpha(b)$ and $\alpha(a^\dagger)=\alpha(a)^\dagger$ for local operators $a$ and $b$. What should replace $*$-automorphisms for the case of non-invertible symmetries? In 1+1d continuum quantum field theories, non-invertible topological lines act on local operators as~\cite{Gaiotto:2014kfa, Bhardwaj:2017xup, Chang:2018iay}
\ie \label{tdl.action}
\f(a) =  \frac{1}{d_\mathcal{L}} ~ 
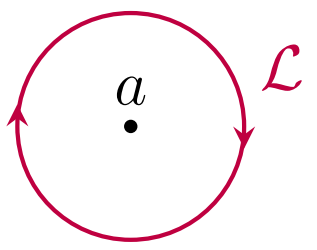
\begin{tikzpicture}[scale=1,baseline=(A.south)]
    \node [draw, fill=black, circle, inner sep=0.75pt] (A) at (0,0)  {};
    \draw (0,0) node [above] {$a$};
    \draw [color=purple, thick, decoration = {markings, mark=at position .5 with {\arrow[scale=1]{stealth[reversed]}}, mark=at position 1 with {\arrow[scale=1]{stealth[reversed]}}}, postaction=decorate] (0,0) circle (0.75);
    \draw (1,0.1) node[above] {\color{purple}$\mathcal{L}$};
    \end{tikzpicture} \,,
\fe
where the normalization is chosen such that $\f(\mathbbm{1})=\mathbbm{1}$. This action is clearly not an automorphism since $\f$ is not multiplicative, namely $\f(ab) \neq \f(a)\f(b)$. 

A more general action on local operators is the lasso action~\cite{Chang:2018iay}
\ie \label{lasso.action}
a ~\longmapsto~ 
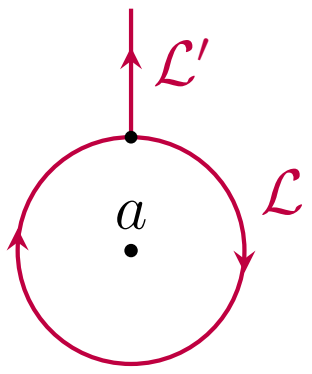
\begin{tikzpicture}[scale=1,baseline=(A.south)]
    \node [draw, fill=black, circle, inner sep=0.75pt] (A) at (0,0)  {};
    \draw (0,0) node [above] {$a$};
    \draw [color=purple, thick, decoration = {markings, mark=at position .5 with {\arrow[scale=1]{stealth[reversed]}}, mark=at position 1 with {\arrow[scale=1]{stealth[reversed]}}}, postaction=decorate] (0,0) circle (0.75);
    \draw (1,0.1) node[above] {\color{purple}$\mathcal{L}$};
    \draw [color=purple, thick, decoration = {markings, mark=at position .7 with {\arrow[scale=1]{stealth}}}, postaction=decorate] (0,0.75) -- (0,1.6);
    \fill (0,0.75) circle (1.2pt);
    \draw (0,1.25) node[right] {\color{purple}$\mathcal{L}'$};
    \end{tikzpicture} \,,
\fe
which is crucial for understanding the representation theory of non-invertible symmetries~\cite{Lin:2022dhv}. However, since the lasso action is non-local, i.e., it generally maps a local operator to a non-local one, we do not use it to define non-invertible symmetries. An important, and somewhat surprising, result is that the symmetry can be completely reconstructed from the action \eqref{tdl.action} alone, which maps local operators to local ones. Thus, the data of the lasso action can be derived given the action \eqref{tdl.action} on all local operators.

As emphasized in \cite{Okada:2024qmk}, and first shown in the context of conformal nets by Bischoff and collaborators \cite{Bischoff:2016jmy, Bischoff:2016rpu, Bischoff:2020iac, Bischoff:2021snn, Bischoff:2022fxf}, non-invertible symmetries act locally by \emph{quantum channels}.\footnote{See also~\cite{Lootens:2023wnl, Jones:2024lws, Evans:2025msy} for realizations of fusion category symmetries by quantum channels on spin chains, and~\cite{Jones:2026dcb} for non-invertible ``self-duality'' symmetries acting as automorphisms on the corresponding symmetric subalgebra.} Specifically, the map $\f$ in \eqref{tdl.action} is a unital completely positive (UCP) map that describes a quantum channel in the Heisenberg picture. See Section \ref{sec:background} for background on quantum channels and operator algebras. However, as we will demonstrate through the examples in Section \ref{sec:def}, not every such quantum channel qualifies as a non-invertible symmetry. Our main goal is to find the additional conditions that such a quantum channel must satisfy.

We will argue that a quantum channel $\f: \cA \to \cA$, acting on the algebra of local operators $\cA$, generates a non-invertible symmetry if it satisfies the following three conditions:
\begin{enumerate}
    \item \textbf{Locality:} $\f$ is a finite-range quantum channel in the sense that $\f(\cA_I) \subseteq \cA_{I^{+\ell}}$.
    \item \textbf{Dualizability:} the dual map $\f^\star$ is also a locality-preserving quantum channel.
    \item \textbf{Purity:} if $\f = \f_1 + \f_2$ for completely positive maps $\f_1$ and $\f_2$, then both are multiples of $\f$.
\end{enumerate}
Here, $\cA_I$ denotes the operator algebra supported on an interval $I$, and the dual map is uniquely determined by the relation $\tr(\f(a)b)=\tr(a\f^\star(b))$, where $\tr\!:\cA \to \bC$ is the infinite-temperature state. We motivate and discuss these conditions in detail in Section \ref{sec:def}.

We refer to quantum channels satisfying the three conditions above as \emph{non-invertible quantum cellular automata} (QCAs).\footnote{In~\cite{richter1996ergodicity}, QCAs were defined as UCP maps on the quasi-local algebra that need not be reversible ($*$-automorphisms)~\cite{Schumacher:2004hix}. Our non-invertible QCAs differ in that we further require dualizability and purity, while imposing a weaker locality-preserving condition. In particular, the Kramers--Wannier transformation \eqref{intro.kw} does not obey their locality-preserving condition.} This terminology is motivated by their rigid structure and associated index theory, which will be explored in~\cite{Seifnashri:QCAStructure}. Non-invertible symmetries are then generated by non-invertible QCAs that preserve the Hamiltonian, in the precise sense described in Section~\ref{sec:purity}.

We will show in Section \ref{sec:def} that the Kramers--Wannier (KW) transformation $\dd: \cA \to \cA$ acting on the qubit chain satisfies all three conditions above. Its action is given by
\ie \label{intro.kw}
    \dd(X_j) = Z_{j-1} Z_{j} \,, \qquad \dd(Z_{j}Z_{j+1})= X_{j} \,,
\fe
where it acts multiplicatively on the algebra of $\bZ_2$ even operators (for the $\bZ_2$ spin-flip symmetry), and annihilates all $\bZ_2$ odd operators. For example, $\dd(Z_j)=0$ and $\dd(X_jX_{j+1}) = Z_{j-1}Z_{j+1}$. This map is locality-preserving and dualizable since $\dd$ and its dual, given by $\dd^\star(\mathcal{O}_j)=\dd(\mathcal{O}_{j+1})$, both have range $\ell=1$. Its purity will be demonstrated in Section \ref{sec:purity}. It satisfies the fusion relations
\ie
    \dd^\star \dd = \frac{1}{2}\left( 1+\eta \right)\,, \qquad \dd\dd=\frac{1}{2}T^{-1}\left( 1+\eta \right)\,, \qquad \eta \dd = \dd \eta\,,
\fe
where $\eta$ is the $\bZ_2$ spin-flip automorphism acting as $\eta(X_j)=X_j$ and $\eta(Z_j)=-Z_j$, and $T$ is the lattice translation acting as $T(\mathcal{O}_j) = \mathcal{O}_{j+1}$.

We emphasize that our locality requirement for non-invertible QCAs is weaker than the standard one for invertible QCAs. We require $\f(\cA_R) \subseteq \cA_{R^{+\ell}}$ only for topologically trivial regions $R$. Whereas ordinary QCAs must satisfy this condition for arbitrary regions, including disconnected ones. The Kramers--Wannier map illustrates this distinction since $\dd(Z_1Z_j)=X_1X_2 \cdots X_{j-1}$ is supported on the entire interval rather than at its endpoints.

In higher dimensions, this weaker locality condition allows transformations that are invertible but are not ordinary QCAs. For example, the Wegner duality discussed in Section~\ref{sec:wegner} is an automorphism of the constrained observable algebra in infinite volume, yet maps a loop operator to a membrane operator filling the loop. It therefore satisfies our locality requirement but fails the stronger condition for regions with nontrivial topology. Thus, the term ``non-invertible QCA'' does not necessarily imply that the induced map on local operators is non-invertible. Rather, the terminology is motivated by the non-invertibility of the associated defects and their boundary algebras.

\subsection*{Symmetry operator vs.\ symmetry transformation}

There is an important dichotomy for working with symmetry operators and symmetry transformations that we want to emphasize here. When working with a symmetry operator $\op: \cH \to \cH$, it is important to stay on a finite lattice so that there is a canonical (and separable) Hilbert space $\cH$. On the other hand, for symmetry transformations described by a quantum channel $\dd: \cA \to \cA$, it is important to stay on the infinite lattice, or on a lattice that has trivial topology.

For example, while the KW \textit{operator} $\mathsf{D}$ is well-defined on a finite circle, the corresponding KW \textit{transformation} $\cal D$ is not. This is because on a periodic chain with $L$ sites, \eqref{intro.kw} leads to $\dd(Z_1 Z_{L/2}) = X_1 X_2 \cdots X_{L/2-1}$ and also at the same time to $\dd(Z_1 Z_{L/2}) = X_{L/2} X_{L/2+1} \cdots X_{L}$. In general, the symmetry transformation \eqref{tdl.action} is not globally well-defined on regions with non-trivial topology such as the circle.

Relatedly, the symmetry operator $\op$ can annihilate states, whereas the symmetry transformation $\dd$ does not. In particular, $\op$ annihilates any state that is odd under the $\bZ_2$ symmetry $\eta = \prod_j X_j$ since $\op^\dagger\op = 1+\eta$. In contrast, $\dd$ is a quantum channel and maps normalized states to normalized states (see Section \ref{sec:qc}). 
For instance, on an infinite lattice, $\dd$  maps the state $\ket{\cdots ++-++\cdots}\bra{\cdots ++-++\cdots}$ with a single $\bZ_2$ charge into the mixed state
\ie
    \frac{1}{2}\ket{\cdots 00011\cdots}\bra{\cdots 00011\cdots} + \frac{1}{2}\ket{\cdots 11100\cdots}\bra{\cdots 11100\cdots}
\fe
of single domain-wall states. On a periodic lattice, there is no state with a single domain wall. Relatedly, $\ket{\cdots ++-++\cdots}$ is $\bZ_2$-odd and is annihilated by the operator $\mathsf{D}$.

Focusing on a local region $I$, the relation between the symmetry operator $\op$ and the symmetry transformation $\dd$ is given by
\ie
    \sum_i a_i \op b_i = 0 \qquad \Leftrightarrow \qquad \sum_i a_i \dd(b_i x) = 0 \text{ for all } x \in \cA \,,
\fe
where $a_i,b_i \in \cA_I$. Pictorially,
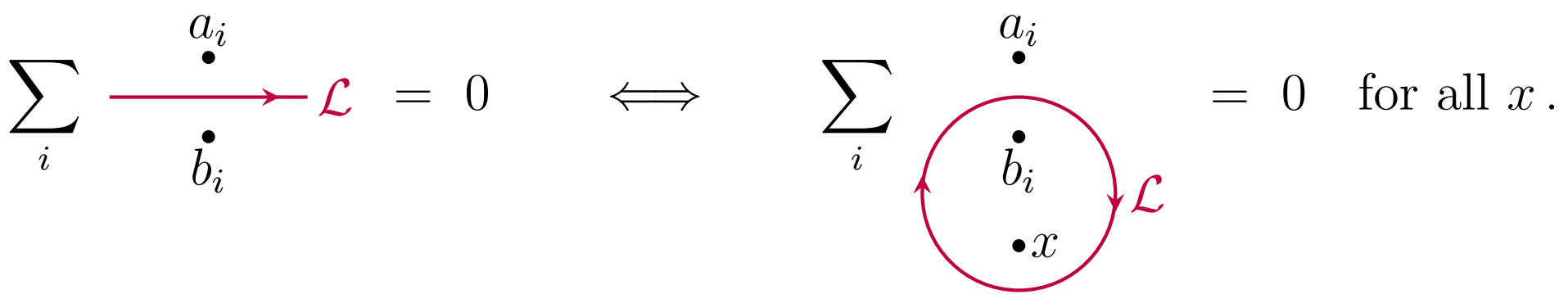
\ie \label{op.vs.map}
\sum_i ~
\begin{tikzpicture}[scale=1,baseline=-.1cm]
    \draw [color=purple, thick, decoration = {markings, mark=at position .86 with {\arrow{stealth}}}, postaction=decorate] (-.8,0) -- (.8,0);
    \draw (.8,0) node[right, inner sep=2pt] {\color{purple}$\mathcal{L}$};
    \fill (0,.32) circle (1.5pt); \draw (0,.32) node[above, inner sep=2.5pt] {$a_i$};
    \fill (0,-.32) circle (1.5pt); \draw (0,-.32) node[below, inner sep=2.5pt] {$b_i$};
\end{tikzpicture}
~=~ 0
\qquad \Longleftrightarrow \qquad
\sum_i ~
\begin{tikzpicture}[scale=1,baseline=-.1cm]
    \draw [color=purple, thick, decoration = {markings, mark=at position .5 with {\arrow[scale=1]{stealth[reversed]}}, mark=at position 1 with {\arrow[scale=1]{stealth[reversed]}}}, postaction=decorate] (0,-.78) circle (0.78);
    \draw (.78,-.78) node[right, inner sep=3pt] {\color{purple}$\mathcal{L}$};
    \fill (0,.32) circle (1.5pt); \draw (0,.32) node[above, inner sep=2.5pt] {$a_i$};
    \fill (0,-.32) circle (1.5pt); \draw (0,-.32) node[below, inner sep=2.5pt] {$b_i$};
    \fill (0,-1.2) circle (1.5pt); \draw (0,-1.2) node[right, inner sep=2.5pt] {$x$};
\end{tikzpicture}
~=~ 0 \quad \text{for all } x \,.
\fe
When $\dd$ is pure, the space $\{ \sum_i a_i \otimes b_i \in \cA \otimes \cA \mid \sum_i a_i \dd(b_i x) = 0 \text{ for all } x \in \cA \}$ uniquely determines $\dd$~\cite{Seifnashri:QCAStructure}. For example, the KW symmetry operator satisfies $\op X_j = Z_{j-1}Z_j \op$ and $\op Z_j Z_{j+1} = X_j \op$.

\subsection*{Symmetry defects and sequential circuits}

As mentioned above, the ultimate way to ensure the locality of a non-invertible symmetry is for it to admit a local topological defect realization. In a quantum lattice system, a topological\footnote{Since all the defects we consider here are topological, we often drop the adjective topological.} defect corresponds to a modification of the Hilbert space and the Hamiltonian that can be moved via local unitary operators~\cite{Hauru:2015abi, Aasen:2016dop, Aasen:2020jwb, Cheng:2022sgb, Seifnashri:2023dpa, Sinha:2023hum, Seiberg:2024gek, Sinha:2025jhh}. Moving the defect across the space via these local unitary operators generates a \emph{sequential circuit}~\cite{Chen:2023qst, Vanhove:2024lmj, Tantivasadakarn:2025txn, Inamura:2026hif, Lu:2026rhb, Sato:2026rpj}. Conversely, having a sequential circuit, we can define defect Hamiltonians and defect Hilbert spaces.\footnote{See, for instance, the defect creation homomorphisms discussed in \cite{Seifnashri:2025fgd}.}

In Section \ref{sec:defects}, we construct non-invertible QCAs from topological defects, or equivalently sequential circuits, that are \emph{dualizable} and \emph{irreducible}. The dualizability of the defect is essential to get a locality-preserving quantum channel. Otherwise, as demonstrated by the example in Section \ref{sec:kw.deformed}, the associated sequential circuit need not lead to a locality-preserving quantum channel. 

In the forthcoming work~\cite{Seifnashri:QCAStructure}, we will demonstrate that any non-invertible QCA with a finite quantum dimension admits a dualizable and irreducible topological defect. This will show that our conditions are both necessary and sufficient, and hence non-invertible QCAs generate non-invertible symmetries.

The construction of the defect allows us to discuss the index theory of non-invertible QCAs that classifies them up to blend equivalence. Two non-invertible QCAs are said to be blend equivalent if there is another one interpolating between them across a finite codimension-one region in space~\cite{Freedman:2019ncn}. The indices include a generalization of the GNVW index~\cite{Gross:2011yvb} and the quantum dimension, which we will show are blend invariants in~\cite{Seifnashri:QCAStructure}. This will also generalize the correspondence between invertible QCAs and invertible subalgebras~\cite{Haah:2022abh, Jones:2026gmd}. Indices of non-invertible symmetry operators in 1+1 dimensions have recently been studied assuming the existence of topological defects~\cite{Inamura:2026hif, Wen:2026ncw}. However, in these works the indices are not shown to be blend invariants for quantum channels, as they concern only defects and MPOs. For automorphisms acting on certain subalgebras of local operators, a generalization of the GNVW index and the non-invertible symmetry they generate have been discussed in~\cite{Jones:2023imy, Ma:2024ypm, Jones:2024lws}.

A non-invertible symmetry can have three different realizations: as a local transformation, a defect, and an operator. To distinguish them, we use the following notation:
\ie
\underbrace{\dd}_{\text{UCP map}},
\qquad
\underbrace{\defect}_{\text{symmetry defect}},
\qquad
\underbrace{\op}_{\text{symmetry operator}},
\fe
where we use $\op^{\dagger}$ for the operator adjoint, $\overline{\defect}$ for the dual defect, and $\dd^{\star}$ for the dual map.

\subsection*{Summary of results and the outline}

In Section \ref{sec:background}, we review background on operator algebras, particularly C$^*$-algebras, relevant to discussing quantum states and quantum channels on infinite lattice systems. In particular, we review the notion of the multiplicative domain of quantum channels, which is extremely useful for non-invertible symmetries.

Section \ref{sec:def} contains our main result, which is the definition of non-invertible QCAs. We motivate the defining properties of non-invertible QCAs physically, and demonstrate their necessity by discussing examples that violate them. Two bonus results of this section are as follows. First, we clarify and relate different notions of symmetric Hamiltonians on infinite spin chains. Second, we show that if $\f_1+\f_2$ is a symmetry, then $\f_1$ and $\f_2$ must individually be symmetries.

In Section \ref{sec:onsite}, we discuss various examples of on-site non-invertible symmetries. Surprisingly, we find on-site non-invertible symmetries that are maximally anomalous in the sense that the only symmetric Hamiltonian is the constant one. We also construct a large class of non-invertible symmetries characterized by a finite collection of unitary matrices, with $\mathrm{Rep}(G)$ arising as a special case.
Section \ref{sec:defects} discusses topological defects and sequential circuits and their relation to non-invertible QCAs. In Section \ref{sec:higher.d}, we discuss examples in higher dimensions. Appendices \ref{app:choi} and \ref{app:tp} discuss results on the multiplicative domain of quantum channels. Appendices \ref{app:symm.H}, \ref{app:lemma}, and \ref{app:repG} contain technical derivations of some of the results in the main text. Appendix \ref{app:dipolar} contains another example in 1+1d. Finally, Appendix \ref{app:dsMPO} relates our perspective on non-invertible symmetries on the lattice to that of Tantivasadakarn--Liu--Chen in \cite{Tantivasadakarn:2025txn}.

\section{Background on operator algebras \label{sec:background}}

In this section, we review the background on the operator-algebraic formulation of quantum systems, which is relevant to our discussion of infinite quantum spin chains.

\subsection{Infinite quantum spin chains}

We work mainly with spin chains on the infinite lattice $\Lambda = \bZ$. In this setting, there is no canonical separable Hilbert space unless we choose a state, or more precisely, its superselection sector. Therefore, instead of working with Hilbert spaces, we work with the C$^*$-algebra of (quasi-)local operators, which we denote by $\cA$.\footnote{In our convention, we use $a^\dagger$ for the adjoint of $a$ in the C$^*$-algebra. We write $\f^\star$ for the adjoint/dual of a map $\f : \cA \to \cA$ (see Section \ref{sec:dual}).} For references, see \cite{bratteli2012operator, Naaijkens:2013vqa, Tasaki:2022gka}.

We assume that there is a finite-dimensional Hilbert space $\cH_j = \bC^{d_j}$ at each site $j \in \bZ$. The on-site operator algebra is given by the matrix algebra $\cA_j = B(\cH_j) = \bC^{d_j \times d_j}$. The operator algebra of the infinite chain is defined by first taking the union of all finitely supported operators to obtain $\cA_\text{loc}:=\bigcup_{I \subset \bZ} \cA_I$, where $\cA_I = \bigotimes_{j \in I} \cA_j$ is the operator algebra supported on the finite interval $I \subset \bZ$. Then, one completes this space in the operator-norm topology to get the (uniformly hyperfinite)
C$^*$-algebra of \emph{quasi-local} operators
\ie
    \cA = \overline{ \bigcup_{I \subset \bZ} \bigotimes_{j\in I} \cA_j }^{\| \cdot \|} \,,
\fe
where $I \subset \bZ$ runs over finite intervals. The operator norm $\|\cdot\|$ is the usual matrix spectral norm $\|\mathcal{O}\| = \sup_{\psi \in \cH_I} \sqrt{\langle \mathcal{O}\psi|\mathcal{O}\psi\rangle/\langle\psi|\psi\rangle}$ for a local operator $\mathcal{O} \in B(\cH_I)$.
As a result of the completion, the algebra $\cA$ contains quasi-local operators, which are not strictly local and have tails going to infinity. The distinction between the local and quasi-local algebras will matter only in a few places, such as Sections~\ref{sec:dual} and~\ref{sec:kw.deformed}. However, we ensure that our formalism is meaningful in the thermodynamic limit in the presence of quasi-local operators.

\subsubsection*{Quantum states of operator algebras}

In our setup, quantum states are defined by the expectation values of local operators rather than by vectors in a Hilbert space. That is, a quantum state is defined as a normalized positive functional $\psi: \cA \to \bC$. In other words, $\psi$ maps a local operator $a \in \cA$ to its expectation value $\psi(a)=\langle a \rangle_\psi$ such that the map is
\begin{itemize}
    \item \textbf{Normalized:} $\psi(1)=1$.
    \item \textbf{Positive:} $\psi(a^\dagger a) \geq 0$ for all $a \in \cA$.
\end{itemize}
In a quantum mechanical system with a finite-dimensional Hilbert space $\cH$, the choice of $\psi $ is equivalent to a density matrix $\rho: \cH \to \cH$, where $\psi(\mathcal{O}) = \Tr_\cH(\mathcal{O} \rho)$ for any operator $\mathcal{O}$. The state $\psi$ being normalized and positive is equivalent to the standard conditions of the density matrix having trace one and being positive: $\Tr_\cH(\rho)=1$ and $\bra{v}\rho \ket{v} \geq 0$ for all $\ket{v} \in \cH$.

To define quantum states, we do not need a Hilbert space and only need the C$^*$-algebra $\cA$. However, choosing a quantum state $\psi: \cA \to \bC$, one can construct a canonical Hilbert space using the Gelfand–Naimark–Segal (GNS) construction. This Hilbert space describes the superselection sector of $\psi$, i.e., the states that are related to it by quasi-local perturbations. More precisely, states in the GNS Hilbert space are given by acting with a quasi-local operator $a \in \cA$ on the state $\psi$ to find $\ket{a \psi}$, equipped with the inner product $\langle a\psi | b\psi \rangle = \psi(a^\dagger b)$. We then mod out by the null states satisfying $\langle a\psi | a\psi \rangle = \psi(a^\dagger a) = 0$ and then Cauchy complete this vector space with respect to the inner product.

\subsubsection*{The infinite-temperature state}

An interesting feature of spin chains with a finite-dimensional Hilbert space at each site is the existence of the infinite-temperature state, or equivalently, the trace on $\cA$ denoted by
\ie \label{tr.state}
    \tr:\cA \to \bC \,.
\fe
Here, $\tr:\cA \to \bC$ denotes the normalized trace, also referred to as the tracial state. Physically, this corresponds to the infinite-temperature state. It is first defined on local operators and then uniquely extended to quasi-local operators by continuity. Specifically, for a local operator $\mathcal{O} \in \cA_I = B(\cH_I)$, the trace is given by $\tr(\mathcal{O}) = \Tr_{\!\cH_I}(\mathcal{O}) / \text{dim}(\cH_I)$. The trace extends continuously to $\cA$ since $ |\tr(a) | \leq \| a \| $, where $\| \cdot \|$ denotes the operator norm.

Applying the GNS construction to the tracial state \eqref{tr.state}, we obtain its GNS Hilbert space $\cH_\mathrm{tr}$ and an injective representation $\pi_{\tr}:\cA \to B(\cH_{\tr})$. A useful algebra that will appear in our discussion later is the hyperfinite type II$_1$ factor $\mathscr{R}=\pi_{\tr}(\cA)'' \subset B(\cH_{\tr})$ acting on the GNS Hilbert space of the tracial state. Operators in $\mathscr{R}$ can be described as limits of uniformly operator-norm bounded sequences of local operators. More precisely, let a sequence $a_1,a_2, a_3,\cdots \in \cA_\text{loc}$ satisfy
\ie \label{convergence.in.R}
    \sup_n \|a_n\| < \infty \qquad \text{and} \qquad \lim_{n,m\to\infty}\|a_n-a_m\|_2 = 0 \,,
\fe
where $\|a\|_2 = \sqrt{\tr(a^\dagger a)}$ denotes the $L^2$-norm of $a \in \cA$. Then $\pi_{\tr}(a_n)$ converges in the strong operator topology on $B(\cH_{\mathrm{tr}})$ to an operator in $\mathscr{R}$, and every operator in $\mathscr{R}$ arises in this way. Two such sequences $(a_n)$ and $(b_n)$ represent the same operator in $\mathscr{R}$ precisely when $\|a_n-b_n\|_2 \to 0$.

\subsection{Quantum channels in the Heisenberg picture} \label{sec:qc}

Unitary transformations are the most general linear operations that act on pure states and preserve probabilities. If we also allow for non-pure quantum states that may be coupled to an environment, then the most general linear operation that preserves probability is a quantum channel. They are often discussed in the Schr\" {o}dinger picture, where they act on density matrices. However, the locality properties of a quantum channel are obscured in the Schr\" {o}dinger picture. Thus, we work in the Heisenberg picture, in which quantum channels act on observables and map local operators to local operators. Moreover, in the thermodynamic limit, there are no density matrices, and quantum channels are most naturally described in the Heisenberg picture.

Quantum channels in the Heisenberg picture are described by unital completely positive (UCP) maps acting on an operator algebra $\cA$. More precisely, a quantum channel is a linear map $\f : \cA \to \cA$ that is
\begin{itemize}
    \item \textbf{Unital:} It preserves the identity: $\f(1)=1$.
    \item \textbf{Completely positive:} $\f \otimes \mathbbm{1}_{N \times N} : \cA \otimes \bC^{N\times N} \to \cA \otimes \bC^{N\times N}$ is a positive map for all $N \in \mathbb{N}$. 
\end{itemize}
The unital condition guarantees that the normalization of states is preserved under the action of the quantum channel $\f$ on states given by
\ie
    \psi \mapsto \psi' = \psi \circ \f\,,
\fe
since $\psi'(1) = \psi(\f(1)) = \psi(1)$. Because any positive operator in a C$^*$-algebra takes the form $a^\dagger a$, the positivity of $\f \otimes \mathbbm{1}_{N \times N}$ is equivalent to requiring $\sum_{m,n =1}^N\f(a_m^\dagger a_n) \otimes \ket{m}\!\bra{n} \in \cA \otimes \bC^{N\times N}$ to be a positive operator. Namely,
\ie
    \sum_{m,n = 1}^N b_m^\dagger \f(a_m^\dagger a_n) b_n \geq 0 \qquad \text{for }a_1, \cdots, a_N, b_1, \cdots, b_N \in \cA \,.
\fe

In a quantum mechanical system with a finite-dimensional Hilbert space $\cH$, quantum channels can be described in the Schr\" {o}dinger picture as completely-positive trace-preserving (CPTP) linear maps that act on density matrices. Namely, given a UCP map $\f$ describing a quantum channel in the Heisenberg picture, there exists a unique CPTP map $\f^\star$ which describes the same quantum channel in the Schr\" {o}dinger picture. The action of the quantum channel on states in the Schr\" {o}dinger picture is given by $\rho \mapsto \rho' = \f^\star(\rho)$. Thus the relation between $\f$ and $\f^\star$ is given by
\ie
    \Tr_{\mathcal{H}} \big( \f(\mathcal{O}) \rho \big) = \Tr_\cH \big( \mathcal{O} \f^\star (\rho) \big) \qquad \text{for all density matrices } \rho\,.
\fe
Indeed, $\f$ is unital if and only if $\f^\star$ is trace-preserving.

\subsubsection*{Invertible symmetries}

A unitary symmetry of an infinite quantum spin chain is described by a $*$-automorphism ($\dagger$-automorphism) of the operator algebra, which is an invertible linear map $\alpha:\cA \to \cA$ such that
\ie
    \alpha(ab) = \alpha(a) \alpha(b) \qquad \text{and} \qquad \alpha(a^\dagger) = \alpha(a)^\dagger\,,
\fe
for all $a,b\in \cA$. As we will show below, $*$-automorphisms are the same as invertible quantum channels whose inverse is also a quantum channel. In a quantum mechanical system with a Hilbert space $\cH$, a unitary symmetry corresponds to a unitary operator $U: \cH \to \cH$ acting by the $*$-automorphism $\alpha_U$ where
\ie \label{unitary.symm}
 \alpha_U(a) = U a U^\dagger \,.
\fe

\subsection{Multiplicative domain of quantum channels}

A quantum channel generally does not preserve multiplication, but it acts multiplicatively on a distinguished subalgebra called its \emph{multiplicative domain}~\cite{choi1974schwarz} (see also \cite{paschke1972completely} and \cite{Paulsen_2003, rahaman2017multiplicative} for reviews). As we will see, this notion is extremely useful for studying the action of non-invertible symmetries on local operators.

Given a UCP map $\f : \cA \to \cA$, the Kadison--Schwarz inequality states that~\cite{choi1974schwarz}
\ie \label{kd.inequality}
    \f(a^\dagger a) \geq \f(a^\dagger) \f(a) \qquad \text{for } a\in \cA\,.
\fe
If equality holds for $a \in \cA$, then $\f(ba) = \f(b) \f(a)$ for all $b \in \cA$.\footnote{The Kadison–Schwarz inequality and this statement hold more generally for unital 2-positive maps.} Thus, $a$ and $a^\dagger$ saturate the inequality if and only if $a$ is inside the \emph{multiplicative domain} of $\f$ denoted by
\ie
    \cA_\f = \big\{ a \in \cA \mid \f(a b) = \f(a) \f(b) \text{ and } \f(b a) = \f(b) \f(a) \text{ for all } b \in \cA \big\} \,.
\fe
For a proof, see Appendix \ref{app:choi}.

When $\f$ is associated with a topological line defect $\mathcal{L}$, the multiplicative domain has the following nice geometric picture. A local operator $\mathcal{O}$ is inside the multiplicative domain of $\f$ if and only if
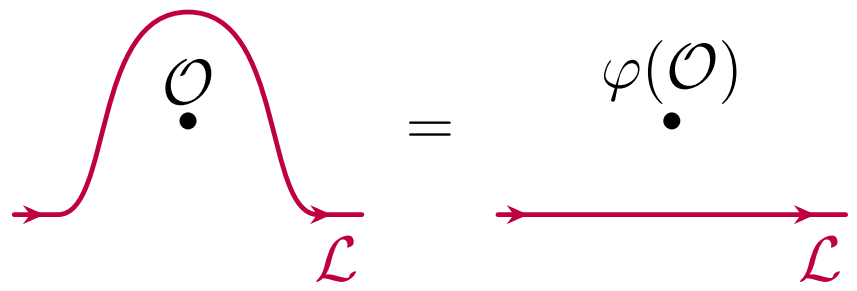
\begin{equation}
\begin{tikzpicture}[
    x=1cm, y=1cm,
    baseline=(eq.base),
    mdline/.style={
        draw=purple,
        line width=.9pt,
        line cap=round,
        line join=round,
        postaction={decorate},
        decoration={
            markings,
            mark=at position .20cm with
                {\arrowreversed{Stealth[length=4pt,width=3.5pt]}},
            mark=at position -.20cm with
                {\arrowreversed{Stealth[length=4pt,width=3.5pt]}}
        }
    },
    oplabel/.style={above=3pt,inner sep=0pt},
    linelabel/.style={purple,below=4pt,inner sep=0pt}
]
    \begin{scope}[xshift=-1.6cm]
        \draw[mdline]
            (1.15,-.62) -- (.86,-.62)
            .. controls (.50,-.62) and (.64,.72) .. (0,.72)
            .. controls (-.64,.72) and (-.50,-.62) .. (-.86,-.62)
            -- (-1.15,-.62);

        \node[linelabel] at (.98,-.62) {$\mathcal L$};
        \fill (0,0) circle (1.6pt);
        \node[oplabel] at (0,0) {$\mathcal O$};
    \end{scope}

    \node (eq) at (0,-.08) {$=$};

    \begin{scope}[xshift=1.6cm]
        \draw[mdline] (1.15,-.62) -- (-1.15,-.62);

        \node[linelabel] at (.98,-.62) {$\mathcal L$};
        \fill (0,0) circle (1.6pt);
        \node[oplabel] at (0,0) {$\varphi(\mathcal O)$};
    \end{scope}
\end{tikzpicture} ~.
\end{equation}
Consequently, when $\f$ is represented by a non-invertible symmetry operator $\mathsf{L}:\cH \to \cH$ acting on a Hilbert space $\cH$, a local operator $\mathcal{O}$ is inside the multiplicative domain of $\f$ iff
\ie
    \mathsf{L} \, \mathcal{O} = \f(\mathcal{O}) \, \mathsf{L} \,.
\fe
This is to be contrasted with the general case, where
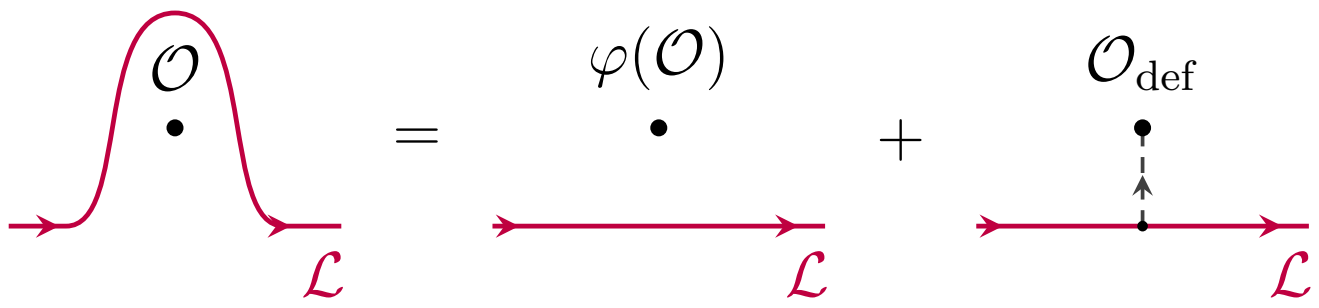
\begin{equation}
\begin{tikzpicture}[
  x=1cm, y=1cm, baseline=-.12cm,
  symmetry/.style={
    purple, line width=.9pt,
    postaction={decorate},
    decoration={
      markings,
      mark=at position .08 with {\arrowreversed{stealth}},
      mark=at position .92 with {\arrowreversed{stealth}}
    }
  },
  defect/.style={
    black!75, densely dashed, line width=.7pt,
    postaction={decorate},
    decoration={
      markings,
      mark=at position .52 with {\arrow{stealth}}
    }
  }
]

\begin{scope}[xshift=-3.2cm]
  \draw[symmetry]
    (1.10,-.65) -- (.72,-.65)
    .. controls (.30,-.65) and (.54,.76) .. (0,.76)
    .. controls (-.54,.76) and (-.30,-.65) .. (-.72,-.65)
    -- (-1.10,-.65);
  \node[purple,below] at (.98,-.68) {$\mathcal L$};
  \fill (0,0) circle (1.6pt);
  \node[above=3pt] at (0,0) {$\mathcal O$};
\end{scope}

\node at (-1.60,-.08) {$=$};

\begin{scope}
  \draw[symmetry] (1.10,-.65) -- (-1.10,-.65);
  \node[purple,below] at (.98,-.68) {$\mathcal L$};
  \fill (0,0) circle (1.6pt);
  \node[above=3pt] at (0,0) {$\varphi(\mathcal O)$};
\end{scope}

\node at (1.60,-.08) {$+$};

\begin{scope}[xshift=3.2cm]
  \draw[symmetry] (1.10,-.65) -- (-1.10,-.65);
  \node[purple,below] at (.98,-.68) {$\mathcal L$};
  \draw[defect] (0,-.65) -- (0,0);
  \fill (0,-.65) circle (1pt);
  \fill (0,0) circle (1.6pt);
  \node[above=3pt] at (0,0) {$\mathcal O_{\mathrm{def}}$};
\end{scope}
\end{tikzpicture} ~.
\end{equation}

As an application, we show that an invertible quantum channel whose inverse is also a quantum channel is a $*$-automorphism. First, note that positivity and linearity imply that $\f$ is $*$-preserving. This is because any operator $x \in \cA$ can be written as $x = (a-b)+i(c-d)$, where $a,b,c,d \in \cA$ are all positive operators, thus $\f(x^\dagger) = \f(a-b-ic+id) = \f(a)-\f(b)-i\f(c)+i\f(d) = \f(x)^\dagger$. The non-trivial step is to show that such a channel is multiplicative. To show the latter, we only need to show $\f(a^\dagger a) = \f(a^\dagger) \f(a)$ for all $a\in \cA$. Using the Kadison–Schwarz inequality for $\f$ and its inverse $\f^{-1}$, we find
\ie
    \f(a^\dagger a) \geq \f(a^\dagger) \f(a) = \f \f^{-1}(\f(a^\dagger) \f(a)) \geq  \f \left(\f^{-1}\f(a^\dagger) \f^{-1}\f(a) \right) = \f(a^\dagger a) \,.
\fe
Thus, all the inequalities above must in fact be equalities. Therefore $\f$ is multiplicative, and hence a $*$-automorphism.

\section{Definition of non-invertible QCA}\label{sec:def}

In this section, we discuss the definition of non-invertible QCAs as \emph{locality-preserving quantum channels} that are \emph{dualizable} and \emph{pure}. We discuss each of these three conditions in separate subsections: we motivate them, discuss how invertible symmetries and the Kramers--Wannier symmetry fit them, and finally discuss non-examples that violate them.

\subsection{Locality-preserving quantum channel}\label{sec:LQC}

Quantum channels are the most general linear maps on quantum states that preserve probability, i.e., map normalized states to normalized states. We consider quantum channels in the Heisenberg picture, where they act on local operators rather than density matrices. 

Given the algebra of local operators $\cA$, a locality-preserving quantum channel is a linear unital completely positive (UCP) map $\f: \cA \to \cA$ that maps local operators to local operators with a finite range $\ell \in \bZ_{\geq 0}$. Namely,
\ie
    &\sum_{m,n = 1}^{N} b_m^\dagger \f(a_m^\dagger a_n) b_n \geq 0 \qquad &&\text{for all } a_1,\cdots ,a_N,b_1,\cdots,b_N \in \cA\,, \\
    &\f(1)=1 \,,\\
    &\f(\cA_I) \subseteq \cA_{I^{+\ell}} \qquad &&\text{for all finite intervals } I\subset \bZ \,.
\fe
The first condition is complete positivity, which is equivalent to positivity of the map $\f \otimes \mathbbm{1}: \cA \otimes \bC^{N \times N} \to \cA \otimes \bC^{N \times N}$ for each $N \in \mathbb{N}$. The second condition is unitality, which ensures that the normalization of quantum states/probabilities is preserved. Finally, the last condition is the finite-range condition, where $\cA_I \subset \cA$ denotes the algebra of local operators supported on a finite interval $I = [j_1,j_2]$, and $I^{+\ell}:=[j_1 -\ell,j_2+\ell]$ is its $+\ell$ neighborhood. We refer to a finite-range quantum channel with $\ell=0$ as acting \textit{on-site}. Interestingly, on-site non-invertible symmetries can be anomalous, as illustrated by the example in Section \ref{anomalous.onsite}. In higher dimensions, finite intervals $I$ must be replaced with finite regions that have the topology of a ball, of any shape.

Importantly, such quantum channels need \emph{not} preserve \emph{short-range correlation} and may not be globally well-defined on the circle: 
\begin{itemize}
   \item \textbf{Locality:} We emphasize that our locality-preserving condition holds only for intervals and need not hold for disconnected regions. For instance, as we will see later, for the Kramers--Wannier non-invertible symmetry we have $\dd(Z_{j_1} Z_{j_2}) = X_{j_1} X_{j_1+1} \cdots X_{j_2-1}$. The image of $Z_{j_1} Z_{j_2}$ is not supported on a finite neighborhood of sites $j_1$ and $j_2$; rather, it is supported on a finite neighborhood of the entire interval $I=[j_1,j_2]$.

Because the locality-preserving condition only holds for intervals, such quantum channels can generate long-range correlations. In particular, they can map states with finite correlation lengths to states with infinite correlation lengths. For instance, the Kramers--Wannier symmetry maps the product state with $\langle X_j \rangle_{+} = 1$ to the GHZ state, characterized by $\langle Z_j \rangle_{\text{GHZ}} = 0$ and $\langle Z_j Z_{j+1} \rangle_{\text{GHZ}} = 1$, which is long-range correlated.\footnote{The authors of Ref.~\cite{Piroli:2020jqb} define a ``causality-preserving quantum channel" as a channel that maps any operator supported in a region $R$ to an operator supported within a thickened region $R^{+\ell}$. In contrast, as discussed above, we impose this requirement only for connected intervals, not for arbitrary regions. We also note that their notion of a ``locality-preserving quantum channel" differs from ours as it does not generate long-range entanglement. Thus, our locality-preserving condition is weaker than both their locality-preserving and causality-preserving conditions.
}

    \item \textbf{Global aspects:} The action of non-invertible symmetries on local operators is not globally well-defined on a circle.\footnote{A globally well-defined quantum channel on the circle was constructed in \cite{Bartsch:2026wqq} by incorporating the twisted sectors through the action of the tube algebra \cite{Lin:2022dhv}. However, this quantum channel does not map local operators to local operators.} Thus, the quantum channels that we are interested in are only well-defined on each local patch, or on the infinite lattice $\bZ$. We can consider them as UCP maps acting on the net of local algebras, namely as maps from $\cA_I$ to $\cA_{I^{+\ell}}$.
    
    Alternatively, we can view them as a single UCP map acting on the C$^*$-algebra of quasi-local operators associated with the entire infinite chain (see Section~\ref{sec:background})
    \ie
    \cA = \overline{ \bigcup_{I \subset \bZ} \cA_I}^{\| \cdot \|} ~.
\fe
The action of $\f$ on the local algebra $\cA_\text{loc}=\bigcup_{I \subset \bZ} \cA_I$ extends uniquely to the quasi-local algebra $\cA$ by the continuity of $\f$. Note that every UCP map $\f$ is continuous because it does not increase the operator (spectral) norm (it is contractive): $\| \f(a) \| \leq \| a \|$. The contractivity follows from the Kadison--Schwarz inequality \eqref{kd.inequality}.\footnote{We also use $0\leq a \leq b \Rightarrow \|a\| \leq \|b\|$. Namely, $\| \f(a) \|^2 = \| \f(a)^\dagger \f(a) \| \leq \|\f(a^\dagger a)\| \leq \| \f(\|a\|^2 \mathbbm{1}) \| = \|a\|^2$.}
\end{itemize}

\subsubsection*{Motivation}

A motivation for considering quantum channels is that these operations are the most general \emph{linear} operations on quantum states, with the finite-range/locality-preserving condition following from locality. Note that the locality-preserving condition implies $\f(1) \in \mathbb{C}$, so up to an overall normalization, the locality-preserving condition implies unitality. Another motivation comes from continuum field theory, where non-invertible topological defects act locally by quantum channels. This was first shown in the context of conformal nets in 1+1d continuum by Bischoff and collaborators \cite{Bischoff:2016jmy, Bischoff:2016rpu, Bischoff:2020iac, Bischoff:2021snn, Bischoff:2022fxf}, and more recently pointed out in \cite{Lootens:2023wnl, Okada:2024qmk}.

More precisely, a non-invertible topological line defect $\mathcal{L}$ acts on local operators by the following encircling action:
    \ie \label{encircling}
\f(a) =  \frac{1}{d_\mathcal{L}} ~ 
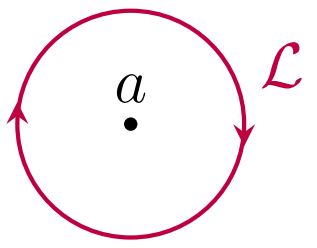
\begin{tikzpicture}[scale=1,baseline=(A.south)]
    \node [draw, fill=black, circle, inner sep=0.75pt] (A) at (0,0)  {};
    \draw (0,0) node [above] {$a$};
    \draw [color=purple, thick, decoration = {markings, mark=at position .5 with {\arrow[scale=1]{stealth[reversed]}}, mark=at position 1 with {\arrow[scale=1]{stealth[reversed]}}}, postaction=decorate] (0,0) circle (0.75);
    \draw (1,0.1) node[above] {\color{purple}$\mathcal{L}$};
    \end{tikzpicture} \,.
    \fe
    We have included the normalization factor $d_{\mathcal{L}} \geq 1$, the quantum dimension of $\mathcal{L}$, to ensure that the map $\f$ is unital.

\subsubsection*{Invertible symmetries}

Invertible symmetries correspond to unitary transformations, i.e., $*$-automorphisms. For instance, a unitary operator $U$ generates a $*$-automorphism from the conjugation action \eqref{unitary.symm}. Such actions are clearly a special case of quantum channels. Thus, our definition includes ordinary invertible symmetries as a special case.

\subsubsection*{Kramers--Wannier transformation}

To verify that the Kramers--Wannier transformation indeed acts locally by a quantum channel $\dd: \cA \to \cA$, we express it as a composition of two locality-preserving quantum channels:
\ie \label{KW.map}
    &\dd =  \alpha_\text{KW} \circ E : \cA \to \cA^\text{even} \subset \cA \,,\\
    &E(a) = \frac{1}{2} \Big( a + \eta(a) \Big) \,, \\
    &\alpha_\text{KW}(X_j) = Z_{j-1} Z_{j}\,, \qquad  \alpha_\text{KW}(Z_{j} Z_{j+1}) = X_{j} \,.
\fe
Here, $E: \cA \to \cA^\text{even}$ is a conditional expectation\footnote{A conditional expectation is a UCP map $E:\cA \to \mathcal{B} \subseteq \cA$ such that $E(b) = b$ for all $b \in \mathcal{B}$ and $E(b_1 a b_2 ) = b_1 E(a) b_2$ for all $a \in \cA$ and $b_1,b_2 \in \mathcal{B}$.} projecting an operator onto its $\bZ_2$ even component:
\ie
    E(a) = \frac{1}{2} \Big( a + \eta(a) \Big) = \frac{a}{2} + \left( \prod_{j \in \text{supp}(a)} X_j \right) \frac{a}{2} \left( \prod_{j \in \text{supp}(a)} X_j \right) \,.
\fe
The map $\eta:\cA \to \cA$ is the automorphism associated with the $\bZ_2$ spin-flip symmetry and $\alpha_\text{KW}: \cA^\text{even} \to \cA^\text{even}$ is an automorphism of the algebra of $\bZ_2$ even operators $\cA^\text{even}$. Since both $E$ and $\alpha_\text{KW}$ are locality-preserving quantum channels, their composition $\dd=\alpha_\text{KW} E$ is also a locality-preserving quantum channel.

In summary, the Kramers--Wannier transformation is a linear map $\dd:\cA \to \cA$ satisfying
\ie \label{KW.map2}
	&\dd(X_j) = Z_{j-1} Z_{j} ~, \qquad &\dd(Z_{j}Z_{j+1})= X_{j} \,, \\
    &\dd(a a') = \dd({a}) \dd({a'}) & \text{for } {a},{a'} \in {\cA^\text{even}}  \,, \\
    &\dd({b}) = 0 & \text{for } {b} \in {\cA^\text{odd}} \,,
\fe
where ${\cA^\text{even}}$ and ${\cA^\text{odd}}$ are the subspaces of $\bZ_2$ even and odd local operators, respectively. 
This determines $\dd$ by linearity since $\cA = {\cA^\text{even}} \oplus {\cA^\text{odd}}$.
In particular, $\dd(Z_j)=0$.

\subsubsection*{Non-example}

Let us also describe a transformation that is not a quantum channel. Consider \emph{non-unitary} but invertible transformations. For example, consider an invertible operator $U$ and its action on local operators given by
\ie
    \f_U(a) := U a U^{-1}\,.
\fe
Since positive maps preserve the adjoint, i.e., $\f(a^\dagger) = \f(a)^\dagger$, $\f_U$ is a positive map if and only if $U^\dagger U = \lambda \mathbbm{1}$ for some $\lambda > 0$. Thus, if $U$ is not proportional to a unitary operator, it corresponds to a non-unitary transformation that would not be a positive map, and in particular will not be a UCP map. An example is $U=e^{\theta Q}$ (as opposed to $U=e^{i\theta Q}$), where $Q$ is a Hermitian conserved charge and $\theta \in \mathbb{R}$.

\subsection{Dualizability \label{sec:dual}}

The second axiom is dualizability. We require that the dual quantum channel, as described below, be well-defined in the thermodynamic limit and locality-preserving.

Let $\f: \cA \to \cA$ be a UCP map acting on an infinite spin chain. Its dual (when it exists\footnote{In general, the map $\f^\star$ might not exist. However, assuming $\f$ is trace-preserving, the dual always exists as a map $\f^\star: \cA \to \mathscr{R} \supset \cA$, where $\mathscr{R}=\pi_{\tr}(\cA)''$ is the hyperfinite type II$_1$ factor.}) is the completely positive map $\f^\star: \cA \to \cA$ characterized by
\ie \label{def.dual}
    \tr\!\big( \f(a) b \big) = \tr\!\big( a \f^\star(b) \big) \quad \text{for all } a,b \in \cA\,.
\fe
Here, $\tr:\cA \to \bC$ is the tracial (infinite-temperature) state \eqref{tr.state}.

While the relation between $\f$ and $\f^\star$ in \eqref{def.dual} resembles the correspondence between the Heisenberg and Schrödinger pictures of a quantum channel, the interpretation here is different. There is no underlying Hilbert space in the present setting, and $\f$ and $\f^\star$ instead represent distinct quantum channels, both in the Heisenberg picture.

Whenever the dual map $\f^\star$ exists, it is uniquely determined by the relation \eqref{def.dual} because the trace is faithful, i.e.\ $\tr(a^\dagger a) =0 \Rightarrow a=0$. Moreover, the dual map is also completely positive. Dualizability requires $\f^\star$ to be locality-preserving. In particular, $\f^\star$ must be unital, which is equivalent to $\f$ being trace-preserving: $\tr(\f(a)) = \tr(a)$. Physically, the latter condition requires the infinite-temperature state to be symmetric/preserved under $\f$.

\subsubsection*{Motivation}

A physical motivation for the dualizability condition is the fact that if $\f$ is a symmetry of a Hamiltonian $H$, then the dual quantum channel $\f^\star$ would also be a symmetry of $H$. Namely,
\ie
    \alpha_t \circ \f = \f \circ \alpha_t \quad \Rightarrow \quad \alpha_t \circ \f^\star = \f^\star \circ \alpha_t \,,
\fe
where $\alpha_t(a)=e^{itH}ae^{-itH}$ is the time-evolution automorphism.

To show this, first note that the condition above is equivalent to
\ie
    \f\big( [H,a] \big) = [H, \f(a)] \qquad \text{for all } a\in \cA_\mathrm{loc}\,.
\fe
Given the condition above, and using the faithfulness and cyclicity of the trace, we find
\ie
   \tr\!\left( \f\big( [H,a] \big) b \right)  &= \tr\!\left( [H, \f(a)] b \right) \quad &\Rightarrow && \quad \tr\!\left( [H,a] \f^\star(b) \right)  &= \tr\!\left( [b, H ] \f(a) \right) \\
   \Rightarrow \quad  \tr\!\left( [ \f^\star(b), H ] a \right)  &=  \tr\!\left( \f^\star([b,H ]) a \right) \quad & \Rightarrow && \quad  [H, \f^\star(b)] &= \f^\star\big( [H,b] \big) \,.
\fe
Thus, if $\f$ commutes with time evolution, then $\f^\star$ does as well. Therefore, it is natural to expect that $\f^\star$ remains well defined in the thermodynamic limit and is also locality-preserving.

\subsubsection*{Invertible symmetries}

For an invertible symmetry associated with an automorphism $\alpha$,  the dual symmetry is simply its inverse: $\alpha^\star = \alpha^{-1}$.

We note that automorphisms of spin chains are automatically trace-preserving since the trace on $\cA$ is unique and automorphisms map tracial states to tracial states. Given the trace-preserving property, it is straightforward to see that the dual equals the inverse. It remains to show that the dual is locality-preserving. We will show that $\alpha^{-1}(\cA_I) \subseteq \cA_{I^{+\ell}}$, given an automorphism $\alpha$ of range $\ell \in \bZ_{\geq 0}$.

Let $a \in \cA_I$ and $b \in \cA_{(I^{+\ell})^\text{c}}$, where $(I^{+\ell})^\text{c}$ denotes the complement of the interval $I^{+\ell}$. Since $\alpha$ has range $\ell$, $a$ commutes with $\alpha(b)$. Applying $\alpha^{-1}$, we conclude that $\alpha^{-1}(a)$ commutes with $b$. As this holds for every $b \in \cA_{(I^{+\ell})^\text{c}}$, it follows that $\alpha^{-1}(\cA_I) \subseteq (\cA_{(I^{+\ell})^\text{c}})'$. By Haag duality, $(\cA_{(I^{+\ell})^\text{c}})' = \cA_{I^{+\ell}}$, and therefore $\alpha^{-1}(\cA_I) \subseteq \cA_{I^{+\ell}}$. This shows that $\alpha^{-1}$ is automatically locality-preserving, a property that follows from Haag duality \cite{Nill:1995nk, Naaijkens:2013vqa, Jones:2023ptg}.

\subsubsection*{Kramers--Wannier transformation}

For the Kramers--Wannier transformation, its dual is given by composition with lattice translation. Specifically, the Kramers--Wannier transformation \eqref{KW.map2} acts as a ``half-translation" to the left, and its dual generates a ``half-translation" in the opposite direction. In particular,
\ie
     \dd^\star(X_j) = Z_{j}Z_{j+1}\,, \qquad \dd^\star(Z_{j} Z_{j+1}) = X_{j+1} \,.
\fe

The dual KW transformation is given by
\ie 
    \dd^\star =  \alpha_\text{KW}^{-1} \circ E \,,
\fe
where the conditional expectation $E : \cA \to \cA^\text{even} \subset \cA$ and the automorphism $\alpha_\text{KW} : \cA^\text{even} \to \cA^\text{even}$ are defined in \eqref{KW.map}. This can be verified explicitly. Indeed,
\ie
    \tr\!\big( \dd(a) \, b \big) &= \tr\!\big( \alpha_\text{KW}(E(a)) \, E(b) \big) = \tr\!\big( E(a) \, \alpha_\text{KW}^{-1}(E(b)) \big) = \tr\!\big( a \, \alpha_\text{KW}^{-1}(E(b)) \big) \\ &= \tr\!\big( a \, \dd^\star(b) \big) \,.
\fe
Here, the first and third equalities follow from the identity $\tr(x y) = \tr(x E(y))$ for $x \in \cA^\text{even}$ and $y \in \cA$, and the second equality uses the fact that the dual of $\alpha_\text{KW}: \cA^\text{even} \to \cA^\text{even}$ is given by its inverse.

\subsubsection*{Non-examples} 

A locality-preserving quantum channel may fail to be dualizable for several reasons. First, it may fail to be trace-preserving. Second, even if it is trace-preserving, its dual may not exist as a well-defined UCP map in the thermodynamic limit. Finally, it may possess a dual UCP map that is not locality-preserving. We will present examples illustrating each of these cases.

\textbf{Non-trace-preserving examples:} The first non-example is the non-trace-preserving UCP map acting on the qubit chain
\ie
    \f_0(a) = \ket{0}\!\!\bra{0}_0 a \ket{0}\!\!\bra{0}_0 + \ket{1}\!\!\bra{0}_0 a \ket{0}\!\!\bra{1}_0 =\bra{0} a \ket{0}_0 \otimes \mathbbm{1}_0 \,,
\fe
whose dual $\f_0^\star(a) = 2 \,\tr_{\!0}(a) \otimes \ket{0}\!\!\bra{0}_0$ is not unital and in particular not locality-preserving. Another example is the quantum channel $\f_\omega:\cA \to \cA$ defined as 
\ie
    \f_\omega(a) = \omega(a) \mathbbm{1} \,,
\fe
where $\omega: \cA \to \mathbb{C}$ is an arbitrary (normalized) state different from the tracial state. The quantum channel above is locality-preserving since it maps every operator into a $\mathbb{C}$-number, but it is not trace-preserving unless $\omega=\tr$.

\textbf{Trace-preserving without UCP dual:} For the second case, let $\cA$ be the quasi-local algebra of the infinite qubit chain, where $X_j, Y_j, Z_j$ are Pauli operators at site $j$. We construct a UCP map $\phi: \cA \to \cA$ as a composition of two locality-preserving quantum channels. Namely, $\phi = f \circ E$ where $E: \cA \to \cA_Z$ is a conditional expectation onto the commutative subalgebra $\cA_Z \subset \cA$ generated by Pauli-$Z$ operators and $f: \cA_Z \to \cA_Z$ is a unital $*$-endomorphism. Specifically, $E = \bigotimes_j E_j$ is an on-site UCP map constructed from local quantum channels
\ie
    E_j(a) = \frac{a+Z_j a Z_j}{2} \,.
\fe
The second map $f: \cA_Z \to \cA_Z$ is a unital $*$-endomorphism of the commutative C$^*$-algebra $\cA_Z$, and can be identified with a (non-reversible) classical cellular automaton acting on the classical configurations of the spins. Its action on Pauli-Z operators at even and odd sites is given by
\ie \label{non.example}
    f(Z_{2j+1})=Z_{2j+1}\,, \qquad f(Z_{2j}) = Z_{2j} \mathsf{CZ}_{2j+1,2j+2}\,,
\fe
where $\mathsf{CZ}_{2j+1,2j+2}=\frac{1+Z_{2j+1}}{2}+Z_{2j+2}\frac{1-Z_{2j+1}}{2}$ is the Controlled-$Z$ gate acting on sites $2j+1$ and $2j+2$. The non-reversibility of $f$ is essential ($f$ is not surjective). Indeed, by the Curtis–Hedlund–Lyndon theorem, the inverse of a reversible and translationally invariant classical cellular automaton is itself locality-preserving, which would furnish a well-defined dual map for $\phi$.

Now let us show that the dual of $\phi$ does not exist as a map from $\cA$ to $\cA$. First note that $\phi=f \circ E$ is trace-preserving since both $E$ and $f$ are trace-preserving. We will show that $\phi^\star(Z_0)$ cannot exist as a quasi-local operator satisfying
\ie
    \tr\!\big( \phi^\star(Z_0) \, a \big) = \tr\!\big( Z_0 \, \phi(a) \big) \qquad \text{for all }a \in \cA\,.
\fe
First, it is clear that $\phi^\star(Z_0) \in \cA_Z$ since $\tr\!\big( \phi^\star(Z_0) (a-E(a)) \big) = 0$ for all $a \in \cA$. Thus,
\ie \label{phi.star.z0}
    \phi^\star(Z_0) &= \sum_{S} \tr\!\big( Z_0 \, \phi(S) \big) \, S \\
    &= \sum_{J=0}^{\infty}
\left( \prod_{j=0}^{J-1} \frac{1 - Z_{2j+1}}{2} \right)
\frac{1 + Z_{2J+1}}{2} Z_0 Z_2 Z_4 \cdots Z_{2J} \,,
\fe
where the sum in the first line runs over all Pauli-$Z$ strings $S$. Only strings of the form $S= \prod_{j=0}^{J} Z_{2j}(Z_{2j+1})^{\epsilon_{2j+1}}$ with $\epsilon_{2j+1} \in \{0,1\}$ contribute to the above expansion. For such strings,
\ie
    \tr\!\big( Z_0 \, \phi(S) \big) = \frac{1}{2} \prod_{j=0}^{J-1}\frac{(-1)^{\epsilon_{2j+1}}}{2}\,,
\fe
which follows from
\ie
    f(Z_0Z_2 \cdots Z_{2J}) = Z_0 Z_{2J+2} \prod_{j=0}^{J} \left( Z_{2j+2} \frac{1+Z_{2j+1}}{2}+\frac{1-Z_{2j+1}}{2} \right) \,.
\fe
We note that $\phi^\star(Z_0)$ is not a quasi-local operator since the sum \eqref{phi.star.z0} does not converge in the operator norm topology. This is because the ``tail''
\ie \label{tail}
    \left( \prod_{j=0}^{J-1} \frac{1 - Z_{2j+1}}{2} \right)
\frac{1 + Z_{2J+1}}{2} Z_0 Z_2 Z_4 \cdots Z_{2J}
\fe
has an operator norm equal to 1 and is hence nonvanishing as $J \to \infty$.\footnote{The normalized Hilbert-Schmidt ($L^2$) norm $\|a\|_2=\sqrt{\tr(a^\dagger a)} \leq \| a \|$ of the tail \eqref{tail} vanishes as $J \to \infty$. Thus, $\phi^\star(Z_0)$ is a well-defined operator inside $\mathscr{R}=\pi_{\tr}(\cA)''$.}

\textbf{UCP dual with infinite range:} Finally, we discuss a locality-preserving quantum channel whose dual exists as a well-defined UCP map that is not locality-preserving. This example is constructed as a truncation of the Kramers--Wannier transformation. Let $E_{\geq 0} : \cA \to \cA^{\eta_{\geq 0}}$ be the conditional expectation onto the subalgebra $\cA^{\eta_{\geq 0}} = \cA_{< 0} \otimes (\cA_{\geq 0})^\text{even}$ invariant under the $\bZ_2$ symmetry $\eta_{\geq 0} = X_0 X_1 X_2 \cdots$ acting on the non-negative sites, where
\ie
    E_{\geq 0}(a) = \frac{1}{2}\left( a + \Big(\prod_{j\geq 0}X_j\Big) a \Big(\prod_{j\geq 0}X_j\Big) \right)\,.
\fe
Consider the $*$-isomorphism $\alpha_{\geq 0}: \cA^{\eta_{\geq 0}} \to \cA$ acting trivially on negative sites and as the Kramers--Wannier transformation on non-negative sites:
\ie
    \alpha_{\geq 0} : \quad \begin{aligned}	
    \mathcal{O}_j &\mapsto \mathcal{O}_j     &&\text{for } j <0 \\
    X_0 &\mapsto Z_0   && \\
    X_{j} &\mapsto Z_{j-1} Z_{j}   &&\text{for } j >0  \\
    Z_j Z_{j+1} &\mapsto X_{j}   &~~&\text{for } j \geq 0
	\end{aligned} ~, 
\fe
where $\mathcal{O}_j$ is any operator supported at site $j$. Using these two UCP maps, we can construct a truncated KW transformation $\dd_{\geq0} = \alpha_{\geq 0} \circ E_{\geq 0}$. The dual UCP map is given by $(\dd_{\geq0})^\star = (\alpha_{\geq 0})^{-1} : \cA \to \cA^{\eta_{\geq 0}} \subset \cA$, which is in fact a homomorphism. Its action is given by
\ie
   (\dd_{\geq0})^\star : \quad \begin{aligned}	
    \mathcal{O}_j &\mapsto \mathcal{O}_j     &&\text{for } j < 0 \\
    X_{j} &\mapsto Z_j Z_{j+1}   &&\text{for } j \geq 0 \\
    Z_{j} &\mapsto X_{0} X_1 \cdots X_{j-1} X_j   &~~&\text{for } j \geq 0
	\end{aligned} ~. \label{DgeqJ}
\fe
As we see, the action of the dual UCP map on $Z_j$ as $j \to +\infty$ is not locality-preserving even though the map $\dd_{\geq0}$ is locality-preserving.

\subsection{Purity}\label{sec:purity}

The final condition we impose is \emph{CP-purity}. Namely, we require $\f: \cA \to \cA$ to be an extremal ray of the convex cone of completely positive (CP) maps from $\cA$ to $\cA$.\footnote{For brevity, we will sometimes drop the prefix ``CP'' and use purity to mean CP-purity.} That is, $\f$ is pure if every decomposition $\f = \f_1 +\f_2$ into CP maps $\f_1, \f_2$ is trivial with $\f_1 = t_1\f$ and $\f_2 = t_2\f$ for numbers $t_1,t_2 \geq 0$ and $t_1+t_2 = 1$.

Note that the purity condition is only meaningful in the thermodynamic limit, where $\cA$ is the algebra of the infinite chain. A quantum channel $\f_I: \cA_I \to \cA_I$ acting on a finite interval $I$ would not be pure unless it is an automorphism. This is because UCP maps acting on finite-dimensional algebras admit a Kraus operator representation, and CP-purity implies that there is only a single Kraus operator. Namely, $\f_I(a)=K a K^\dagger$. However, the unital condition implies that $K$ must be a unitary operator. Thus, a CP-pure UCP map acting on a finite-dimensional algebra is an automorphism. Relatedly, there are no genuine non-invertible symmetries in quantum mechanics without spatial dimensions and locality. Given any pure quantum channel acting on a finite-dimensional Hilbert space, the associated Kraus operator and its Hermitian conjugate are conserved quantities and generate an automorphism.

It is useful to also discuss the notion of \emph{Choi-purity}, which is stronger than CP-purity. For a trace-preserving UCP map $\f$, let
\ie \label{choi.state}
    \omega_\f : \cA \otimes \cA^\mathrm{op} \to \bC \,, \qquad
    \omega_\f(a \otimes b^\mathrm{op}) = \tr\!\big( a \, \f(b) \big) \,,
\fe
be the \emph{Choi state} of $\f$. Here, $\cA^\mathrm{op}$ is the opposite algebra of $\cA$. After choosing a basis, we can identify $\cA$ with $\cA^\mathrm{op}$ via transposition and write $\omega_\f : \cA \otimes \cA \to \bC$ with $\omega_\f(a \otimes b) = \tr\!\big( a \, \f(b^\mathsf{T}) \big)$. The positivity of the state $\omega_\f$ on the doubled algebra is equivalent to the complete positivity of $\f$. The assignment $\f \mapsto \omega_\f$ is linear and also injective since $\tr:\cA \to \bC$ is faithful. Moreover, Choi-purity implies CP-purity:
\ie \label{choi.purity}
    \omega_\f \ \text{is a pure state} \qquad \Longrightarrow \qquad \f \ \text{is CP-pure} \,.
\fe
Indeed, a decomposition $\f = \f_1 + \f_2$ into CP maps gives a decomposition $\omega_\f = \omega_{\f_1} + \omega_{\f_2}$ into positive functionals. Thus, purity of $\omega_\f$ forces $\omega_{\f_i} = t_i \omega_\f$, and then $\tr\big( a(\f_i(b) - t_i \f(b)) \big) = 0$ for all $a,b \in \cA$, so $\f_i = t_i \f$. We call a UCP map Choi-pure if its Choi state is pure.\footnote{One can show that $\f:\cA \to \cA$ is Choi-pure if and only if it remains CP-pure when considered as a UCP map $\widehat{\f}:\cA \to \mathscr{R}$ from the quasi-local algebra $\cA$ to the hyperfinite type II$_1$ factor $\mathscr{R}=\pi_{\tr}(\cA)''$.}

\subsubsection*{Motivation}

The primary motivation for the purity condition comes from continuum field theory, where purity corresponds to the simplicity of topological defects: non-invertible symmetries correspond to simple topological defects, defined as those that cannot be decomposed as a direct sum of other defects.

A second motivation, intrinsic to the lattice setting, is the statement that $\f = \f_1 + \f_2$ is a symmetry only if both $\f_1$ and $\f_2$ are symmetries. This motivates us to focus on pure maps and to regard them as the fundamental symmetries. If this were not the case, we would risk overlooking some interesting symmetries by restricting our attention to pure quantum channels.
 
\begin{tcolorbox}[title=Purity of Fundamental Symmetries]
Let $\f = \f_1 +\f_2$ be a locality-preserving and dualizable quantum channel, where $\f_1$ and $\f_2$ are CP maps. Then $\f$ preserves a Hamiltonian $H$ with finite-range and uniformly bounded interactions only if $\f_1$ and $\f_2$ do. Namely, $\f(H) = H$ and $\f^\star(H) = H$ only if $\f_1(H)=\f_1(1)H=H\f_1(1)$ and $\f_2(H)=\f_2(1)H=H\f_2(1)$.
\end{tcolorbox}

The proof of this statement in finite volume follows from the algebraic properties of the multiplicative domain of UCP maps. In finite volume, the Hamiltonian is a local operator $H \in \cA$. First, as shown in Appendix \ref{app:contractivity}, $\f(H) = H$ implies $H \in \cA_\f$ since $\f$ is a trace-preserving UCP map. Combining this with Choi's theorem on the multiplicative domain of $\f = \f_1 + \f_2$, proven in Appendix \ref{app:md.composite}, we find
\ie \label{h.choi}
    H \in \cA_\f = \Big\{a\in \cA \mid \f_i(ax) = \f(a)\f_i(x) \;\text{ and }\; \f_i(xa) = \f_i(x)\f(a) \;\text{ for all } x\in \cA \Big\}\,.
\fe
Thus given $\f(H)=H$ we conclude $\f_1(H)=\f_1(1)H = H \f_1(1)$ and also $\f_2(H)=\f_2(1)H = H\f_2(1)$. This establishes the statement in finite volume.

The infinite-volume proof is more involved and relies on the extra assumptions of uniform boundedness of the interactions of the Hamiltonian and also $\f^\star(H) = H$. Moreover, in infinite volume, $H$ is not a local operator and is a formal sum of local operators. Thus, equalities such as $\f_i(H)=\f_i(1)H=H\f_i(1)$ make sense only after introducing some boundary cutoff. The idea of the infinite-volume proof is to introduce an appropriate smooth IR cutoff for the Hamiltonian and show that the corresponding Hamiltonian is approximately inside the multiplicative domain of $\f$. 

Let us demonstrate with two examples that both the uniform boundedness assumption and $\f^\star(H) = H$ are necessary. For the uniform boundedness assumption, consider the example with
\ie
    \f = \frac{1}{2}(T+T^{-1}) \qquad \text{and} \qquad H = \sum_j{jZ_j}\,,
\fe
where $T$ is the lattice translation symmetry. Clearly $\f(H)=H$ and $\f^\star(H)=H$ since $\f^\star=\f$, but $T(H)\neq H$. Indeed, the interactions are not uniformly bounded since $\|jZ_j \|=|j|$. For the necessity of the assumption $\f^\star(H) = H$, consider the truncation of lattice translation given in \eqref{dual.shift.truncation}, where
\ie
    \f = (T_{\geq 0})^\star \qquad \text{and} \qquad H = \sum_{j \geq 0} Z_jZ_{j+1} \,.
\fe
We have $\f(H)=H$ because $\tr(Z_j)=0$. However, its CP summands do not preserve $H$ since $(T_{\geq 0})^{-1}(\bra{0} H \ket{0}_0) = H + Z_0 \neq H$. Indeed, this example does not satisfy the condition $\f^\star(H) = H$ and instead $\f^\star(H) = H - Z_0Z_1$, where $\f^\star = T_{\geq 0}$ is given in \eqref{shift.truncation.def}.

The infinite-volume statement follows from Lemma~\ref{lemma.symm.h} proven in Appendix \ref{app:symm.H}. The lemma states that given $\f(H)=H$ and $\f^\star(H)=H$ for a Hamiltonian $H = \sum_j h_j$ with finite-range and uniformly bounded interactions, we have
\ie \label{approximate.choi.theorem}
    \lim_{n \to \infty} \big \| \f_i(a H_n) - \f_i(a) H_n \big \|_2 = \lim_{n \to \infty} \big \| \f_i(H_n a) -  H_n \f_i(a) \big \|_2 = 0 \quad \text{for } a \in \cA \,,
\fe
where $H_n$ is the following cutoff local Hamiltonian
\ie
    H_n = \sum_{j=-n}^n h_j + \sum_{n < |j| < n+n^2} \left(1 - \frac{|j|-n}{n^2} \right)h_j \,. 
\fe
This is basically an infinite-volume version of Choi's theorem \eqref{h.choi}. Setting $a=1$ in \eqref{approximate.choi.theorem} implies that $\f_i(H)=\f_i(1)H=H\f_i(1)$, hence $\f_1$ and $\f_2$ are both symmetries of $H$.

A further consequence of \eqref{approximate.choi.theorem} relates two natural notions of invariance of a Hamiltonian under a quantum channel. Let $H$ be a finite-range Hamiltonian with uniformly bounded interactions, and $\f$ be a locality-preserving, dualizable quantum channel. If both $\f$ and its dual $\f^\star$ preserve $H$, then $\f$ commutes with the time evolution generated by $H$:
\begin{tcolorbox}[title=Symmetric Hamiltonians]
\vspace{-\abovedisplayskip}
\ie
    \f(H)=\f^\star(H)=H \qquad \Longrightarrow \qquad \f([H,a])=[H,\f(a)]\quad\text{for all } a\in\cA_\mathrm{loc} \,.
\fe
\end{tcolorbox}
This result is a consequence of \eqref{approximate.choi.theorem} applied to the case where $\f_1=\f$ and $\f_2=0$:
\ie
    \big\| \f([H,a]) - [H,\f(a)] \big\|_2 &= \lim_{n \to \infty} \big\| \f([H_n,a]) - [H_n,\f(a)] \big\|_2 \\
    &\leq \lim_{n \to \infty} \left( \big\| \f(H_na) - H_n\f(a) \big\|_2 + \big\| \f(aH_n) - \f(a)H_n \big\|_2 \right) \\ &=0\,,
\fe
where in the second line we have used the triangle inequality. Thus we find $\| \f([H,a]) - [H,\f(a)] \|_2=0$, and therefore $\f([H,a]) = [H,\f(a)]$ for all $a \in \cA_\mathrm{loc}$. Because of this result, from now on we say that a Hamiltonian $H$ is symmetric/invariant under a non-invertible QCA $\f$ if $\f(H)=H$ and $\f^\star(H)=H$.

\subsubsection*{Invertible symmetries}

The purity condition is automatic for invertible symmetries, since every automorphism is CP-pure. This is an immediate consequence of Choi's theorem, as we will now demonstrate.

The multiplicative domain of an automorphism $\alpha$ is the entire algebra, so $\cA_\alpha=\cA$. Suppose $\alpha=\f_1+\f_2$, where $\f_1$ and $\f_2$ are completely positive maps. By Choi's theorem, $\f_1(a)=\f_1(1)\alpha(a)=\alpha(a)\f_1(1)$ and $\f_2(a)=\f_2(1)\alpha(a)=\alpha(a)\f_2(1)$ for $a \in \cA_\alpha = \cA$. It follows that $\f_i(1)$ commutes with every element of $\alpha(\cA)=\cA$, and hence $\f_i(1)\in\cA'=\mathbb{C}$. Therefore, each $\f_i$ is a scalar multiple of $\alpha$, proving that $\alpha$ is CP-pure.

\subsubsection*{Kramers--Wannier transformation}

Here, we prove that the Kramers--Wannier UCP map \eqref{KW.map2} is CP-pure.

Suppose that $\dd = \f_1 + \f_2$ where $\f_1,\f_2$ are completely-positive maps. Applying \eqref{gen.choi.th} from Appendix \ref{app:md.composite}, we find
\ie
    \cA^\text{even} = \Big\{ a\in \cA \mid \f_i(ax) = \dd(a) \f_i(x) \;\text{ and }\; \f_i(xa) = \f_i(x) \dd(a) \; \text{ for all } x \in \cA \Big\} \,.
\fe
In particular, this implies that $\f_i(1)$ commutes with every element inside $\dd(\cA^\text{even})=\cA^\text{even}$, and therefore $\f_i(1) \in \mathbb{C}$ since $\cA^\text{even}$ has a trivial commutant. Thus we find that $\f_i = \f_i(1) \dd_i$, where $\dd_i$ agrees with $\dd$ on all $\bZ_2$-even operators. Since the multiplicative domain of $\dd_i$ includes $\cA^\text{even}$, to show $\dd_i = \dd$ it is enough to show that they agree on a single $\bZ_2$-odd operator such as the Pauli operator at site $0$, $Z_0$.

Using the linearity of $\dd_i$ and the fact that $X_{-n} \cdots X_{n-1} X_n$ is inside the multiplicative domain of $\dd_i$, we find
\ie
    \dd_i(X_{-n} \cdots X_n) \dd_i(Z_0) &= - \dd_i(Z_0) \dd_i(X_{-n} \cdots X_n) \\
    \implies \quad Z_{-n-1}Z_{n} \dd_i(Z_0) &= - \dd_i(Z_0) Z_{-n-1}Z_{n}~.
\fe
Since $\dd_i(Z_0)$ is a quasi-local operator and the relation above should hold for any non-negative integer $n$, we conclude that $\dd_i(Z_0)=0$. Thus, $\dd_i$ agrees with $\dd$ on $Z_0$, and by Choi's theorem we conclude that they are equal. Hence, the KW map is pure. 

\subsubsection*{Non-example} 

Here, we discuss an example of a locality-preserving and dualizable UCP map that is not pure.

The non-example is given by truncating the lattice translation automorphism, also known as the shift. Consider the $*$-homomorphism
\ie \label{shift.truncation.def}
    T_{\geq 0}(\cO_j) = \begin{cases}
        \cO_j & j <0 \\
        \cO_{j+1} & j \geq 0
    \end{cases} ~.
\fe
This map is not pure since it can be decomposed into CP maps as
\ie \label{shift.truncation}
    T_{\geq 0}(a) = \ket{0}\!\bra{0}_0 \, T_{\geq 0}(a) + \ket{1}\!\bra{1}_0 \, T_{\geq 0}(a) \,.
\fe
Since $\ket{0}\!\bra{0}_0$ and $\ket{1}\!\bra{1}_0$ commute with the image of $T_{\geq 0}$, each summand above defines a CP map. Moreover, even though $T_{\geq 0}$ is locality-preserving, its CP summands are not locality-preserving.

The homomorphism $T_{\geq 0}$ is dualizable, and its dual is given by the locality-preserving UCP map
\ie
    (T_{\geq 0})^\star = T_{\geq 0}^{-1} \circ E_0\,,
\fe
where $E_0 = \tr_{0} : \cA \to \cA_{<0} \otimes \cA_{>0}$ is the normalized partial trace at site $j=0$ and $T_{\geq 0}^{-1} : \cA_{<0} \otimes \cA_{>0} \to \cA$ is the left inverse of $T_{\geq 0}$. In particular,
\ie \label{dual.shift.truncation}
    (T_{\geq 0})^\star(\cO_j) = \begin{cases}
        \cO_j & j <0 \\
        \tr(\cO_0)\cdot \mathbbm{1} ~~& j = 0 \\
        \cO_{j-1} & j > 0
    \end{cases} ~.
\fe
The dual map is not pure and decomposes as
\ie
    (T_{\geq 0})^\star(a) =  T_{\geq 0}^{-1}\left( \frac{1}{2} \bra{0} a \ket{0}_0 +\frac{1}{2} \bra{1} a \ket{1}_0 \right)\,.
\fe
The multiplicative domain of $(T_{\geq 0})^\star$ is equal to $\cA_{(T_{\geq 0})^\star} = \cA_{<0} \otimes \cA_{>0}$. It is not irreducible, since it has a non-trivial commutant consisting of operators supported on site $j=0$. 

Note that any Hamiltonian that is preserved by $T_{\geq 0}$ must identically vanish on all non-negative sites. Moreover, such Hamiltonians must be preserved by the non-locality-preserving CP summands in \eqref{shift.truncation}.

\section{On-site non-invertible QCAs} \label{sec:onsite}

In this section, we present several examples of on-site non-invertible QCAs, namely quantum channels $\f:\cA\to\cA$ for which both $\f$ and $\f^\star$ have range $\ell=0$ (see Section~\ref{sec:LQC}).

\subsection{Cosine symmetry}\label{sec:cosine}

Here, we discuss the continuous cosine non-invertible symmetry \cite{Cao:2025qnc, Seifnashri:2025fgd}, which is one of the simplest non-invertible symmetries on the lattice. It includes Rep(D$_{2n}$) for any $n$ as a subcategory, which in particular includes Rep(S$_3$) and Rep(D$_{8}$). 
The cosine symmetry was
first discussed in the context of the orbifold branch of the $c = 1$ compact boson CFT in \cite{Chang:2020imq, Thorngren:2021yso}.

\subsubsection*{MPO presentation} Let us start with the Matrix Product Operator (MPO) presentation of the symmetry. On a periodic chain of $L$ qubits, the cosine non-invertible symmetry operators are MPOs constructed from the tensor
\ie \label{cosine.mpo.tensor}
    \mathbb{U}^\theta_j = e^{i\theta \frac{\mathbb{Z}}{2}} \otimes \ket{+}\!\bra{+}_j + \mathbb{X}\, e^{i\theta \frac{\mathbb{Z}}{2}} \otimes \ket{-}\!\bra{-}_j \,, \qquad \theta \in [0,2\pi) \,.
\fe
Above, $\mathbb{X}$ and $\mathbb{Z}$ are Pauli operators acting on the two-dimensional virtual Hilbert space $\mathcal{V} = \bC^2$, and $\ket{\pm}_j$ are the $X_j$ eigenstates at site $j$. Tracing over the virtual Hilbert space gives the MPO\footnote{While $\tr:\cA \to \bC$ denotes the \emph{normalized} trace acting on the physical operators, $\Tr = \Tr_{\!\mathcal{V}}$ is the \emph{unnormalized} trace over the finite-dimensional virtual Hilbert space $\mathcal{V}$ where $\Tr \! (\mathbbm{1}_\mathcal{V}) = \mathrm{dim}(\mathcal{V})$.}
\ie \label{cosine.mpo}
    \mathsf{L}_\theta &= \Tr \!\!\left[ \mathbb{U}^\theta_L \cdots \mathbb{U}^\theta_2 \mathbb{U}^\theta_1 \right] = \Tr \!\!\left[ \eta^{\frac{1-\mathbb{X}}{2}} e^{i\theta \mathbb{Z} Q} \right] \\
    &= \frac{1+\eta}{2} \left( e^{i\theta Q} + e^{-i\theta Q} \right) = (1+\eta) \cos(\theta Q) \,,
\fe
where
\ie
    Q &= \frac{1}{2} \left( 1+X_1+X_1X_2 + \cdots + X_1X_2 \cdots X_{L-1} \right) \,, \\
    \eta &= X_1X_2 \cdots X_L \,.
\fe
The operators $\mathsf{L}_\theta$ are non-invertible and obey the cosine-like fusion relations
\ie \label{cosine.fusion}
    \mathsf{L}_{\theta_1} \mathsf{L}_{\theta_2} = \mathsf{L}_{\theta_1+\theta_2} + \mathsf{L}_{\theta_1-\theta_2} \,, \qquad \mathsf{L}_\theta^\dagger = \mathsf{L}_\theta = \mathsf{L}_{-\theta} \,, \qquad \mathsf{L}_{\theta+2\pi} = (-1)^L\, \mathsf{L}_\theta \,,
\fe
which explain the name of the symmetry.

\subsubsection*{The UCP map}
On the infinite chain, the MPO tensors \eqref{cosine.mpo.tensor} define on-site UCP maps $\f_\theta : \cA \to \cA$. For a local operator $a \in \cA_I$ supported on the interval $I=[1,L]$,
\ie \label{cosine.ucp}
    \f_\theta(a) &= \frac{1}{2} \Tr \!\Big[\mathbb{U}^\theta_L \cdots \mathbb{U}^\theta_1 \; a \; (\mathbb{U}^\theta_1)^\dagger \cdots (\mathbb{U}^\theta_L)^\dagger \Big] \\
        &= \frac{1}{2} \left( e^{i\theta Q}\, \frac{a + \eta a \eta}{2}\, e^{-i\theta Q} + e^{-i\theta Q}\, \frac{a + \eta a \eta}{2}\, e^{i\theta Q} \right) \,,
\fe
where $(\mathbb{U}^\theta_j)^\dagger = V \mathbb{U}^{-\theta}_j V^\dagger$ for the gauge transformation $V = e^{-i\theta \frac{\mathbb{Z}}{2}}$ of the virtual space. The unitarity of the tensors $\mathbb{U}^\theta_j$, together with the cyclicity of the trace, ensures that \eqref{cosine.ucp} is independent of the choice of the interval containing the support of $a$. In particular, $\f_\theta(a)$ is supported in the same interval as $a$, so $\f_\theta$ acts on-site.

\subsubsection*{Locality, dualizability, and fusion}
Let us now verify the three conditions of Section \ref{sec:def}. Being on-site UCP maps, the $\f_\theta$ are locality-preserving quantum channels. They satisfy fusion relations similar to \eqref{cosine.fusion},
\ie \label{cosine.ucp.fusion}
   \f_{\theta_1} \circ \f_{\theta_2} = \frac{1}{2}\, \f_{\theta_1+\theta_2} + \frac{1}{2}\, \f_{\theta_1-\theta_2} \,, \qquad \f_\theta^\star = \f_\theta = \f_{-\theta} = \f_{\theta+2\pi} \,.
\fe
In particular, $\f_\theta$ is self-dual and hence dualizable. The self-duality follows from \eqref{cosine.ucp}. Taking the dual corresponds to replacing the Kraus operators $\frac{1}{2}e^{\pm i\theta Q}$ and $\frac{1}{2}\eta e^{\pm i\theta Q}$ by their Hermitian conjugates $\frac{1}{2}e^{\mp i\theta Q}$ and $\frac{1}{2}\eta e^{\mp i\theta Q}$, which form the same set.

\subsubsection*{Purity} The maps $\f_\theta$ are CP-pure for $\theta \neq 0,\pi$. At these two special angles, they decompose into pure maps as
\ie \label{cosine.nonpure}
    \f_0(a)=\frac{1}{2}\left( a + \eta  a  \eta \right) \,, \qquad \f_\pi(a)=\frac{1}{2}\left( \eta_\mathrm{e}\, a\, \eta_\mathrm{e} + \eta_\mathrm{o}\, a\, \eta_\mathrm{o} \right) \,,
\fe
with $\eta_\mathrm{e} = \prod_{j \text{ even}} X_j$ and $\eta_\mathrm{o} = \prod_{j \text{ odd}} X_j$. Purity at the remaining angles can be shown using the Choi criterion \eqref{choi.purity}. The idea is to show that the Choi state $\omega_\theta: \cA \otimes \cA^\text{op} \to \bC$ of the cosine UCP map $\f_\theta$ is an injective Matrix Product State (MPS), hence pure.

Bending the input leg of the tensor \eqref{cosine.mpo.tensor} turns it into the following MPS tensor on the doubled chain with the same bond dimension
\ie \label{cosine.choi.mps}
    \mathbb{A}_j^\theta = \frac{1}{\sqrt{2}} \left( e^{i\theta \frac{\mathbb{Z}}{2}} \otimes \ket{++}_j + \mathbb{X}\, e^{i\theta \frac{\mathbb{Z}}{2}} \otimes \ket{--}_j \right) \,.
\fe
The associated MPS is precisely the Choi state of the cosine UCP map. This is because for local operators $a,b \in \cA_{[1,L]}$ supported on the interval $[1,L]$, we have
\ie
    \frac{1}{2} \Tr \!\Big[ (\mathbb{A}^\theta_1)^\dagger \cdots (\mathbb{A}^\theta_L)^\dagger \left( a \otimes b^\mathsf{T} \right) \mathbb{A}^\theta_L \cdots \mathbb{A}^\theta_1 \Big] &= \frac{1}{2} \, \tr \circ \Tr \!\Big[ (\mathbb{U}^\theta_1)^\dagger \cdots (\mathbb{U}^\theta_L)^\dagger  \; a \;  \mathbb{U}^\theta_L \cdots \mathbb{U}^\theta_1  \; b \, \Big]  \\
    &= \tr\! \big( a \, \f_\theta(b) \big) = \omega_\theta \! \left( a \otimes b^\mathsf{T} \right) \,.
\fe
Because $\mathbb{U}^\theta_j$ is unitary, the transfer matrix (or more precisely the transfer channel)
\ie \label{transfer.matrix}
    \mathcal{E}_\theta( x ) = \frac{1}{2}\sum_{s=\pm} u_s^\dagger \, x \, u_s \,, \qquad u_+ = e^{i\theta \frac{\mathbb{Z}}{2}} \,, \qquad u_- = \mathbb{X}\, e^{i\theta \frac{\mathbb{Z}}{2}} \, ,
\fe
is unital and trace-preserving, so the MPS tensor is in canonical form. As was shown in \cite{Fannes:1990ur} and stated in the MPS terminology in \cite{Perez-Garcia:2006nqo, Cirac:2020obd}, the MPS tensor defines a pure state in the thermodynamic limit if the transfer channel \eqref{transfer.matrix} is \emph{primitive}, meaning that $1$ is a simple eigenvalue and the only one on the unit circle. In that case, the tensor is called normal, which is equivalent to injectivity after blocking. The blocked tensor for the block $[1,L]$ is injective when the matrices $u_{s_L}\cdots u_{s_2} u_{s_1} \in \bC^{2\times 2}$ with $s_j = \pm$ span the entire $\bC^{2\times 2}$. Two sites suffice since the four unitary matrices
\ie \label{cosine.two.site.span}
    u_+u_+ = e^{i\theta \mathbb{Z}} \,, \qquad u_+ u_- = \mathbb{X} \,, \qquad
    u_- u_+ = \mathbb{X}\, e^{i\theta \mathbb{Z}} \,, \qquad u_-u_- = \mathbbm{1} \,,
\fe
span $\bC^{2\times 2}$ if $\sin\theta \neq 0$. Hence, for $\theta \neq 0,\pi$ the Choi state of $\f_\theta$ is an injective MPS, and therefore $\f_\theta$ is CP-pure by \eqref{choi.purity}.

The two special angles are visible in the same computation. At $\theta = 0,\pi$ the span in \eqref{cosine.two.site.span} degenerates to $\langle \mathbbm{1},\mathbb{X} \rangle \subset \bC^{2\times 2}$, whose minimal projections $\frac{1}{2}(\mathbbm{1}\pm\mathbb{X})$ produce exactly the two CP summands in \eqref{cosine.nonpure}. Equivalently, the transfer channel $\mathcal{E}_\theta$ acts by $\mathbbm{1}\mapsto\mathbbm{1}$, $\mathbb{X}\mapsto \cos\theta\,\mathbb{X} + \sin\theta\,\mathbb{Y}$, and $\mathbb{Y},\mathbb{Z}\mapsto 0$, so its spectrum is $\{1,\cos\theta,0,0\}$. The peripheral spectrum (eigenvalues on the unit circle) consists of the simple eigenvalue 1 alone exactly for $\theta \neq 0,\pi$. The subleading eigenvalue also sets the correlation length of the Choi state given by $\xi(\theta)=-1/\ln(|\cos\theta|)$.

It is instructive to contrast the purity of $\f_\theta: \cA \to \cA$ with the non-purity of its restriction to a finite interval. The four operators $\frac{1}{2}e^{\pm i\theta Q}$ and $\frac{1}{2}\eta e^{\pm i\theta Q}$ appearing in \eqref{cosine.ucp} are Kraus operators for the restriction of $\f_\theta$ to $[1,L]$, so that restriction is a convex combination of four UCP maps and is never pure. None of the four, however, survives the thermodynamic limit for $\theta \neq 0,\pi$. They depend on the choice of interval and fail to converge as $L \to \infty$. For instance,
\ie \label{cosine.action}
    \f_\theta(Z_jZ_{j+1}) = \frac{ e^{i\theta X_1 \cdots X_j} + e^{-i\theta X_1 \cdots X_j} }{2}\, Z_jZ_{j+1} = \cos(\theta)\, Z_jZ_{j+1} \,,
\fe
where the two individual terms above are non-local, but their sum is local.

\subsubsection*{Multiplicative domain}

We now determine the multiplicative domain of $\f_\theta$. Since $\f_\theta$ is trace-preserving, by Appendix \ref{app:md.tp} it coincides with the fixed-point subalgebra of $\f_\theta^\star \circ \f_\theta$, which by \eqref{cosine.ucp.fusion} equals $\frac{1}{2}\left( \f_0 + \f_{2\theta} \right)$. A fixed point of this convex combination is necessarily a common fixed point of $\f_0$ and $\f_{2\theta}$.\footnote{If $a = \frac{1}{2}\big(\f_0(a) + \f_{2\theta}(a)\big)$, the $L^2$-contractivity of Appendix \ref{app:contractivity} and the parallelogram law $\| x+y \|_2^2 + \|x-y\|_2^2 = 2\|x\|_2^2 + 2\|y\|_2^2$ force $\f_0(a) = \f_{2\theta}(a) = a$. This also follows from \eqref{l2.saturation} and \eqref{gen.choi.th}.} Thus, on a finite interval the multiplicative domain $\cA_\theta$ at angle $\theta$ is
\ie \label{cosine.md}
    \cA_{\theta} \cap \cA_{[1,L]} = \left\{ a \in \cA_{[1,L]} ~\middle|~ [e^{2i\theta Q},a] = [\eta, a] = 0 \right\} \,.
\fe

For an irrational angle $\theta_\mathrm{ir} \notin 2\pi \mathbb{Q}$, the condition $[e^{2i\theta_\mathrm{ir} Q},a]=0$ forces $[Q,a]=0$, so the multiplicative domain becomes independent of the angle.\footnote{This follows by writing $a = \sum_{q,q'}a_{q,q'} \ket{q}\!\bra{q'}$, where $\{\ket{q}\}$ is an eigenbasis of the quantized charge $Q$.} Thus, $\cA_{\theta_\mathrm{ir}}$ consists of the $\bZ_2$-even operators commuting with the charge $Q$. Therefore,
\ie \label{cosine.md.ir}
    &\cA_{\theta_\mathrm{ir}} \cap \cA_{[1,L]} = \left\{ a \in \cA_{[1,L]} ~\middle|~ \big[ Q, a \big] = [\eta ,a ] = 0 \right\} \,.
\fe
Moreover, $\f_{\theta_\mathrm{ir}}$ acts as the identity map on this subalgebra.

Two consequences are worth recording. First, distinct pure symmetries $\f_{\theta_1}$ and $\f_{\theta_2}$, for irrational angles $\theta_1 \neq \theta_2$, share the same multiplicative domain and act identically on it. Therefore, a non-invertible QCA is \emph{not} determined by its action on its multiplicative domain. However, the set of symmetric Hamiltonians only depends on the multiplicative domain $\cA_{\theta_\mathrm{ir}}$. By constructing the conditional expectation
\ie
    E = \int_{0}^{2\pi} \frac{\mathrm{d}\theta}{2\pi} \f_\theta ~:~ \cA \to \cA_{\theta_\mathrm{ir}} \subset \cA \, ,
\fe
we find that any symmetric Hamiltonian satisfies $E(H)=H$ and consequently can be written as $H = \sum_j E(h_j)$. Thus, a Hamiltonian $H$ is invariant under the cosine symmetry if and only if its interactions can be chosen to be inside $\cA_{\theta_\mathrm{ir}}$. Second, the inclusion $\cA_{\theta_\mathrm{ir}} \subset \cA$ has infinite (Watatani) index \cite{watatani1990index} even though the quantum dimension of $\f_{\theta_\mathrm{ir}}$ is $2$. This is in contrast with the Kramers--Wannier symmetry, whose multiplicative domain $\cA^\text{even}$ is an index-$2$ subalgebra of $\cA$.

\subsection{Anomalous on-site non-invertible symmetries \label{anomalous.onsite}}

Here, we discuss on-site non-invertible QCAs that are maximally anomalous in the sense that they admit no symmetric Hamiltonian besides $H=$ const. This demonstrates that, for non-invertible symmetries, the anomaly-free condition and on-siteability are distinct.

Our first example is
\ie
    \mathcal{N} = \mathsf{H} \f_{\pi/2}\,,
\fe
where $\f_{\pi/2}$ is the generator of the Rep(D$_8$) symmetry from the cosine example and $\mathsf{H}$ is the on-site invertible symmetry given by the product of Hadamard unitary operators $\mathsf{H}_j = (X_j+Z_j)/\sqrt{2}$ at each site $j$. 
We emphasize that this is a valid non-invertible QCA since it is the composition of a $\bZ_2$ on-site symmetry and a standard non-invertible symmetry (which, e.g., is realized in the cluster model \cite{Seifnashri:2024dsd}).
Its multiplicative domain ${\cal A}_{\cal N}$ consists of the local operators invariant under $\eta_\mathrm{e} = \prod_{j \text{ even}} X_j$ and $\eta_\mathrm{o} = \prod_{j \text{ odd}} X_j$, and its action on ${\cal A}_{\cal N}$ is determined by
\ie
	&\mathcal{N}(X_j) = Z_j \,, \qquad &&\mathcal{N}(Z_{j-1}Z_{j+1}) = -X_{j-1}Z_jX_{j+1} \,, \\
    &\mathcal{N}(a a') = \mathcal{N}({a}) \mathcal{N}({a'}) \qquad\quad && \text{for } {a},{a'} \in {\cA_\mathcal{N}}  \,.
\fe
For any other operator $b$ with $\eta_\mathrm{e}b\eta_\mathrm{e} = -b$ or $\eta_\mathrm{o}b\eta_\mathrm{o} = -b$, we have $\mathcal{N}({b}) = 0$.

The map $\mathcal{N}$ is manifestly locality-preserving, dualizable, and CP-pure, with $\mathcal{N}^\star = \f_{\pi/2} \mathsf{H}$. Strikingly, the only symmetric Hamiltonian is the trivial Hamiltonian, and the only symmetric state is the infinite-temperature state $\tr\!:\cA \to \bC$. This follows from the fact that, for any local operator $a$ supported on an interval of length $L$, we have
\ie
    \mathcal{N}^{2\lceil L/2\rceil}(a) = \tr(a) \mathbbm{1}\,,
\fe
which we show in Appendix~\ref{app:lemma}.
For instance, $\mathcal{N}^2(X_j)=\mathcal{N}(Z_j)=0$, $\mathcal{N}^2(Z_j)=\mathcal{N}^2(Y_j)=0$, and $\mathcal{N}^4(Z_jX_{j+1}Z_{j+2})=-\mathcal{N}^3(X_jX_{j+2})=-\mathcal{N}^2(Z_jZ_{j+2})=\mathcal{N}(X_jZ_{j+1}X_{j+2})=0$. 
Indeed, applying $\mathcal{N}$ repeatedly thermalizes every state to infinite temperature, so a state that is invariant under $\mathcal{N}$ must be the infinite-temperature state. Namely, for any state $\omega$
\ie
    \omega \circ \mathcal{N}^{L+1} (a) = \omega(\mathcal{N}^{L+1}(a)) = \tr\!(a)\,.
\fe
Thus, the reduced density matrix on an interval of length $L$ becomes maximally mixed after acting with $\mathcal{N}$ for $L+1 \geq 2\lceil L/2\rceil$ times on any state $\omega$.

This anomalous on-site non-invertible symmetry is not in contradiction with the result of~\cite{Evans:2025msy}, where it was shown that on-site fusion category symmetries have to be anomaly-free. This is because the powers of $\mathcal{N}$ are not built from finitely many simple symmetry transformations, and hence $\mathcal{N}$ does not generate a fusion category with finitely many simple objects. Indeed, in~\cite{Evans:2025msy} the operators invariant under a fusion category symmetry form a subalgebra of finite index in $\cA$, whereas the only operators invariant under $\mathcal{N}$ are scalars.

We note that this anomalous on-site symmetry is also considered on-site by the definition in \cite{Meng:2024nxx}, which defines on-site symmetries as those whose quantum dimension agrees with the minimum bond dimension of their corresponding MPO. The non-invertible QCA $\mathcal{N} = \mathsf{H} \f_{\pi/2}$ has minimum bond dimension 2, since the cosine symmetry $\f_{\pi/2}$ has bond dimension 2. Moreover, its quantum dimension is also equal to 2 since
\ie
    \mathcal{N}^\star \mathcal{N} = \frac{1}{4}\left( \mathbbm{1} + \eta_\mathrm{e} + \eta_\mathrm{o} + \eta_\mathrm{e}\eta_\mathrm{o} \right)\,.
\fe
Thus, its quantum dimension matches its bond dimension. Here, we use the ``Frobenius--Perron'' definition of the quantum dimension as the smallest positive number $d_\f$ such that $\f^\star \f - \frac{\mathbbm{1}}{d_\f^2}$ is completely positive. This definition is motivated by the factor of the quantum dimension in \eqref{encircling}, which implies that the fusion relation involving $d_\f\f$ matches that of defects.  For all the on-site examples in this work, the quantum dimension and the minimum bond dimension agree.

\subsubsection*{Non-invertible QCA with a trivial multiplicative domain}
A corollary of the example above is another maximally anomalous on-site non-invertible QCA, which has a trivial multiplicative domain. We will show that the map
\ie
    \Gamma = \f_{\pi/4} \mathsf{H} \f_{\pi/4}
\fe
has a trivial multiplicative domain, i.e., $\cA_\Gamma = \mathbb{C}$.

From \eqref{md.composition}, $a \in \cA_\Gamma$ if and only if $a \in \cA_{\f_{\pi/4}}$ and $\mathsf{H} \f_{\pi/4}(a) \in \cA_{\f_{\pi/4}}$, and therefore
\ie
    \f_{\pi/2}(a) = a \qquad \text{and} \qquad \f_{\pi/2} \mathsf{H} \f_{\pi/4}(a) = \mathsf{H} \f_{\pi/4}(a) \,.
\fe
Here, we have used the fact that $\cA_{\f_{\pi/4}}$ is the same as the fixed-point subalgebra of $\f_{\pi/2}$, which follows from \eqref{fixed.subalgebra} and \eqref{choi.theorem}. Denoting $b=\f_{\pi/4}(a)$, we find
\ie
    \mathcal{N}^2(b)=\mathsf{H} \f_{\pi/2} \mathsf{H} \f_{\pi/2}(b) = \mathsf{H} \f_{\pi/2} \mathsf{H}(b) = \mathsf{H}\mathsf{H} (b) = b \,.
\fe
However, as shown above, the fixed-point subalgebra of $\mathcal{N}$ and $\mathcal{N}^2$ is trivial. Therefore, $b$ is a scalar, implying $a$ is also a scalar and $\cA_\Gamma = \mathbb{C}$. Since $\Gamma$ is on-site, it does not admit any non-trivial symmetric Hamiltonian. This is because we can keep acting with $\Gamma$ to project any local term in the Hamiltonian to its multiplicative domain. The on-site condition is necessary because we want to keep the terms local by acting iteratively with $\Gamma$. In other words, since $\Gamma$ acts on-site, the contractivity shown in Appendix \ref{app:contractivity} implies
\ie
    \lim_{n \to \infty} \Gamma^n (a) = \tr\!(a) \mathbbm{1} \,.
\fe
Thus $\Gamma$ is also maximally anomalous.

It remains to show $\Gamma$ is CP-pure. Using the notation of \eqref{cosine.choi.mps} and \eqref{transfer.matrix} with $\theta=\pi/4$, the MPS tensor associated with the Choi state of $\Gamma$ is
\ie 
    \mathbb{A}^\Gamma_j = \sum_{r,s = \pm} \frac{(-1)^{\frac{(1-r)(1-s)}{4}} u_r \otimes u_s}{2} \otimes \ket{r, s}_j \,.
\fe
The previous subsection already establishes that $\mathrm{span}\{u_ru_s\} = \bC^{2 \times 2}$. Thus $\mathrm{span}\{ u_r u_t \otimes u_s u_v\} = \bC^{4 \times 4}$, which shows that $\Gamma$ is Choi-pure and consequently CP-pure.

\subsection{Rep(\texorpdfstring{$G$}{G}) symmetries} \label{sec:repg}

Here, we introduce a general class of on-site non-invertible QCAs that includes the cosine symmetries as a special case. The main advantage of this class is its simplicity, which makes it easy to construct examples exhibiting a wide range of properties. The defining data of these non-invertible QCAs are $N$ unitary matrices, where $N$ is the dimension of the physical Hilbert space at each site.

Consider a spin chain with an $N$-dimensional local Hilbert space $\cH_j = \bC^N$. Choose $N$ unitary $d \times d$ matrices $u_1,u_2,\cdots,u_N \in \mathrm{U}(d)$ and consider the MPO tensor
\ie \label{repg.tensor}
    \mathbb{U}_j = \sum_{s=1}^N u_s \otimes \ket{s}\!\bra{s}_j \,,
\fe
where $\{\ket{s}_j\}_{s=1}^N$ is an orthonormal basis of the physical Hilbert space at site $j$ and $\{u_s\}_{s=1}^N$ act on the $d$-dimensional virtual space $\mathcal{V}=\bC^d$. Note that $\mathbb{U}_j: \mathcal{V} \otimes \cH_j \to \mathcal{V} \otimes \cH_j$ is a unitary operator acting on the tensor product of the virtual and physical spaces. The class of MPOs that have unitary tensors was recently studied in \cite{Xie:2025ccn}. On a finite periodic chain, the corresponding MPO is diagonal in a product state basis:
\ie \label{repg.mpo}
    \mathsf{L} &= \Tr \!\! \left[ \mathbb{U}_L \cdots \mathbb{U}_2 \mathbb{U}_1 \right] \\ &= \sum_{s_1,\cdots,s_L = 1}^N \Tr\left( u_{s_L} \cdots u_{s_2} u_{s_1}\right)\,  \ket{s_1, s_2, \cdots , s_L}\!\bra{s_1, s_2, \cdots , s_L} \,.
\fe
In particular, all eigenvalues of $\mathsf{L}$ are attained on product states, suggesting that $\mathsf{L}$ generates an anomaly-free non-invertible symmetry.

\begin{figure}[th]
\centering
\begingroup
\definecolor{figoneblue}{HTML}{245B78}
\begin{tikzpicture}[
  x=1cm, y=1cm, line cap=round, line join=round,
  virt/.style={draw=figoneblue, line width=1.3pt},
  virtdots/.style={draw=figoneblue, line width=1.3pt, dash pattern=on 0pt off 3.2pt},
  phys/.style={draw=black!70, line width=.6pt},
  midarrow/.style={postaction=decorate, decoration={markings,
      mark=at position .5 with {\arrow{Stealth[length=2.2mm, width=1.9mm]}}}},
  opbox/.style={draw=black!75, fill=black!6, line width=.7pt, rounded corners=1.5pt,
      minimum height=.7cm, inner sep=0pt},
  tens/.style={draw=figoneblue, fill=figoneblue!8, line width=.8pt, rounded corners=1.5pt,
      minimum width=.74cm, minimum height=.74cm, inner sep=1pt, font=\small},
  tensd/.style={tens, fill=figoneblue!22},
  proj/.style={draw=black!60, fill=white, line width=.5pt, rounded corners=1pt,
      inner sep=1.5pt, minimum height=.5cm, font=\footnotesize},
  site/.style={anchor=north, font=\small, text=black!75, inner sep=1pt},
]

  \def\xa{0} \def\xb{1.45} \def\xc{3.3}
  \def\xd{2.375}  
  \def\xg{.38}

  \node[anchor=east] at (-1.15,0) {$\displaystyle \f(a) \;=\; \frac{1}{d}$};
  \node[opbox, minimum width=4.35cm] at (1.65,0) {$a$};
  \foreach \x/\n in {\xa/1, \xb/2, \xc/J} {
    \node[tens]  (U\n) at (\x, 1.4) {$\mathbb{U}_{\n}$};
    \node[tensd] (D\n) at (\x,-1.4) {$\mathbb{U}_{\n}^{\dagger}$};
    \draw[phys] (U\n.north) -- (\x, 2.15);
    \draw[phys] (U\n.south) -- (\x, .35);
    \draw[phys] (\x,-.35) -- (D\n.north);
    \draw[phys] (D\n.south) -- (\x,-2.15);
    \node[site] at (\x,-2.25) {$\n$};
  }
  \node[site, text=black!55] at (\xd,-2.25) {$\cdots$};
  \foreach \y/\l in {1.4/U, -1.4/D} {
    \draw[virt] (\l 1.east) -- (\l 2.west);
    \draw[virt] (\l 2.east) -- (\xd-\xg,\y);
    \draw[virtdots] (\xd-\xg,\y) -- (\xd+\xg,\y);
    \draw[virt] (\xd+\xg,\y) -- (\l J.west);
  }

  \draw[virt, midarrow] (D1.west) -- ++(-.15,0)
      .. controls (-1.05,-1.4) and (-1.05,1.4) .. (-.52,1.4) -- (U1.west);
  \draw[virt, midarrow] (UJ.east) -- ++(.15,0)
      .. controls (\xc+1.05,1.4) and (\xc+1.05,-1.4) .. (\xc+.52,-1.4) -- (DJ.east);
  \node[figoneblue, font=\small, anchor=west] at (\xc+.88,.75) {$\cV$};

  \begin{scope}[xshift=8.05cm]
    \def\ra{0} \def\rb{1.5} \def\rc{3.55} \def\rd{2.525}
    \node[anchor=east] at (-.85,0) {$\displaystyle =\; \sum_{\bm{s},\bm{t}} c_{\bm{s},\bm{t}}$};
    \node[opbox, minimum width=4.65cm] at (1.775,0) {$a$};
    \foreach \x/\n in {\ra/1, \rb/2, \rc/J} {
      \draw[phys] (\x,.35) -- (\x, 2.15);
      \draw[phys] (\x,-.35) -- (\x,-2.15);
      \node[proj] at (\x, 1.4) {$\ket{s_{\n}}\!\bra{s_{\n}}$};
      \node[proj] at (\x,-1.4) {$\ket{t_{\n}}\!\bra{t_{\n}}$};
      \node[site] at (\x,-2.25) {$\n$};
    }
    \node[text=black!55] at (\rd, 1.4) {$\cdots$};
    \node[text=black!55] at (\rd,-1.4) {$\cdots$};
    \node[site, text=black!55] at (\rd,-2.25) {$\cdots$};
  \end{scope}

  \begin{scope}[xshift=2.45cm, yshift=-4.55cm]
    \node[anchor=east] at (-1.15,0) {$\displaystyle c_{\bm{s},\bm{t}} \;=\; \frac{1}{d}$};
    \foreach \x/\n in {\xa/1, \xb/2, \xc/J} {
      \node[tens, minimum width=.84cm]  (u\n) at (\x, .62) {$u_{s_{\n}}$};
      \node[tensd, minimum width=.84cm] (v\n) at (\x,-.62) {$u_{t_{\n}}^{\dagger}$};
    }
    \foreach \y/\l in {.62/u, -.62/v} {
      \draw[virt] (\l 1.east) -- (\l 2.west);
      \draw[virt] (\l 2.east) -- (\xd-\xg,\y);
      \draw[virtdots] (\xd-\xg,\y) -- (\xd+\xg,\y);
      \draw[virt] (\xd+\xg,\y) -- (\l J.west);
    }
    \draw[virt, midarrow] (v1.west) -- ++(-.12,0)
        .. controls (-1.0,-.62) and (-1.0,.62) .. (-.54,.62) -- (u1.west);
    \draw[virt, midarrow] (uJ.east) -- ++(.12,0)
        .. controls (\xc+1.0,.62) and (\xc+1.0,-.62) .. (\xc+.54,-.62) -- (vJ.east);
    \node[anchor=west] at (\xc+1.05,0)
        {$\displaystyle =\; \frac{1}{d}\, \Tr\!\big( U_{\bm{s}} U_{\bm{t}}^{\dagger} \big)$};
  \end{scope}
\end{tikzpicture}
\endgroup
\caption{The UCP map $\f$ of \eqref{repg.ucp} for the Rep($G$) symmetry. Thick blue lines represent the virtual space $\cV = \bC^d$, and thin black lines the physical spaces. The arrows fix the order of the matrix products on $\cV$, and the virtual loop plays the role of the defect loop in \eqref{encircling}. Since $\mathbb{U}_j$ is unitary, the tensors outside the support of $a$ cancel. Expanding each $\mathbb{U}_j$ as in \eqref{repg.tensor}, the virtual loop detaches from the physical legs and gives the coefficient $c_{\bm{s},\bm{t}}$ (bottom).}
\label{fig:ucp}
\end{figure}
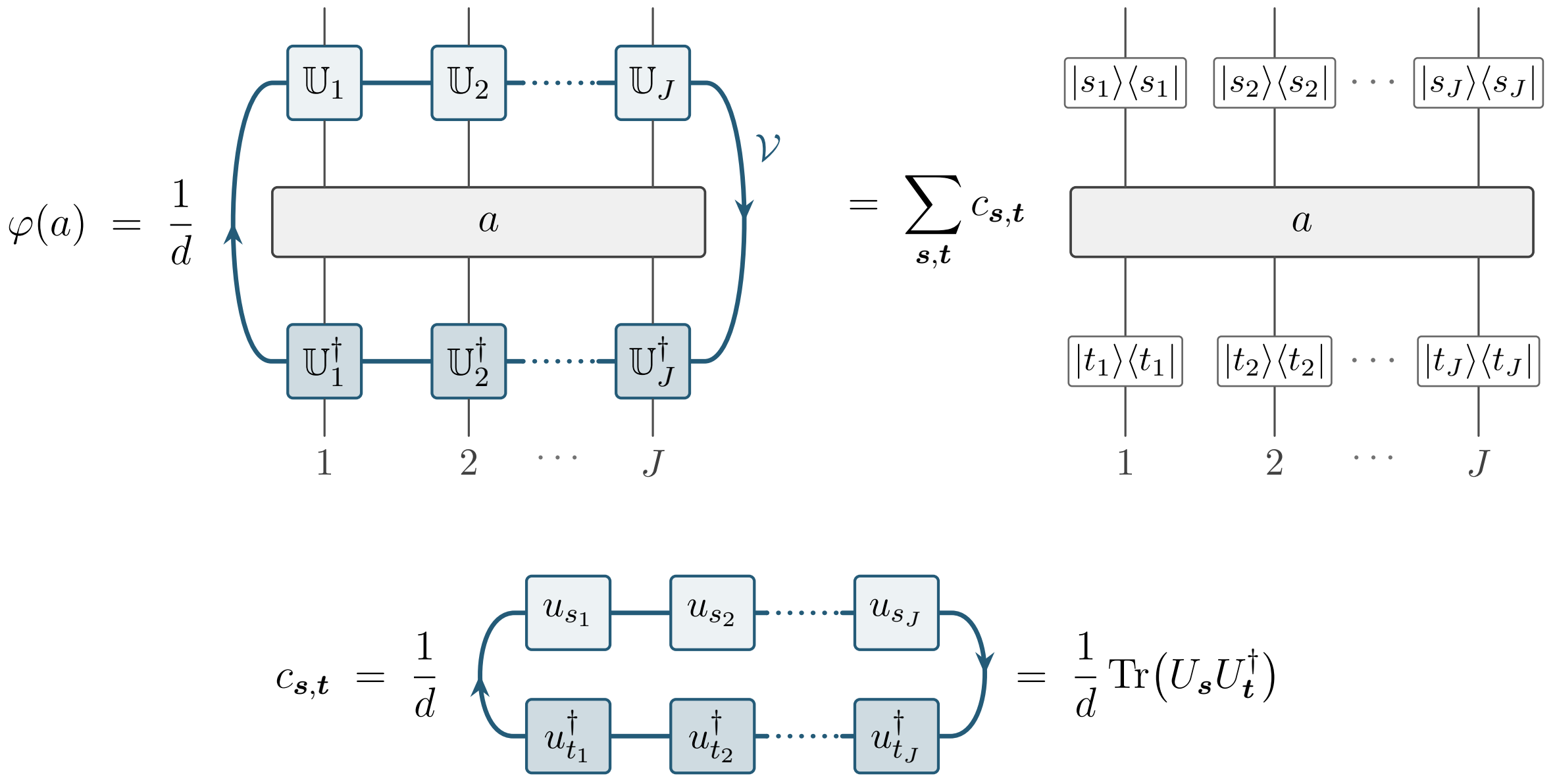

As in \eqref{cosine.ucp}, such an MPO tensor defines an on-site UCP map $\f: \cA \to \cA$ on the infinite chain. For a local operator $a$ supported on the interval $[1, J]$,
\ie \label{repg.ucp}
    \f(a) &= \frac{1}{d} \Tr\Big( \mathbb{U}_J \cdots \mathbb{U}_1  \; a \; \mathbb{U}_1^\dagger \cdots \mathbb{U}_J^\dagger \Big) = \sum_{\bm{s},\bm{t}} c_{\bm{s},\bm{t}}\, \ket{\bm{s}}\!\bra{\bm{s}}  a  \ket{\bm{t}}\!\bra{\bm{t}} \,, \\
    c_{\bm{s},\bm{t}} &= \frac{1}{d} \Tr\!\big( U_{\bm{s}} U_{\bm{t}}^\dagger \big) \,, \qquad \bm{s} = (s_1,\cdots,s_J)\,, \qquad U_{\bm{s}} = u_{s_J} \cdots u_{s_2} u_{s_1} \,.
\fe
As before, the unitarity of the tensors and the cyclicity of the trace ensure that \eqref{repg.ucp} is independent of the choice of the interval containing the support of $a$. In particular, $\f$ is an on-site channel and hence locality-preserving. Moreover, it is manifestly a trace-preserving UCP map. Every product state in this basis is invariant under $\f$ on the infinite chain, since its reduced density matrix $\ket{\bm{s}}\!\bra{\bm{s}}$ on any finite interval $I$ is fixed by $\f^\star$. Specifically, for such states $\langle \f(a) \rangle = \Tr_{\cH_I}( \ket{\bm{s}}\!\bra{\bm{s}} \f(a)) = \Tr_{\cH_I}(\f^\star(\ket{\bm{s}}\!\bra{\bm{s}})a) = \langle a \rangle$, where $a$ is supported on the interval $I$. See Figure \ref{fig:ucp}.

Multiplying matrices $u_1,\cdots,u_N$ with each other and recording all relations among them, we obtain a finitely generated discrete group\footnote{When $G$ is finite, our MPO \eqref{repg.mpo} is a special case of the construction in \cite{Molnar:2022nmh, Lootens:2021tet}.}
\ie \label{eq:g}
    G = \langle u_1, u_2, \cdots, u_N \rangle \subseteq \mathrm{U}(d) \,,
\fe
and its \emph{defining} representation $\rho:G \hookrightarrow \mathrm{U}(d)$, which is a faithful $d$-dimensional unitary representation.  We will show that $\f$ obeys a Rep($G$) fusion algebra. However, this realization need not be faithful as we will discuss below.

\subsubsection*{Dualizability and fusion}

The quantum channel in \eqref{repg.ucp} is equivalently specified by a choice of a group $G$ and its $N$ generators $g_1, g_2, \cdots,g_N$ together with a $d$-dimensional unitary representation $\rho: G \hookrightarrow \mathrm{U}(d)$. Specifically, we have $G=\langle g_1, g_2, \cdots,g_N\rangle$ and $u_s = \rho(g_s)$. In the following, we will identify $u_s \in \mathrm{U}(d)$ with the abstract group element $g_s \in G$. From this point on, we will write the UCP map \eqref{repg.ucp} as $\f_\rho$ to emphasize its dependence on the representation $\rho$. 
In this new notation, the action of $\f_\rho$ on a local operator is
\ie \label{explicit.rep.g}
    \f_\rho\big( \ket{\bm{s}} \! \bra{\bm{t}} \big) = \frac{\chi_\rho(U_{\bm{s}} U_{\bm{t}}^{-1})}{d_\rho} \ket{\bm{s}} \! \bra{\bm{t}} \,, \qquad \chi_\rho(g) = \Tr\!\rho(g) \,, \qquad d_\rho = \mathrm{dim}(\rho) \,,
\fe
where $\bm{s},\bm{t}$ and $U_{\bm{s}},U_{\bm{t}}$ are defined in \eqref{repg.ucp}. More generally, any finite-dimensional unitary representation $\sigma$ of $G$ defines a UCP map $\f_\sigma$ by replacing $u_s$ with $\sigma(u_s)$ in \eqref{repg.tensor}. The dual of $\f_\rho$ in \eqref{explicit.rep.g} is obtained by transposing its matrix coefficients. Since $c_{\bm{t},\bm{s}} = \overline{c_{\bm{s},\bm{t}}}$, the dual channel is given by the complex conjugate representation $\overline{\rho}$:
\ie
    \f_\rho^\star = \f_{\overline{\rho}} \,.
\fe
In particular, the dual is also on-site and hence locality-preserving. Thus, every quantum channel in this class is dualizable.

Since $\f_\rho$ depends on $\rho$ only through its normalized character, these maps satisfy the normalized Rep($G$) fusion algebra
\ie \label{repg.fusion}
    \f_{\rho_1} \circ \f_{\rho_2} = \f_{\rho_1 \otimes \rho_2} \,, \qquad
    \f_{\rho_1 \oplus \rho_2} = \frac{d_{\rho_1}}{d_{\rho_1}+d_{\rho_2}}\, \f_{\rho_1} + \frac{d_{\rho_2}}{d_{\rho_1}+d_{\rho_2}}\, \f_{\rho_2} \,.
\fe
The first relation holds because characters are multiplicative under the tensor product, $\chi_{\rho_1\otimes\rho_2}(g) = \chi_{\rho_1}(g)\chi_{\rho_2}(g)$. The second follows from $\chi_{\rho_1\oplus\rho_2}=\chi_{\rho_1}+\chi_{\rho_2}$ and $d_{\rho_1\oplus\rho_2}=d_{\rho_1}+d_{\rho_2}$.

\subsubsection*{Faithfulness}

The Rep($G$) non-invertible symmetry generated by $\f_\rho$ might not be faithful for two reasons:
\begin{enumerate}
    \item[I.] The quantum channel $\f_\rho$ depends on the representation $\rho$ only through its restriction to the normal subgroup
    \ie \label{eq:g0}
        G_0 = \Big\langle U_{\bm{s}} U_{\bm{t}}^{-1}, U_{\bm{t}}^{-1}U_{\bm{s}} \;:\; |\bm{s}|=|\bm{t}|  \Big\rangle \subseteq G \,,
    \fe
    generated by \emph{balanced words} $U_{\bm{s}} U_{\bm{t}}^{-1}$ and $U_{\bm{t}}^{-1}U_{\bm{s}}$ that appear in \eqref{explicit.rep.g}. More precisely, $\f_\rho$ depends only on the normalized characters of the restricted representation $\rho|_{G_0}$. Note also that $\chi_\rho(U_{\bm{s}} U_{\bm{t}}^{-1}) =\chi_\rho(U_{\bm{t}}^{-1}U_{\bm{s}})$.

    This can also be understood in terms of gradings of $G$. The subgroup $G_0$ can be identified with the trivially graded subgroup of $G$ defined as the intersection of the kernels of all $\bZ_k$-grading homomorphisms $\mathrm{deg}_k: G \to \bZ_k$ under which every $u_s$ has degree $1$.\footnote{Every balanced word has a trivial grading with respect to any homomorphism $\deg_k$, therefore $G_0 \subseteq \mathrm{ker}(\mathrm{deg}_k)$. For the other direction, we need to show that $G_0$ is the kernel of some grading homomorphism. Note that $G/G_0$ is a cyclic group. Thus, the universal grading homomorphism $\deg: G \to G/G_0$ is such an example, where we may need to compose it with projections $G/G_0 \to \bZ_k$ to make it into a finite grading.} Indeed, after blocking sites into blocks of size $k$, the new MPO tensor and hence $\f_\rho$ depend only on the restriction of $\rho$ to the kernels of these homomorphisms.

    \item[II.] When $G_0$ is an infinite group, a faithful representation of $G_0$ may not generate all representations of $G_0$ by tensor products and decomposition. Thus, $\f_\rho$ would not generate the entire Rep($G_0$) symmetry by fusion and decomposition into pure channels. In that case, we should consider the closure of $G_0 \subseteq \mathrm{U}(d)$ denoted by 
    \ie
        \overline{G_0} = \overline{\Big\langle U_{\bm{s}} U_{\bm{t}}^{-1}, U_{\bm{t}}^{-1}U_{\bm{s}} \;:\; |\bm{s}|=|\bm{t}|  \Big\rangle}^{\| \cdot \|} \subseteq \mathrm{U}(d)\,.
    \fe
    Since $\overline{G_0}$ is compact, all of its representations are generated from the faithful representation $\rho$ by fusion, taking the dual, and decomposition into irreducible representations.\footnote{Throughout, we only consider finite-dimensional and continuous representations of compact groups.} Therefore, $\f_\rho$ generates all the other elements of the $\mathrm{Rep}(\overline{G_0})$ symmetry, which is a nice \emph{tensor category} (see for instance \cite{etingof2015tensor}).
\end{enumerate}

We will argue that $\f_\rho$ generates a faithful Rep($\overline{G_0}$) non-invertible symmetry associated with the following quantum channels. For any representation $\pi$ of $\overline{G_0}$, we define $\f_\pi$ similarly to \eqref{explicit.rep.g} as
\ie \label{rep.g.0.maps}
    \f_{\pi}\big( \ket{\bm{s}} \! \bra{\bm{t}} \big) = \frac{\chi_\pi(U_{\bm{s}} U_{\bm{t}}^{-1})}{d_\pi} \ket{\bm{s}} \! \bra{\bm{t}} \,, \qquad \bm{s} = (s_{j_1}, \cdots , s_{j_2})\,,
\fe
which is well-defined only after suitable blocking. The issue is that $\pi$ may not extend to a representation of $G$, so in general $\chi_\pi(u_i \, U_{\bm{s}} U_{\bm{t}}^{-1}u_i^{-1}) \neq \chi_\pi(U_{\bm{s}} U_{\bm{t}}^{-1})$. Therefore, the action on $\ket{\bm{s}} \! \bra{\bm{t}}$ can change as we enlarge the interval $[j_1,j_2]$ supporting it. We block sites into blocks of size $k$, where $k$ is the smallest positive integer such that $\chi_\pi(u_i^k \, U_{\bm{s}} U_{\bm{t}}^{-1}u_i^{-k}) = \chi_\pi(U_{\bm{s}} U_{\bm{t}}^{-1})$ for each generator $u_i$, and every $\bm{s}$ and $\bm{t}$. Namely, in the definition of the UCP map in \eqref{rep.g.0.maps}, we take $j_1$ and $j_2+1$ to be multiples of $k$, which guarantees that it is well-defined. We show in Appendix \ref{app:faithfulness} that such a finite $k$ always exists. Relatedly, the map $\f_{\pi}$ commutes with lattice translation by $k$ sites and need not commute with one-site translation $T$.

Faithfulness of Rep($\overline{G_0}$) amounts to two statements. First, as shown in Appendix \ref{app:faithfulness},
\ie
    \f_{\pi_1}=\f_{\pi_2} \quad \Leftrightarrow \quad \left(\pi_1 \right)^{\oplus d_{\pi_2}} \simeq \left(\pi_2 \right)^{\oplus d_{\pi_1}} \,,
\fe
for representations $\pi_1,\pi_2$ of $\overline{G_0}$. So we have a faithful realization of the normalized fusion algebra of Rep($\overline{G_0}$). Secondly, every $\f_\pi$ above with irreducible $\pi$ appears as a CP summand of some composition of $\f_\rho$ and $\f_{\rho}^\star$. This is because, as mentioned above, every irreducible representation $\pi$ of $\overline{G_0}$ appears in a restriction $(\rho^{\otimes n} \otimes \overline{\rho}^{\otimes m})|_{\overline{G_0}} = \pi \oplus \cdots$ for some $m,n \in \bZ_{\geq 0}$, thus
\ie
    (\f_{\rho})^{n}(\f_{\rho}^\star)^{m} = \f_{\rho^{\otimes n} \otimes \overline{\rho}^{\otimes m}} =\frac{d_\pi}{(d_\rho)^{m+n}}\f_\pi + \cdots \,.
\fe
Therefore, the quantum channel $\f_\rho$ generates an on-site and faithful Rep($\overline{G_0}$) non-invertible symmetry.

For example, $u_0=1$ and $u_1=e^{i\theta}$ with $\theta \notin 2\pi\mathbb{Q}$ give $G_0=G = \bZ$ and $\overline{G_0}=\mathrm{U}(1)$. The corresponding MPO, up to a phase, is the unitary operator $\prod_j e^{-i\theta Z_j/2}$. Since $\theta$ is a fixed number, taking different powers of this unitary operator generates a $\mathrm{Rep}(\overline{G_0})=\bZ$ symmetry, which is a subgroup of $\mathrm{Rep}(G_0)=\mathrm{U}(1)$. Nevertheless, any Hamiltonian or state invariant under this $\bZ$ symmetry is automatically symmetric under the larger $\mathrm{U}(1)$ symmetry.

The symmetry algebraically generated by $\f_\rho$ through fusion, duals, and decomposition is Rep($\overline{G_0}$). However, its invariant Hamiltonians can have a larger symmetry. Since $\rho$ is faithful, \eqref{explicit.rep.g} implies
\ie \label{symm.local.op}
    \f_\rho\big( \ket{\bm{s}} \! \bra{\bm{t}} \big) = \ket{\bm{s}} \! \bra{\bm{t}} \quad \Leftrightarrow \quad U_{\bm{s}} = U_{\bm{t}} \,.
\fe
This equality is a relation in $G$, so it should also hold for every other finite-dimensional unitary representation $\sigma$ of $G$. Thus, every local operator, state, or Hamiltonian invariant under $\f_\rho$ is invariant under all $\f_\sigma$ and their pure components. This larger $\mathrm{Rep}(G)$ symmetry need not be faithful as discussed above. 
When the cyclic group $G/G_0$ is finite, and therefore isomorphic to $\bZ_k$, this larger symmetry is $\mathrm{Rep}(G_0)$. One way to see this is to construct the $\mathrm{Rep}(G_0)$ non-invertible symmetries explicitly by blocking every $k$ consecutive sites and substituting $\pi$ with a representation of $G_0$ in \eqref{rep.g.0.maps}. Indeed, in the example above, $G=G_0=\bZ$ and the enlarged symmetry is $\mathrm{Rep}(G_0)=\mathrm{U}(1)$, which is the closure of $\mathrm{Rep}(\overline{G_0})=\bZ$. In general, the enlarged symmetry is not generated by $\f_\rho$, even after taking the closure of the UCP maps. It also need not exhaust all the symmetries shared by the Hamiltonians invariant under $\f_\rho$, which are classical for generic $u_s$.

In summary, we showed that the faithful symmetry algebraically generated by $\f_\rho$ is $\mathrm{Rep}(\overline{G_0}) \subseteq \mathrm{Rep}(G_0)$. However, the symmetry implied by $\f_\rho$ is larger and includes $\mathrm{Rep}(G_0)$ when $G/G_0 = \bZ_k$.

\subsubsection*{Purity}

We show in Appendix \ref{app:purity} that $\f_\rho$ is pure precisely when the associated representation of $\overline{G_0}$ is a multiple of a single irreducible representation. Namely, we will show
\ie
    \text{$\f_\rho$ is CP-pure} \quad \Longleftrightarrow \quad  \text{$\rho|_{\overline{G_0}} \simeq \pi^{\oplus m}$ for an irreducible representation $\pi$} \,. 
\fe
Such a representation is called \emph{isotypical}. It can be shown that $\rho|_{\overline{G_0}}$ is isotypical iff $\rho|_{G_0}$ is isotypical.

\subsubsection*{Multiplicative domain}

Since $\f_\rho$ is trace-preserving, by Appendix \ref{app:md.tp} its multiplicative domain $\cA_{\f_\rho}$ is the fixed-point subalgebra of the channel $\f_\rho^\star \circ \f_\rho = \f_{\overline{\rho} \otimes \rho}$, which multiplies $\ket{\bm{s}}\!\bra{\bm{t}}$ by $|c_{\bm{s},\bm{t}}|^2=|\Tr(U_{\bm{s}} U_{\bm{t}}^{-1})/d|^2$. Note $|c_{\bm{s},\bm{t}}| = 1$ precisely when $U_{\bm{s}}$ and $U_{\bm{t}}$ are equal up to a phase, thus
\ie \label{repg.md}
    \cA_{\f_\rho} \cap \cA_{[1,J]} = \mathrm{span}\big\{ \ket{\bm{s}}\!\bra{\bm{t}} ~:~ U_{\bm{s}} \propto U_{\bm{t}} \big\}  \,.
\fe
The multiplicative domain always contains the diagonal subalgebra $\cA_\mathrm{diag} \subset \cA$ generated by the projections $\ket{s}\!\bra{s}_j$, on which $\f_\rho$ acts as the identity.

As for the cosine symmetry, 
\ie \label{limit.conditional.expectation}
    E = \lim_{n \to \infty} (\f_\rho^\star \circ \f_\rho)^n ~:~ \cA \to \cA_{\f_\rho}
\fe
defines an on-site conditional expectation onto $\cA_{\f_\rho}$. Indeed $(\f_\rho^\star \circ \f_\rho)^n$ multiplies $\ket{\bm{s}}\!\bra{\bm{t}}$ by $|c_{\bm{s},\bm{t}}|^{2n}$ which converges to $1$ if $U_{\bm{s}} \propto U_{\bm{t}}$ and otherwise to $0$. In general, for any on-site non-invertible QCA, the limit $\lim_{n \to \infty} (\f^\star \circ \f)^n$ exists and defines an on-site conditional expectation onto the multiplicative domain of $\f$. A Hamiltonian $H=\sum_j h_j$ that is preserved by $\f_\rho$ and $\f_\rho^\star$ is also preserved by $(\f_\rho^\star \circ \f_\rho)^n$ and therefore satisfies $E(H)= \sum_j E(h_j) = H$. Thus, a Hamiltonian is symmetric only if its interactions can be chosen inside $\cA_{\f_\rho}$. For generic unitaries $u_1,\cdots,u_N \in \mathrm{U}(d)$ with $d \geq 2$, the images of distinct words in $\mathrm{PU}(d)$ are distinct \cite{epstein1971almost}, so that $\cA_{\f_\rho} = \cA_\mathrm{diag}$. In that case, the only symmetric Hamiltonians are the classical ones, and imposing $\f_\rho$ as a symmetry leads to infinitely many local conserved charges.

On the other hand, when $G_0$ is finite, the multiplicative domain is large and is in fact irreducible. Namely, $(\cA_{\f_\rho})' \cap \cA = \bC$, so the symmetry does not lead to any obvious local conserved charges. To see this, let $x \in \cA$ commute with $\cA_{\f_\rho}$. Since $\cA_{\f_\rho}$ contains the maximal diagonal subalgebra $\cA_\mathrm{diag}$, the operator $x$ is also diagonal and therefore is associated with a continuous function $f$ on the classical configuration space $\{1,\ldots, N\}^{\bZ}$. Let $\omega_0$ be the all-$1$ configuration. For any word $\bm s$ of length $L$, finiteness of $\{U_{\bm s}^n u_1^{-nL}:n\geq0\}\subseteq G_0$ implies $U_{\bm s}^k=u_1^{kL}$ for some $k\geq1$. Thus $(\ket{\bm s}\!\bra{1,\ldots,1})^{\otimes k}$ belongs to $\cA_{\f_\rho}$ by \eqref{repg.md}, and since it commutes with $x$, we have $f(\omega_{\bm s})=f(\omega_0)$, where $\omega_{\bm s}$ consists of $k$ consecutive copies of $\bm s$ and $1$'s elsewhere. Allowing arbitrary words, these configurations contain every finite pattern and hence form a dense set. Continuity therefore makes $f$ constant, proving $(\cA_{\f_\rho})'\cap\cA=\bC$.

\subsection{Rep(SU(2)) symmetry of the dual folded XXZ model}

Here, we study a special case of the previous subsection in which $\overline{G_0}=\mathrm{SU}(2)$. Consider the qubit chain with the MPO tensor \eqref{repg.tensor} with
\ie \label{su2.words}
    u_0 = i\mathbb{H} \,, \qquad u_1 = e^{i\theta \mathbb{Z}}\,, \qquad \theta \notin \pi\mathbb{Q} \,,
\fe
where $\mathbb{H} = (\mathbb{X}+\mathbb{Z})/\sqrt{2}$ is the Hadamard unitary operator acting on the virtual degrees of freedom. The matrices in \eqref{su2.words} have determinant one, hence the group $G$, which is generated by $u_0,u_1$ in \eqref{su2.words} is a subgroup of SU(2). The balanced subgroup $G_0 \subseteq G$, defined in \eqref{eq:g0}, contains
\ie
    u_1^2u_0^{-2} = -e^{2i\theta \mathbb{Z}}\,, \qquad u_0(u_1^2u_0^{-2})u_0^{-1} = -e^{2i\theta \mathbb{X}}\,.
\fe
These two elements generate a dense subgroup of $\mathbb{X}$ and $\mathbb{Z}$ rotations, and therefore a dense subgroup of the entire SU(2). Namely,
\ie
    \overline{G_0}=\mathrm{SU}(2) \,.
\fe
So, based on the discussion in Section \ref{sec:repg}, we expect a faithful Rep(SU(2)) symmetry. See \eqref{G.G0} below for more information on the groups $G$ and $G_0$.

Let $\pi_r$ be the spin-$r$ representation of SU(2), with $r = 0,\frac{1}{2},1,\cdots$ and dimension $d_r=2r+1$. Replacing $u_0$ and $u_1$ in \eqref{repg.tensor} with $\pi_r(u_0)$ and $\pi_r(u_1)$ we find the non-invertible QCA $\f_r$. Explicitly, when $U_{\bm{s}}U_{\bm{t}}^{-1} \in \mathrm{SU}(2)$ has eigenvalues $e^{\pm i\alpha_{\bm{s},\bm{t}}}$
\ie
    \f_r \big( \ket{\bm{s}}\!\bra{\bm{t}} \big) = \frac{\chi_r(U_{\bm{s}}U_{\bm{t}}^{-1})}{2r+1} \ket{\bm{s}}\!\bra{\bm{t}} = \frac{\sin{((2r+1)\alpha_{\bm{s},\bm{t}})}}{(2r+1) \sin{\alpha_{\bm{s},\bm{t}}}} \ket{\bm{s}}\!\bra{\bm{t}} \,.
\fe
All the representations are real (or pseudo-real), thus $\f_r^\star=\f_r$. For the original channel associated with the defining representation, we have
\ie
    \f_{1/2}(X_j) = \frac{\sin{\theta}}{\sqrt{2}} X_j \,, \qquad \f_{1/2}(Y_j) = \frac{\sin{\theta}}{\sqrt{2}} Y_j \,, \qquad \f_{1/2}(Z_j) = Z_j \,.
\fe

Since $G_0$ is dense inside SU(2), the restriction $\pi_r|_{G_0}$ remains an irreducible representation of $G_0$. This follows from Schur's lemma, since $\pi_r(G_0)' = \pi_r(\mathrm{SU}(2))'$. Therefore, the purity criterion in Section \ref{sec:repg} shows that every $\f_r$ is CP-pure. Moreover, all the maps above are distinct and therefore $\f_{1/2}$ generates a faithful Rep(SU(2)) symmetry. Indeed, their fusion rule
\ie
    \f_{r_1} \circ \f_{r_2} = \sum_{\ell = |r_1 - r_2|}^{r_1+r_2} \frac{2\ell + 1}{(2r_1+1)(2r_2+1)} \f_{\ell} \,, \qquad \f_0 = \mathbbm{1} \,,
\fe
is the normalized fusion ring of Rep(SU(2)). For example, $(\f_{1/2})^2 = \frac{1}{4}\mathbbm{1} + \frac{3}{4}\f_1$. The corresponding MPOs on periodic chains are $\mathsf{L}_r = \sum_{\bm{s}} \chi_r(U_{\bm{s}}) \ket{\bm{s}}\!\bra{\bm{s}}$, which obey the unnormalized fusion rule $\mathsf{L}_{r_1} \mathsf{L}_{r_2} = \sum_{\ell = |r_1 - r_2|}^{r_1+r_2} \mathsf{L}_{\ell}$.

To describe all invariant local operators and local Hamiltonians, let us assume that the irrational number $\theta/\pi$ is generic, so that all relations satisfied by $u_0$ and $u_1$ follow from $u_0^4=1$ and $u_0^2 u_1 = u_1u_0^2$. In that case, 
\ie \label{G.G0}
    G_0 &= \langle u_1 u_0^{-1} , u_0^{-1}u_1\rangle \cong F_2 \,, \\ G &= \langle u_0,u_1 \mid u_0^4=1,\; u_0^2 u_1=u_1 u_0^2 \rangle \cong F_2 \rtimes \bZ_4 \,.
\fe
Above, $F_2$ is the free group on two generators, and $u_0 \in \bZ_4$ acts on it by exchanging its two free generators $u_1 u_0^{-1}$ and $u_0^{-1}u_1$.

We focus on generic values of $\theta$, so that $G/G_0 \cong \bZ_4$. Thus, any symmetric state or Hamiltonian has a larger Rep($F_2$) symmetry as discussed in Section \ref{sec:repg}. Since $-1 \notin G_0$, $U_{\bm{s}} \not\propto U_{\bm{t}}$ unless $U_{\bm{s}} = U_{\bm{t}}$. Thus, \eqref{repg.md} implies that the multiplicative domain of $\f_{1/2}$ is the same as its fixed-point subalgebra. Namely,
\ie
    \cA_{\f_{1/2}} \cap \cA_{[1,L]} &= \left\{ a \in \cA_{[1,L]} \,:\, \f_{1/2}(a) = a\right\} = \mathrm{span}\left\{ \ket{\bm{s}}\!\bra{\bm{t}} \,:\, U_{\bm{s}} = U_{\bm{t}}  \right\} \\ &= \left\langle Z_j \,, ~ \ket{001}\!\bra{100}_{j-1,j,j+1} \right\rangle \cap \cA_{[1,L]} \,,
\fe
where $\ket{001}\!\bra{100}_{j-1,j,j+1}$ corresponds to the relation $U_{001}=u_1u_0^2 = u_0^2 u_1 =U_{100}$. Here, we have used the fact that all relations between $u_0$ and $u_1$ follow from $u_0^4=1$ and $u_0^2 u_1 = u_1 u_0^2$.

Now we explicitly construct the enlarged Rep($F_2$) symmetries. Let us choose a finite-dimensional unitary representation $\pi$ of $G_0 \cong F_2$, which we identify with arbitrary unitary matrices $A,B \in \mathrm{U}(d)$, where $A=\pi(u_1u_0^{-1})$ and $B=\pi(u_0^{-1}u_1)$. From the identity $u_{s_2}u_{s_1} = - (u_1u_0^{-1})^{s_2}(u_0^{-1}u_1)^{s_1}$, we find $\pi(u_{s_4}u_{s_3}u_{s_2}u_{s_1}) = A^{s_4}B^{s_3}A^{s_2}B^{s_1}$ and consequently the non-invertible symmetries
\ie \label{RepF2.symmetries}
    &\f_{A,B}\big( \ket{\bm{s}}\!\bra{\bm{t}}\big) = \frac{1}{d} \Tr\!\left(V_{\bm{s}} V_{\bm{t}}^{-1} \right) \ket{\bm{s}}\!\bra{\bm{t}}\,, \\
    &V_{\bm{s}} = A^{s_L} B^{s_{L-1}} \cdots A^{s_2} B^{s_1}\,.
\fe
Here $\bm{s}$ and $\bm{t}$ are supported on an interval $[1,L]$ with $L$ even. These maps preserve the subalgebra $\cA_{\f_{1/2}}$ and obey the Rep($F_2$) fusion rule
\ie
    \f_{A,B} \circ \f_{A',B'}=\f_{A\otimes A',B\otimes B'} \,.
\fe
Finally, the lattice translation symmetry exchanges $A$ and $B$. Namely, $T \f_{A,B} T^{-1} = \f_{B,A}$. The one-dimensional representations correspond to invertible symmetries. Namely, $A=e^{i\alpha}$ and $B=e^{i\beta}$ give a $\mathrm{U}(1)_\mathrm{e} \times \mathrm{U}(1)_\mathrm{o}$ symmetry acting on even and odd sites:
\ie
    \f_{e^{i\alpha},e^{i\beta}}(a) = \left(e^{-i\alpha Q_\mathrm{e}} e^{-i\beta Q_\mathrm{o}}\right) a \left( e^{i\alpha Q_\mathrm{e}} e^{i\beta Q_\mathrm{o}} \right) \,, \qquad Q_{\mathrm{e}/\mathrm{o}} = \frac{1}{2} \sum_{j \text{ even/odd}} (1+Z_j)\,.
\fe

Since the symmetry $\f_{1/2}$ acts on-site, following the reasoning around \eqref{limit.conditional.expectation}, the local terms of any symmetric Hamiltonian can be chosen from $\cA_{\f_{1/2}}$. The most general translation-invariant symmetric Hamiltonian with next-to-nearest neighbor interactions is
\ie
    H = -\sum_j \left( t\, T_j + \overline{t} \, T_j^\dagger + g Z_j + \lambda_2 Z_{j}Z_{j+1} + \lambda_2' Z_{j-1}Z_{j+1} + \lambda_3 Z_{j-1}Z_{j}Z_{j+1} \right)\,,
\fe
where $T_j=\ket{001}\!\bra{100}_{j-1,j,j+1}$, $t\in \bC$, and $g,\lambda_2,\lambda_2',\lambda_3 \in \mathbb{R}$.

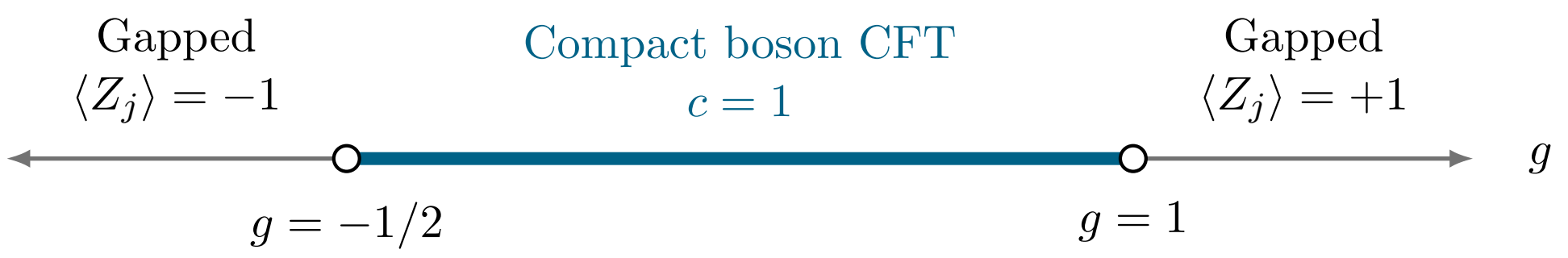
\begin{figure}[t]
\centering
\begin{tikzpicture}[x=4.3cm, y=1cm, font=\small, line cap=round]
    \draw[{Latex[length=2mm]}-{Latex[length=2mm]}, black!55, line width=1pt]
        (-1.15,0) -- (1.65,0);
    \draw[MidnightBlue!85!black, line width=3pt] (-.5,0) -- (1,0);

    \node[align=center] at (-.825,.72) {Gapped\\$\langle Z_j \rangle=-1$};
    \node[align=center, MidnightBlue!85!black] at (.25,.72)
        {Compact boson CFT\\$c=1$};
    \node[align=center] at (1.325,.72) {Gapped\\$\langle Z_j \rangle=+1$};
    \node[anchor=west] at (1.72,0) {$g$};

    \filldraw[fill=white, draw=black, line width=.8pt] (-.5,0) circle (3pt);
    \filldraw[fill=white, draw=black, line width=.8pt] (1,0) circle (3pt);
    \node[align=center][below=6pt] at (-.5,0) {$g=-1/2$};
    \node[align=center][below=6pt] at (1,0) {$g=1$};

\end{tikzpicture}
\caption{Phase diagram of the Hamiltonian \eqref{H.folded}. The radius of the compact boson theory within the gapless interval varies between $R=\sqrt{2}$ as $g \to -1/2$ and $R=1/\sqrt{2}$ as $g\to 1$. The transitions to the gapped phases at $g=1$ and $g=-1/2$ are $z=2$ Lifshitz critical points.}
\label{fig:folded-phase-diagram}
\end{figure}

For $t=1$ and $\lambda_2=\lambda_2'=\lambda_3=0$, we find the integrable dual folded XXZ chain~\cite{Zadnik:2020xcd, Pozsgay:2021dyy}\footnote{Related diagonal matrix product operator symmetries were constructed for the folded XXZ model and its deformations in~\cite{Borsi:2023nyt}.}
\ie \label{H.folded}
    H_\text{folded} &= -\sum_j \left(  T_j + T_j^\dagger + g Z_j \right) \\ &= -\sum_j \left[ \frac{1}{4}\left( X_{j-1}X_{j+1} + Y_{j-1}Y_{j+1} \right)(1+Z_j)+g Z_j \right]\,.
\fe
The phase diagram is known~\cite{Pozsgay:2021dyy}\footnote{Our Hamiltonian is related to~\cite[Eq.~(92)]{Pozsgay:2021dyy} by substituting $\mu = -g$, $\sigma_j^z=-Z_j$, and $\sigma_j^{\pm} = e^{\pm i\pi j/2} (X_j\mp i Y_j)/2$. The phase diagram is discussed in Sec.~IX, Eqs.~(128), (130), at $h=0$.} and shown in Figure~\ref{fig:folded-phase-diagram}. For $g<-1/2$ and $g>1$, the Hamiltonian is gapped, and its ground state is the product state with $Z_j=-1$ and $Z_j=+1$, respectively. For $-1/2<g<1$, it describes a $c=1$ compact boson theory at radius $R$ given by~\cite{alcaraz1999exactly, Verresen:2019jvy}\footnote{At low energies, all strings of $0$'s have even length~\cite{Pozsgay:2021dyy}, and replacing $00\to10$, $1\to0$ maps \eqref{H.folded} to the model of~\cite{alcaraz1999exactly} with their $t=1$ and $\Delta=0$. Their Eqs.~(16) and (29), with $h=2g$ and $\rho=1-R/\sqrt2$, then give \eqref{repsu2.radius}.}
\ie \label{repsu2.radius}
    g =\frac{\cos(\sqrt{2}\pi/R)}{\sqrt{2}R} -\frac{\sin(\sqrt{2}\pi/R)}{2\pi}\,, \qquad \frac{1}{\sqrt{2}}<R<\sqrt{2}\,.
\fe
We are using the convention that $R=1$ is the self-dual radius. The phase transitions at $g=1$ and $g=-1/2$ are Lifshitz critical points with dynamical exponent $z=2$.

The anti-diagonal U(1) symmetry associated with the conserved charge $Q_\mathrm{e}-Q_\mathrm{o}$ acts trivially in the IR, and both $\mathrm{U}(1)_\mathrm{e}$ and $\mathrm{U}(1)_\mathrm{o}$ symmetries flow to the U(1) momentum symmetry of the compact boson CFT in the IR. More generally, the Rep($F_2$) symmetries of the UV in \eqref{RepF2.symmetries} flow to linear combinations of the U(1) momentum symmetry transformations as
\ie
    \mathsf{L}_{A,B} ~&\xrightarrow{\text{IR}}~  \Tr\!\left[ (AB)^{L/2-Q_\mathrm{m}} \right] = \sum_{n=1}^d e^{i\gamma_n (L/2-Q_\mathrm{m})} \quad && \text{(periodic chain)}\,,  \\
    \f_{A,B}(a) ~&\xrightarrow{\text{IR}}~  \frac{1}{d} \sum_{n=1}^d e^{-i\gamma_n Q_\mathrm{m}} \,a\, e^{i\gamma_n Q_\mathrm{m}} \quad && \text{(infinite chain)} \,,
\fe
where $e^{i\gamma_1}, \cdots, e^{i\gamma_d}$ are the eigenvalues of $AB \in \mathrm{U}(d)$, and $Q_\mathrm{m} = \frac{1}{4}\sum_j (1+Z_j)$ is the conserved charge of the U(1) momentum symmetry. At low energies, all strings of $0$'s have even length, hence $Q_\mathrm{e} = Q_\mathrm{o} = Q_\mathrm{m}$. The MPO $\mathsf{L}_{A,B}$ is defined similarly to \eqref{repg.mpo} by $\mathsf{L}_{A,B} = \sum_{\bm{s}} \Tr(A^{s_L} B^{s_{L-1}} \cdots A^{s_2} B^{s_1})\ket{\bm{s}}\!\bra{\bm{s}}$ for even $L$ (cf.\ \eqref{RepF2.symmetries}).

\section{Topological defects and sequential circuits} \label{sec:defects}

Here, we discuss the definition of non-invertible symmetries from \emph{dualizable} and irreducible topological defects, realized by a pair of sequential circuits on the lattice. We will discuss its relation with the non-invertible QCA definition of Section \ref{sec:def}.

\subsection{Topological defects}

A topological defect $\defect$ corresponds to local modifications of the operator algebra $\cA$ to
\ie \label{defect.algebra}
    &\cA_\defect^{(j,j+1)} = \cA_{\leq j} \otimes \mathcal{B}_\defect^{(j,j+1)} \otimes \cA_{> j+\ell} \,, \\ &\mathcal{B}_\defect^{(j,j+1)} \subseteq \cA_{(j,j+\ell]} \otimes \bC^{n \times n} \,,
\fe
for some integers $\ell \geq 0$ and $n \geq 1$, which can be moved via \emph{local} unitary transformations
\ie \label{movement.op}
    &\lambda_j^\defect : \cA_\defect^{(j-1,j)} \to \cA_\defect^{(j,j+1)} \,,
\fe
where $\lambda_j^\defect = \mathbbm{1}_{\cA_{<j}} \otimes \mathcal{U}_j^{\defect} \otimes \mathbbm{1}_{\cA_{>j+\ell}}$ with $\mathcal{U}_j^{\defect}: \mathcal{B}_\defect^{(j-1,j)} \otimes \cA_{j+\ell} \to \cA_{j} \otimes \mathcal{B}_\defect^{(j,j+1)}$.
We will refer to $\cA_\defect^{(j,j+1)}$ as the defect operator algebra and to $\lambda_j^\defect$ as the movement operator. The algebra $\mathcal{B}_\defect^{(j,j+1)}$ contains the defect degrees of freedom, which are quantified by the \emph{lattice dimension} of the defect~\cite{Seiberg:2024gek, Inamura:2026hif}
\ie
    \kappa_{\defect} = \sqrt{\frac{\mathrm{dim}( \mathcal{B}_\defect^{(j,j+1)} )}{\mathrm{dim}( \cA_{(j,j+\ell]} )}} \,.
\fe
For irreducible defects (defined below), the algebra $\mathcal{B}_\defect^{(j,j+1)}$ is simple and therefore $\kappa_{\defect} \in \mathbb{Q}^+$, and the lattice dimension is related to the quantum dimension $d_{\defect}$ by $\kappa_{\defect} \kappa_{\overline{\defect}} = d_{\defect}^2$. For instance, for the Kramers--Wannier symmetry we have $\kappa_{\defect}=1$, $\kappa_{\overline{\defect}}=2$, and $d_{\defect} = \sqrt{2}$.

Using the movement operators, we construct the \emph{defect creation} homomorphisms
\ie \label{defect.creation}
    \pi_{\leq j}^\defect &= \lambda_j^\defect \lambda_{j-1}^\defect \lambda_{j-2}^\defect \cdots &&:\, \cA \to \cA_\defect^{(j,j+1)} \,,\\
    \pi_{> j}^\defect &= (\lambda_{j+1}^\defect)^{-1} (\lambda_{j+2}^\defect)^{-1} \cdots &&:\, \cA \to \cA_\defect^{(j,j+1)} \,,
\fe
which bring a defect to the link $(j,j+1)$ from $-\infty$ and from $+\infty$, respectively. For instance, the defect Hamiltonian can be constructed as $H_\defect^{(j,j+1)}= \pi_{\leq j}^\defect(H) = \pi_{> j}^\defect(H)$ assuming that $H$ is invariant under the generalized symmetry associated with the topological defect $\defect$. The defect is \emph{topological} because it can be moved via unitary transformations. In other words, the defect Hamiltonians are unitarily equivalent since $H_\defect^{(j,j+1)} = \lambda_j^\defect(H_\defect^{(j-1,j)})$. The maps in \eqref{defect.creation} are injective $*$-homomorphisms, which may not be surjective when the corresponding symmetry is not invertible.

We are interested in the relation between topological defects and non-invertible QCAs. Given a defect, we can construct an associated UCP map
\ie \label{ucp.from.defect}
    \f = \left(\pi_{> j}^\defect \right)^\star \circ \pi_{\leq j}^\defect ~:~ \cA \to \mathscr{R} = \pi_{\mathrm{tr}}(\cA)'' \,.
\fe
This map need not be locality-preserving or even take values in the quasi-local operator algebra $\cA$. However, its image is always contained in the hyperfinite type II$_1$ factor $\mathscr{R}$ since it involves the tracial dual of a trace-preserving UCP map. The topological defect presentation guarantees that $\f$ ($\f^\star$) is one-sided locality-preserving from the left (right). Namely,
\ie
    \f(\cA_{[j_1,j_2]}) \subset \pi_{\mathrm{tr}}(\cA_{\geq j_1 - \ell})'' \qquad \text{and} \qquad  \f^\star(\cA_{[j_1,j_2]}) \subset \pi_{\mathrm{tr}}(\cA_{\leq j_2 + \ell})'' \,.
\fe

We say that a non-invertible QCA $\f$ admits a realization by dualizable topological defects if there exists a mutually dual pair of irreducible defects, $\defect$ and $\overline{\defect}$, such that
\ie
      \f = \left(\pi_{> j}^\defect \right)^\star \circ \pi_{\leq j}^\defect = \left(\pi_{\leq j}^{\overline{\defect}} \right)^\star \circ \pi_{> j}^{\overline{\defect}}\,.
\fe
In other words, the UCP map $\f$ can be constructed from sweeping a topological defect from \emph{left to right} and also sweeping the dual defect \emph{from right to left}. From the one-sided locality-preserving observation above, for such a pair of defects, the map $\f$ is automatically locality-preserving and dualizable. The purity of $\f$ follows from the irreducibility of the defect, which will be shown in~\cite{Seifnashri:QCAStructure}. A defect $\defect$ is said to be irreducible if $ (\pi_{\leq j}^\defect(\cA))' \cap  \cA_\defect^{(j,j+1)}= \bC$ and $ (\pi_{> j}^\defect(\cA))' \cap  \cA_\defect^{(j,j+1)}= \bC$. This implies the defect operator algebra $\cA_\defect^{(j,j+1)}$ is simple and consequently $\mathcal{B}_\defect^{(j,j+1)} \cong \bC^{D \times D}$ for some integer $D$.

The converse is the central conjecture of this work:
\begin{tcolorbox}[title=Conjecture]
Every non-invertible QCA admits a realization by dualizable topological defects or, equivalently, by a dualizable sequential circuit \eqref{dual.ucp.from.sequential}.
\end{tcolorbox}
In~\cite{Seifnashri:QCAStructure}, we prove this conjecture for non-invertible QCAs with finite quantum dimension.\footnote{For translation-invariant quantum channels that are locality-preserving from one side, a sequential presentation was established in~\cite{Kretschmann:2005mya}, where the local unitaries are replaced by quantum channels acting on one site and an ancilla that may be infinite-dimensional.}

\subsection{Sequential circuits}

The discussion above, although well-motivated, might sound abstract. Therefore, here we discuss topological defects from the perspective of sequential circuits \cite{Chen:2023qst}. We focus on a simple class of sequential circuits associated with two-site unitary operators
\ie \label{simple.sequential.circuit}
    U_j : \cH_{j} \otimes \cH_{j+1} \to \cH_{j} \otimes \cH_{j+1}
\fe
acting on sites $j$ and $j+1$. They give rise to the sequential circuit $U_{N} U_{N-1} \cdots U_{-N+1} U_{-N}$. 
This corresponds to setting $\ell=1$, $\mathcal{B}_\defect^{(j,j+1)} = \cA_{j+1}$, and $\lambda_j^{\defect}(a) = U_j a U_j^\dagger$ in \eqref{defect.algebra} and \eqref{movement.op}.

More generally, a sequential circuit may have spread $\ell \in \bZ_{\geq0}$ and act on ancillary degrees of freedom, with the local unitary operator $U_j$ acting on $\cH_{[j,j+\ell]} \otimes \bC^n$. Moreover, to obtain irreducible defects, we allow the defect degrees of freedom to be a subspace of $\cH_{(j,j+\ell]} \otimes \bC^n$, specified by an orthogonal projection operator $P_{j,j+1}: \cH_{(j,j+\ell]} \otimes \bC^n \to \cH_{(j,j+\ell]} \otimes \bC^n$. Namely, the movement operators must satisfy $U_j P_{j-1,j} = P_{j,j+1} U_j$ and the defect algebra \eqref{defect.algebra} has $\mathcal{B}_{\defect}^{(j,j+1)} = P_{j,j+1} (\cA_{(j,j+\ell]} \otimes \bC^{n \times n} )P_{j,j+1}$.

Naively, one would define a map on the local operators from the sequential circuit by the conjugation action, i.e., $a \mapsto (U_{N} \cdots U_{-N}) \, a \, (U_{N} \cdots U_{-N})^\dagger$. 
However, this action is generally not well-defined since it may not converge as $N \to \infty$. 

Instead, we define 
\ie \label{ucp.from.sequential}
    \f(a) &= \lim_{N \to \infty} \Phi_N(a)=\lim_{N \to \infty} \tr_{\! N+1} \! \left( U_{N} \cdots U_{-N} \,a\, U_{-N}^\dagger \cdots U_{N}^\dagger \right) \,, \\
    \f^\star(a) &= \lim_{N \to \infty} \overline{\Phi}_N(a)=\lim_{N \to \infty} \tr_{\! -N} \! \left( U_{-N}^\dagger \cdots U_{N}^\dagger \,a\, U_{N} \cdots U_{-N} \right)\,,
\fe
where $\tr_{\!j}$ is the normalized partial trace at site $j$, $\tr_{\!j}(a) = \Tr_{\cH_j}(a)/\Tr_{\cH_j}(1)$. We claim that the sequences above converge in $\mathscr{R} = \pi_{\mathrm{tr}}(\cA)''$. 
To see this, we first note that for $M\geq N$, we have
\ie
    \|\Phi_M(a)-\Phi_N(a)\|_2^2 &= \tr \!\left( \big[\Phi_M(a) - E_{N}(\Phi_{M}(a))\big]\big[\Phi_M(a) - E_{N}(\Phi_{M}(a))\big]^\dagger \right) \\ &=\|\Phi_M(a)\|_2^2-\|\Phi_N(a)\|_2^2 \,,
\fe 
where $E_{N}=\prod_{|j| > N} \tr_{\! j} :\cA \to \cA_{[-N,N]}$ and we have used $E_{N}(\Phi_{M}(a)) = \Phi_N(a) $ for sufficiently large $N$. 
Since $\|\Phi_N(a)\|_2$ is bounded from above by $\|a\|_2$, it follows that $\Phi_N(a)$ is $L^2$-convergent. 
The discussion around \eqref{convergence.in.R} in turn implies that the sequence converges in $\mathscr{R} = \pi_{\mathrm{tr}}(\cA)''$. In fact, since we are interested in sequential circuits for which the sequence above yields a locality-preserving UCP map, this sequence stabilizes for sufficiently large $N$.\footnote{This is because for a map with range $\ell$ and a local operator $a$ supported on the interval $I=(-N,N)$, we have $\f(a) = E_{N+\ell}(\f(a)) = \lim_{M \to \infty} E_{N+\ell}(\Phi_M(a)) = \Phi_{N+\ell}(a)$. \label{footnote.stabilization}}

Below, we will show that \eqref{ucp.from.sequential} coincides with the expression \eqref{ucp.from.defect}. 
For a local operator $a\in \cA_I$ supported on a finite interval $I=[j_1,j_2]$ and $N > -j_1,j_2$, 
\ie \label{phi.n}
    \Phi_N(a) = \tr_{\! N+1} \! \left( U_{N}  \cdots U_{-N} \,a\, U_{-N}^\dagger \cdots U_{N}^\dagger \right) \in \cA_{[j_1-1, N]} \,.
\fe
Note that for $a,b\in \cA_{[-N+1,N-1]}$
\ie
    \tr\big( \Phi_N(a) \, b \big) &= \tr\!\left( U_{N}  \cdots U_{-N} \,a\, U_{-N}^\dagger \cdots U_{N}^\dagger \, b\right) = \tr\!\left( L_N a L_N^\dagger \, R_N^\dagger b R_N \right) \\
    &= \tr\!\left(  \pi_{\leq 0}^\defect(a) \, \pi_{>0}^\defect(b) \right) = \tr\big( \f(a)\, b \big)\,,
\fe
where $L_N = U_0 \cdots U_{-N}$ and $R_N = U_N \cdots U_{1}$. The first equality above follows because $b$ does not have support on site $j=N+1$ and therefore $\tr_{\! N+1}(x) b = \tr_{\! N+1}(xb)$. For any fixed local operators $a,b$, this identity holds for sufficiently large $N$. Using the faithfulness of the trace, we conclude $\lim_{N\to \infty}\Phi_N(a) = \f(a)$. 

From \eqref{phi.n}, we see that $\Phi_N$ is one-sided locality-preserving from the left, as it expands the support by at most one site to the left, while its rightward spread need not be bounded. Similarly, the map
\ie \label{phi.n.star}
    \overline{\Phi}_N(a) = \tr_{\! -N} \! \left( U_{-N}^\dagger \cdots U_{N}^\dagger \,a\, U_{N} \cdots U_{-N} \right) \in \cA_{[-N+1, j_2+1]}
\fe
is also one-sided locality-preserving and expands the support by at most one site to the right, while its leftward spread need not be bounded. 
In general, the sequential circuit arising from a generic choice of unitary operators \eqref{simple.sequential.circuit} leads to a $*$-automorphism of $\mathscr{R}$ that is not locality-preserving.
We will discuss a family of such examples in Section \ref{sec:kw.deformed}.

\subsubsection{Dualizable sequential circuits}

We are interested in sequential circuits that are \textit{dualizable} in the sense that the map \eqref{ucp.from.sequential} can be constructed from a dual sequential circuit in the other direction from right to left. Even when we start with local unitary operators in \eqref{simple.sequential.circuit} that do not act on ancillary degrees of freedom, the dual sequential circuit generally requires ancillary degrees of freedom (see the Kramers--Wannier example in Section~\ref{sec:kw.deformed}). For this reason, we include ancillary degrees of freedom in this discussion. 

\begin{figure}[tp]
\centering
\resizebox{\linewidth}{!}{
\begingroup
\definecolor{figtwoblue}{HTML}{245B78}
\definecolor{figtwoteal}{HTML}{327466}
\begin{tikzpicture}[
 x=1cm,y=1cm,font=\normalsize,
 wire/.style={draw=black!45,line width=.5pt},
 gate/.style={line width=.85pt,rounded corners=1.2pt,
   minimum width=4.20cm,minimum height=.76cm, inner sep=3pt},
 direction/.style={-{Stealth[length=2mm,width=1.4mm]},line width=.8pt},
 heading/.style={font=\small\sffamily,inner sep=0pt},
 site/.style={anchor=north,font=\small,text=black!80,inner sep=1pt}
]
\node[font=\large] at (-1.05,2.66) {$\varphi\;=$};
\node[font=\large] at (7.60,2.66) {$=$};

\node[heading,text=figtwoblue] at (3.25,6.08) {Left-to-right sweep};
\draw[direction,figtwoblue] (1.60,5.64)--(4.90,5.64);

\foreach \x in {0,.65,1.30,1.95,2.60,3.90,4.55,5.20,5.85,6.50} {
 \draw[wire] (\x,.12)--(\x,5.21);
}
\foreach \x/\y/\idx in {1.95/.70/{j-2},2.60/1.68/{j-1},3.25/2.66/{j},3.90/3.64/{j+1},4.55/4.62/{j+2}} {
 \node[gate,draw=figtwoblue,fill=figtwoblue!7] at (\x,\y)
  {$\operatorname{Ad}_{U_{\idx}}$};
}

\node[site] at (1.30,-.03) {$j$};
\node[site] at (5.20,-.03) {$j+\ell$};
\node[text=black!55,font=\small] at (3.25,-.23) {$\cdots$};
\node[text=black!55,font=\large] at (-.38,-.20) {\reflectbox{$\ddots$}};
\node[text=black!55,font=\large] at (6.93,5.34) {\reflectbox{$\ddots$}};

\begin{scope}[xshift=8.70cm]
\node[heading,text=figtwoteal] at (3.25,6.08) {Right-to-left sweep};
\draw[direction,figtwoteal] (4.90,5.64)--(1.60,5.64);
\foreach \x in {0,.65,1.30,1.95,2.60,3.90,4.55,5.20,5.85,6.50} {
 \draw[wire] (\x,.12)--(\x,5.21);
}
\foreach \x/\y/\idx in {4.55/.70/{j+2},3.90/1.68/{j+1},3.25/2.66/{j},2.60/3.64/{j-1},1.95/4.62/{j-2}} {
 \node[gate,draw=figtwoteal,fill=figtwoteal!7] at (\x,\y)
  {$\operatorname{Ad}_{\overline U_{\idx}^{\dagger}}$};
}
\node[site] at (1.30,-.03) {$j$};
\node[site] at (5.20,-.03) {$j+\bar\ell$};
\node[text=black!55,font=\small] at (3.25,-.23) {$\cdots$};
\node[text=black!55,font=\large] at (6.88,-.20) {$\ddots$};
\node[text=black!55,font=\large] at (-.43,5.34) {$\ddots$};
\end{scope}
\end{tikzpicture}
\endgroup
}
\caption{A sequential circuit is dualizable if its associated quantum channel $\f$ admits two equivalent representations, one from sweeping a defect left to right, and the other from right to left. Here, $\text{Ad}_U$ stands for conjugation by the unitary operator $U$, i.e., $\text{Ad}_U(a)= UaU^\dagger$.} 
\label{fig:dualizable}
\end{figure}
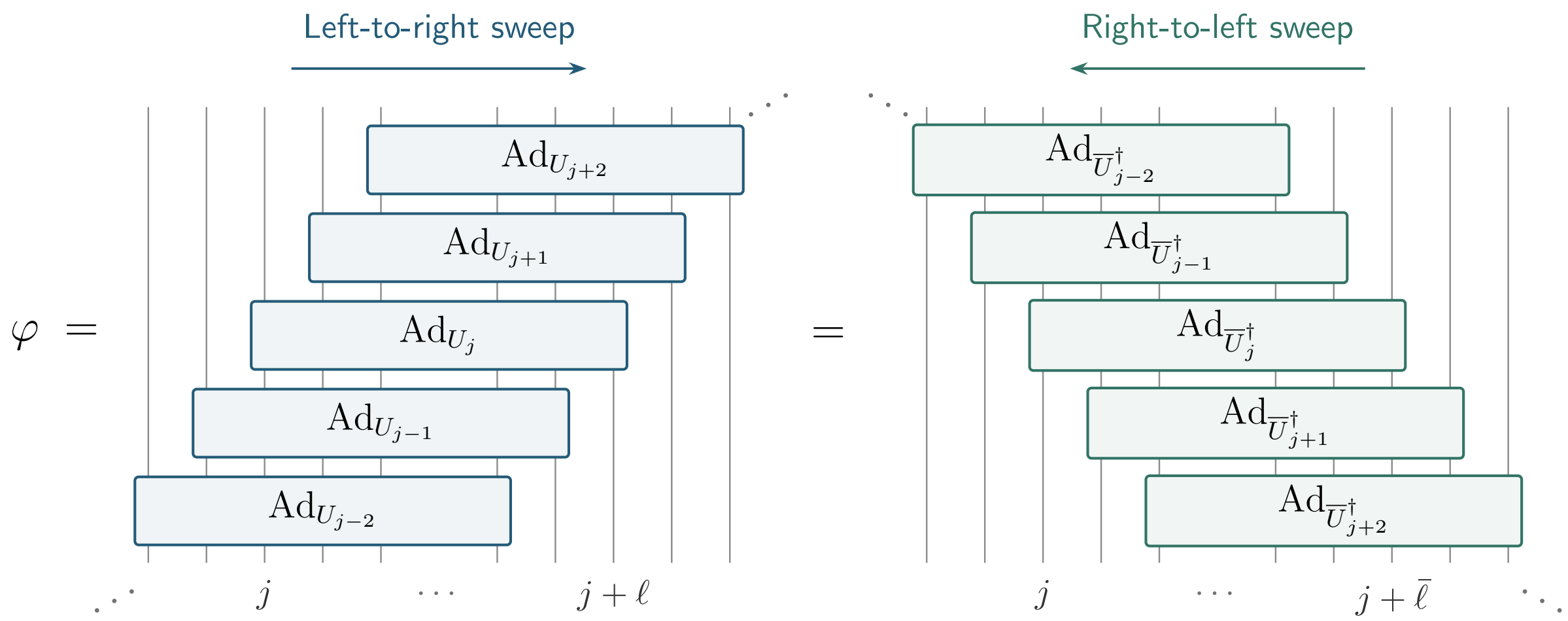

A dualizable sequential circuit consists of local unitary operators
\ie \label{dual.sequential.circuit}
    &U_j : \cH_{[j,j+\ell]} \otimes \bC^n \to \cH_{[j,j+\ell]} \otimes \bC^n \,, \\
    &\overline{U}_j: \cH_{[j,j+\bar{\ell}]} \otimes \bC^{\bar{n}} \to \cH_{[j,j+\bar{\ell}]} \otimes \bC^{\bar{n}}\,,
 \fe
such that for every local operator $a$ (see Figure \ref{fig:dualizable})
\ie \label{dual.ucp.from.sequential}
    &\lim_{N \to \infty} \tr_{\! \mathcal{B}^{(N,N+1)}} \!
    \left( U_N \cdots U_{-N} \,a\,
    U_{-N}^\dagger \cdots U_N^\dagger \right) =  \\&
    \lim_{N \to \infty} \tr_{\! \bar{\mathcal{B}}^{(-N-1,-N)}} \!
    \left( \overline{U}_{-N}^\dagger \cdots \overline{U}_N^\dagger \,a\,
    \overline{U}_N \cdots \overline{U}_{-N} \right) \,.
\fe
Here, $\tr_{\! \mathcal{B}^{(j,j+1)}}$ is a partial trace over the defect degrees of freedom $P_{j,j+1} (\cH_{(j,j+\ell]} \otimes \bC^{n})$ at link $(j,j+1)$ normalized as $\tr_{\! \mathcal{B}^{(j,j+1)}}(X) = {\Tr_{\cH_{(j,j+\ell]} \otimes \bC^n}(X  P_{j,j+1})}/{\Tr_{\cH_{(j,j+\ell]} \otimes \bC^n}(P_{j,j+1})}$, and similarly for the other trace. Recall $U_j P_{j-1,j} = P_{j,j+1} U_j$, thus the operator inside the trace in the first line of \eqref{dual.ucp.from.sequential} commutes with $P_{N,N+1}$ for large enough $N$, because $a$ commutes with $P_{-N-1,-N}$ for large enough $N$.\footnote{In \eqref{dual.ucp.from.sequential}, `$a$' should be understood as $a \otimes \mathbbm{1}_n$ and $a \otimes \mathbbm{1}_{\bar{n}}$ acting trivially on the ancillary space.}

The expressions in \eqref{dual.ucp.from.sequential} define a quantum channel that is locality-preserving with range $\max\{\ell, \bar{\ell}\}$. Following the same argument as in footnote \ref{footnote.stabilization}, the sequences in \eqref{dual.ucp.from.sequential} stabilize for sufficiently large $N$. For a local operator $a \in \cA_{[j_1,j_2]}$,
\ie \label{finite.ucp.from.sequential}
    \f(a)
    &= \tr_{\! \mathcal{B}^{(j_2+\bar{\ell},j_2+\bar{\ell}+1)} } \!
    \left( U_{j_2+\bar{\ell}} \cdots U_{j_1-\ell} \,a\,
    U_{j_1-\ell}^\dagger \cdots U_{j_2+\bar{\ell}}^\dagger \right) \\
    &= \tr_{\! \bar{\mathcal{B}}^{(j_1-\ell-\bar{\ell}-1,j_1-\ell-\bar{\ell})}} \!
    \left( \overline{U}_{j_1-\ell-\bar{\ell}}^\dagger \cdots
    \overline{U}_{j_2}^\dagger \,a\,
    \overline{U}_{j_2} \cdots \overline{U}_{j_1-\ell-\bar{\ell}} \right)
    \in \cA_{[j_1-\ell,j_2+\bar{\ell}]} \,.
\fe
Similarly, the dual map $\f^\star$ can be written as
\ie \label{dual.ucp.from.sequential.star}
    \f^\star(a)
    &= \tr_{\! \bar{\mathcal{B}}^{(j_2+{\ell},j_2+{\ell}+1)} } \!
    \left( \overline{U}_{j_2+\ell} \cdots
    \overline{U}_{j_1-\bar{\ell}} \,a\,
    \overline{U}_{j_1-\bar{\ell}}^\dagger \cdots
    \overline{U}_{j_2+\ell}^\dagger \right) \\
    &= \tr_{\! \mathcal{B}^{(j_1-\bar\ell-{\ell}-1,j_1-\bar\ell-{\ell})} } \!
    \left( U_{j_1-\bar{\ell}-\ell}^\dagger \cdots U_{j_2}^\dagger \,a\,
    U_{j_2} \cdots U_{j_1-\bar{\ell}-\ell} \right)
    \in \cA_{[j_1-\bar{\ell},j_2+\ell]} \,,
\fe
which is also locality-preserving. We have therefore shown that $\f$ constructed from a dualizable sequential circuit is locality-preserving and dualizable as a UCP map. 

\subsection{Kramers--Wannier sequential circuit and its deformation \label{sec:kw.deformed}}

Here, we discuss a one-parameter family of sequential circuits that includes the Kramers--Wannier defect at a special point. As we will see, this one-parameter family is neither dualizable nor locality-preserving except at special points.

Consider the sequential circuit associated with
\ie \label{U.theta}
    U_j^\theta = e^{i\pi X_j/4} e^{-i\theta Y_jZ_{j+1}} = e^{i\theta Z_jZ_{j+1}}e^{i\pi X_j/4} \,,
\fe
whose action on local operators is given by
\ie \label{kw.gate.action}
    U_j^\theta : ~ \begin{aligned}
        X_j &\mapsto \cos(2\theta)X_j-\sin(2\theta)Y_jZ_{j+1} \, , &\quad X_{j+1} &\mapsto \cos(2\theta)X_{j+1}-\sin(2\theta)Z_jY_{j+1} \,, \\
         Y_j &\mapsto -Z_j \, , & Y_{j+1} &\mapsto \cos(2\theta)Y_{j+1}+\sin(2\theta)Z_jX_{j+1} \,, \\
        Z_j &\mapsto \cos(2\theta)Y_j+\sin(2\theta)X_jZ_{j+1} \, , & Z_{j+1} &\mapsto Z_{j+1} \,.
    \end{aligned}
\fe
Following \eqref{ucp.from.sequential}, the corresponding UCP map is
\ie \label{deformed.kw.map}
    \dd_\theta(X_j) &= \sin(2\theta)Z_{j-1}Z_j + i\cos(2\theta) Z_j  \dd_\theta(Z_j) \,,\\
        \dd_\theta(Z_j) &= \cos(2\theta)\sum_{n=0}^{\infty}\sin(2\theta)^nX_jX_{j+1}\cdots X_{j+n-1}Y_{j+n} \,.
\fe
Note that the parameter $\theta$ is $\pi$-periodic. 

To derive \eqref{deformed.kw.map}, first note that all gates $U_k^\theta$ with $k<j$ leave $Z_j$ unchanged, while
\ie \label{derivation}
    &U_N^\theta \cdots U_j^\theta \, Z_j \, (U_j^\theta)^\dagger \cdots (U_N^\theta)^\dagger \\
    &\qquad = \cos(2\theta)\sum_{n=0}^{N-j}\sin(2\theta)^n X_j \cdots X_{j+n-1}Y_{j+n}
    + \sin(2\theta)^{N-j+1}X_j \cdots X_N Z_{N+1} \,,
\fe
which can be verified recursively using \eqref{kw.gate.action}. Taking the partial trace over site $N+1$ removes the second term in \eqref{derivation} and we recover the second line of \eqref{deformed.kw.map}. Note that when $|\sin(2\theta)| \neq 1$, the partial trace is not needed since the second term vanishes as $N \to \infty$. Hence, the trace is needed only for $\theta = \pi/4 \pmod{\pi/2}$. The first line of \eqref{deformed.kw.map} follows from
\ie
    U_j^\theta U_{j-1}^\theta \, X_j \, (U_{j-1}^\theta)^\dagger (U_j^\theta)^\dagger &= \sin(2\theta)Z_{j-1}Z_j+\cos(2\theta)^2X_j-\cos(2\theta)\sin(2\theta)Y_jZ_{j+1} \\
    &= \sin(2\theta)Z_{j-1}Z_j+i\cos(2\theta)Z_j \, U_j^\theta Z_j(U_j^\theta)^\dagger \,.
\fe

For $\theta \notin \pi \bZ/4$, this map is locality-preserving from the left (as guaranteed by the sequential circuit construction), but not from the right.\footnote{The maps in \eqref{deformed.kw.map} are special since they are \emph{approximately} locality-preserving in the sense of \cite{Ranard:2020jkb, Ranard:2026hqu}.} Hence, the sequential circuit \eqref{U.theta} and its corresponding defect are not dualizable.  
The image of $\dd_\theta$ is nonetheless quasi-local since the sum in \eqref{deformed.kw.map} converges in the operator norm.\footnote{Note that $\| \sum_{n=J}^{\infty}\sin(2\theta)^nX_jX_{j+1}\cdots X_{j+n-1}Y_{j+n} \| \leq \frac{|\sin(2\theta)|^J}{1 - |\sin(2\theta)|}$, which vanishes as $J \to \infty$.} Moreover, $\dd_\theta$ is invertible for $\theta \notin \pi \bZ/4$ and defines a $*$-automorphism of $\cA$. In those cases, $\dd_\theta$ acts multiplicatively and is uniquely determined by \eqref{deformed.kw.map}. The dual UCP map, which coincides with its inverse when $\cos(2\theta) \neq 0$, is given by
\ie \label{deformed.kw.map.dual}
    \dd_\theta^\star(X_j) &= \sin(2\theta)Z_jZ_{j+1} - i \cos(2\theta) Z_j  \dd_\theta^\star(Z_j)  \,, \\
    \dd_\theta^\star(Z_j) &= -\cos(2\theta) \sum_{n=0}^{\infty} \sin(2\theta)^n Y_{j-n}X_{j-n+1}\cdots X_j \,.
\fe
We emphasize that even though $\dd_\theta^\star$ is the dual of $\dd_\theta$, the underlying sequential circuit does not admit a dual, i.e., $\overline{U}^\theta_j$ does not exist. This is because $\dd_\theta$ cannot be constructed by sweeping any defect from \emph{right to left} as that would imply that $\dd_\theta$ must be locality-preserving.

For $\theta \in \pi \bZ/4$, the UCP map \eqref{deformed.kw.map} and its dual \eqref{deformed.kw.map.dual} are locality-preserving, and the sequential circuits become dualizable. 
For $\theta \in \pi \bZ/2$, we get an on-site $\bZ_4$ symmetry generated by $e^{i\pi X_j/4}$ at each site. 
The dual circuits are generated by $\overline{U}_j^0 = (\overline{U}_j^{\pi/2})^\dagger = e^{-i \pi X_j/4}$.

At $\theta = \pi/4$, we recover the sequential circuit representation \cite{Chen:2023qst,Seiberg:2023cdc,Seiberg:2024gek} of the non-invertible Kramers--Wannier map $\cal D$ in \eqref{KW.map2}. The dual sequential circuit requires an ancillary qubit $\cH_\mathrm{a}=\mathbb{C}^2$ as pointed out in \cite{Seiberg:2024gek}, with Pauli operators $X_{\mathrm a}$ and $Z_{\mathrm a}$. The dual sequential circuit is generated by\footnote{The sequential circuit presentation of a UCP map is not unique. For instance, for any unitary operators $V_{j,j+1}$ acting on the degrees of freedom near link $(j,j+1)$, the quantum channels associated with $U_j$ and $U'_{j}=V_{j,j+1} U_{j} (V_{j-1,j})^\dagger$ are the same.}
\ie
    \overline{U}_j^{\pi/4} = \mathsf{CZ}_{j,\mathrm{a}} \, \mathsf{H}_j \, \mathsf{SWAP}_{j, \mathrm{a}} \,.
\fe
The quantum channel at $\theta=3\pi/4$ is related to the Kramers--Wannier transformation by $\dd_{3\pi/4} = (\prod_j\mathrm{Ad}_{Z_j}) \dd (\prod_j\mathrm{Ad}_{Z_j})$, whose dual sequential circuit is $\overline{U}_j^{3\pi/4} = Z_{j}\overline{U}_j^{\pi/4}Z_{j}$. 

To conclude, only the following quantum channels with $\theta\in \pi\mathbb{Z}/4$ are valid QCAs satisfying all of our conditions:
 \ie
    \dd_{\pi/4} = \dd \,, \quad \dd_{3\pi/4} = \left( \prod_j \mathrm{Ad}_{Z_j} \right) \dd \left( \prod_j \mathrm{Ad}_{Z_j} \right) \,, \quad \dd_{\pi/2} = \dd_{0}^{-1} =\prod_j \mathrm{Ad}_{e^{-i\pi X_j/4}}\,.
\fe

\subsubsection*{Cases with $\theta \notin \pi \bZ/4$}
Let us examine cases with $\theta \notin \pi \bZ/4$ more carefully and see why they do not lead to valid generalized symmetries. First of all, the map $\dd_\theta$ is invertible. Based on its action on local operators, it defines an invertible symmetry that is not locality-preserving. For $0< \theta < \pi/4$, the corresponding unitary operator is
\ie \label{v.n}
    \mathcal{V}_N =  \exp\left( i \sum_{-N \leq j_1 < j_2 \leq N} \frac{\sin(2\theta)^{j_2-j_1}}{2(j_2-j_1)} Z_{j_1} X_{j_1+1} \cdots X_{j_2-1} Y_{j_2} \right) \exp\left( \frac{i \pi}{4}\sum_{j=-N}^N X_j \right)\,,
\fe
in the sense that for every local operator $a$,\footnote{The limit here is taken in operator norm, i.e., $\| \dd_\theta(a)- \mathcal{V}_N a\mathcal{V}_N^\dagger \| \to 0$.}
\ie
    \dd_\theta(a) = \lim_{N \to \infty} \mathcal{V}_N \, a \, \mathcal{V}_N^\dagger  \,.
\fe

Interestingly, these non-local transformations are symmetries of the transverse-field Ising model away from the critical point (in the disordered phase). Specifically, the map $\dd_\theta$ preserves the Hamiltonian
\ie \label{H.TFIM}
    H(\theta) = -\sum_{j}\left( X_j + \sin(2\theta) Z_{j-1} Z_{j} \right)\,,
\fe
because of the relations
\ie \label{deformed.kw.energy}
    &\dd_\theta(Z_jZ_{j+1}) = \sin(2\theta) X_j+\cos(2\theta)^2\sum_{r=1}^{\infty}\sin(2\theta)^{r-1}Y_jX_{j+1}\cdots X_{j+r-1}Y_{j+r} \, , \\
    &\dd_\theta\big(X_j+\sin(2\theta)Z_jZ_{j+1}\big) = X_j+\sin(2\theta)Z_{j-1}Z_j \,.
\fe
We can even construct the defect Hamiltonians as
\ie
    H^{(J-1,J)}_{\defect_\theta} = -\sum_{j \neq J}\left( X_j + \sin(2\theta) Z_{j-1} Z_{j} \right) -\cos(2\theta)X_J + \sin(2\theta) Z_{J-1} Y_J \,,
\fe
satisfying $U_j^\theta H_{\defect_\theta}^{(j-1,j)} (U_j^\theta)^\dagger=H_{\defect_\theta}^{(j,j+1)}$. However, these defects are not dualizable in the sense that there is no dual defect $\overline{\defect_\theta}$ with the fusion rule $\defect_\theta \otimes \overline{\defect_\theta} = \mathbbm{1} \oplus \cdots$. 

On a periodic chain, the sequential circuit yields a Sequential Matrix Product Operator (sMPO) following \cite{Tantivasadakarn:2025txn}. See \cite{Seiberg:2024gek} for the construction of the MPO from topological defects in the case of the Kramers--Wannier non-invertible symmetry. Writing the movement operator as $U_j^\theta = \mathbbm{1}_{j+1} \otimes \cos(\theta) e^{i\pi X_j/4}  + Z_{j+1} \otimes (-i \sin(\theta) e^{i\pi X_j/4} Y_j) $, on a periodic chain with $L$ sites, we find the sMPO\footnote{The MPO is injective for $\theta \notin \frac{\pi}{2}\bZ$ since $e^{i\pi X_j/4}, X_je^{i\pi X_j/4}, Y_je^{i\pi X_j/4}, Z_je^{i\pi X_j/4}$ span $\bC^{2\times2}$. For $\theta \in \frac{\pi}{2}\bZ$, the MPO is an on-site unitary symmetry and can be reduced to an injective tensor with bond dimension one.}
\ie \label{kw.deformed.mpo}
    \op_{\theta} = \Tr\!\!\left[ \mathbb{M}_L^\theta \cdots \mathbb{M}_2^\theta \mathbb{M}_1^\theta \right]\,, \qquad  \mathbb{M}_j^\theta = \begin{pmatrix}
    \cos(\theta)\mathbbm{1}_j  && \cos(\theta) Y_j\\
    i\sin(\theta)Z_j  &~~& \sin(\theta)X_j \end{pmatrix} e^{i\pi X_j/4}\,.
\fe
This MPO satisfies the fusion algebra
\ie \label{d.theta.fusion}
    \op_{\theta}^\dagger \op_{\theta} = \op_{\theta} \op_{\theta}^\dagger = 1 + \sin(2\theta)^L X_1 X_2 \cdots X_L \,.
\fe
The fusion coefficients are not non-negative integers for $\theta \notin \pi \bZ/4$, unlike the usual fusion category symmetries. The MPO $\op_{\theta}$ commutes with the Hamiltonian \eqref{H.TFIM} on a periodic chain since 
\ie
    &\op_{\theta} \, X_j = \sin(2\theta)Z_{j-1}Z_{j} \, \op_{\theta} + i\cos(2\theta) Z_j \,\op_{\theta}\, Z_j \,, \\
    &\op_{\theta} \, \sin(2\theta)Z_jZ_{j+1} = X_{j} \, \op_{\theta} - i\cos(2\theta) Z_j \,\op_{\theta}\, Z_j \,.
\fe

Note that the operator $\op_{\theta}$ is invertible but not unitary. We can instead consider the operator
\ie \label{v.theta}
    \mathsf{V}_\theta = \op_{\theta}  \left( 1+\sin(2\theta)^L X_1 \cdots X_L \right)^{-1/2}\,,
\fe
which is unitary and commutes with the Hamiltonian. However, $\mathsf{V}_\theta$ is not locality-preserving and does not have a finite range. The unitary operator $\mathsf{V}_\theta$ in \eqref{v.theta} is closely related to $\mathcal{V}_N$ in \eqref{v.n}, as their actions on local operators agree in the thermodynamic limit. The fusion algebra of the sMPO $\op_{\theta}$ does not match that of the corresponding UCP map since $\dd_\theta$ is invertible and therefore
\ie
    \dd_\theta^\star \circ \dd_\theta = \dd_\theta \circ \dd_\theta^\star = \mathbbm{1}\,.
\fe

In summary, the sequential circuit built from a generic two-site unitary operator \eqref{simple.sequential.circuit} leads to an automorphism of $\mathscr{R}$ that is not locality-preserving. For special choices of the unitary operator, the sequential circuit yields a locality-preserving, non-invertible symmetry, such as the Kramers--Wannier symmetry. We compare this example with the sMPO framework of~\cite{Tantivasadakarn:2025txn} in Appendix~\ref{app:dsMPO}.

\section{Higher dimensions} \label{sec:higher.d}

So far, we have focused on non-invertible symmetries in 1+1d systems. We now turn to higher dimensions, where non-invertible symmetries typically involve higher-form global symmetries \cite{Gaiotto:2014kfa}.

In the first two examples below, the action on local operators is locality-preserving only when the relevant higher-form symmetry is exactly topological. This requires local constraints on the operator algebra and therefore cannot be realized on the full operator algebra of a lattice system with a tensor-product Hilbert space. Accordingly, in Sections~\ref{sec:2+1d} and~\ref{sec:wegner}, we impose these constraints and work with the resulting constrained observable algebras.

In Section~\ref{sec:subsystem}, we discuss the subsystem non-invertible symmetry of \cite{Cao:2023doz}, which only requires subsystem symmetries and can be realized on a tensor-product Hilbert space. However, it satisfies our locality-preserving condition only for regions of special shape, and is therefore not a non-invertible QCA. This reflects the fact that its associated non-invertible defect is not fully topological.

\subsection{Non-invertible and 1-form symmetries in 2+1d}\label{sec:2+1d}

Here we consider a lattice non-invertible symmetry in a 2+1d $\mathbb{Z}_2$ gauge theory coupled to Ising matter fields studied in \cite{Choi:2024rjm}. This symmetry involves a $\mathbb{Z}_2^{(0)}$ 0-form global symmetry and a $\mathbb{Z}_2^{(1)}$ 1-form global symmetry. The space is a 2d square lattice, with a qubit on every site and on every link. We denote the Pauli operators that act on site $s$ and link $\ell$ by $X_s,Z_s$ and $\sigma^x_\ell,\sigma^z_\ell$, respectively.

On a finite lattice, the Hilbert space is a tensor product $\mathcal{H} = \bigotimes_s {\cal H}_s \otimes \bigotimes_\ell {\cal H}_\ell$ where each ${\cal H}_s$ and ${\cal H}_\ell$ is isomorphic to $\mathbb{C}^2$. 
The non-invertible operator $\op$ (which is self-adjoint, $\op^\dagger=\op$) acts on the local operators as
\ie
\op X_s = \left(\prod_{\ell \ni s}\sigma^x_\ell\right) \op\,,~~~~~~
\op Z_s Z_{s'} =\left(\prod_{\ell\in \gamma_{ss'}} \sigma^z_\ell\right) \op ,
\fe
where $\gamma_{ss'}$ is any curve that connects $s$ and $s'$. 
Because $\op \left(\prod_{\ell \in \gamma} \sigma^z_\ell\right)=\left(\prod_{\ell \in \gamma} \sigma^z_\ell\right)\op = \op$ for any closed curve $\gamma$, the choice of the open curve $\gamma_{ss'}$ does not matter. 
This non-invertible symmetry operator arises from gauging the $\mathbb{Z}_2^{(0)}$ symmetry $U = \prod_s X_s$ and the $\mathbb{Z}_2^{(1)}$ 1-form symmetry $\eta(\gamma) = \prod_{\ell \in \gamma} \sigma^z_\ell$. 
The 1-form symmetry operator depends not only on the homology class of the closed curve $\gamma$, but also on its detailed shape. 
In this sense, it is referred to as a non-topological 1-form symmetry \cite{Seiberg:2019vrp, Qi:2020jrf, Oh:2023bnk, Choi:2024rjm, Gorantla:2024ocs, Liu:2026tcl, Brito:2026eqi}.

This non-invertible operator $\op$ in finite volume cannot be extended to a locality-preserving map on the full algebra of local operators of an infinite lattice. To see this, suppose there were such a map $\dd$ in infinite volume. Then
\ie
\dd(Z_s Z_{s'})
\overset{?}{=}\prod_{\ell\in\gamma_{ss'}}\sigma^z_\ell \,.
\fe
However, there is no canonical way to choose the open curve $\gamma_{ss'}$. 
This is unlike the Kramers--Wannier map in 1+1d, where there is a unique path connecting two points on an infinite line. 
Moreover, one can take the open curve $\gamma_{ss'}$ to be arbitrarily far away from $s,s'$, and the image operator can be made arbitrarily non-local.

\begin{figure}
\centering
\begin{tikzpicture}[baseline=(current bounding box.center)]

  \def\seglen{0.8}   
  \def\gap{0.09}    
  \def\h{0.8}       

  \begin{scope}[shift={(-1,0)}]
    \coordinate (a1) at (0,0);
    \coordinate (a2) at (2.8,0);
    \fill (a1)[blue] circle (1.5pt);
    \fill (a2)[blue] circle (1.5pt);

    \draw[gray!50, opacity=0.35, line width=1.2pt]
      (0.10,0) -- ++(\seglen,0);
    \draw[gray!50, opacity=0.35, line width=1.2pt]
      ({0.10+\seglen+\gap},0) -- ++(\seglen,0);
    \draw[gray!50, opacity=0.35, line width=1.2pt]
      ({0.10+2*(\seglen+\gap)},0) -- ++(\seglen,0);

    \node[blue,below=2pt] at (a1) {$Z_s$};
    \node[blue,below=2pt] at (a2) {$Z_{s'}$};
  \end{scope}

  \draw[-{Latex[length=3mm,width=2mm]}, thick]
    (2.8,0) -- (4.7,0)
    node[midway, above] {$\dd$};

  \begin{scope}[shift={(5.5,0)}]

    \draw[thick,red] (0.10,0) -- ++(\seglen,0);
    \draw[thick,red] ({0.10+\seglen+\gap},0) -- ++(\seglen,0);
    \draw[thick,red] ({0.10+2*(\seglen+\gap)},0) -- ++(\seglen,0);

    \node [red] at ({0.10+0.5*\seglen},-0.25) {$\sigma^z$};
    \node [red] at ({0.10+\seglen+\gap+0.5*\seglen},-0.25) {$\sigma^z$};
    \node [red] at ({0.10+2*(\seglen+\gap)+0.5*\seglen},-0.25) {$\sigma^z$};
  \end{scope}

  \node at (9.5,0) {$=$};

  \begin{scope}[shift={(10.5,0)}]

    \draw[thick,red] (0.10,0) -- ++(\seglen,0);
    \draw[thick,red] ({0.10+\seglen},0) -- ({0.10+\seglen},\h);

    \draw[thick,red] ({0.10+\seglen},\h) -- ++(\seglen,0);
    \draw[thick,red] ({0.10+2*\seglen},\h) -- ({0.10+2*\seglen},0);

    \draw[thick,red] ({0.10+2*\seglen},0) -- ++(\seglen,0);

    \node [red]at ({0.2+0.5*\seglen},0.5) {$\sigma^z$};
    \node [red]at ({0.10+\seglen+0.5*\seglen},\h+0.25) {$\sigma^z$};
    \node [red]at ({2*\seglen+0.5*\seglen},0.5) {$\sigma^z$};
    \node [red]at ({0.10+2*\seglen+0.5*\seglen},-0.25) {$\sigma^z$};
    \node [red]at ({0.10+0.5*\seglen},-0.25) {$\sigma^z$};
  \end{scope}

\end{tikzpicture}
\caption{The non-invertible map $\dd$ for the non-invertible symmetry in the 2+1d $\mathbb{Z}_2$ gauge theory coupled to Ising matter fields  \cite{Choi:2024rjm}. In infinite volume, this map is only unambiguous on the algebra ${\cal A}^\text{top}$ satisfying the local constraints in \eqref{flat}, which make the $\mathbb{Z}_2^{(1)}$ 1-form symmetry topological.}\label{fig:2+1d}
\end{figure}
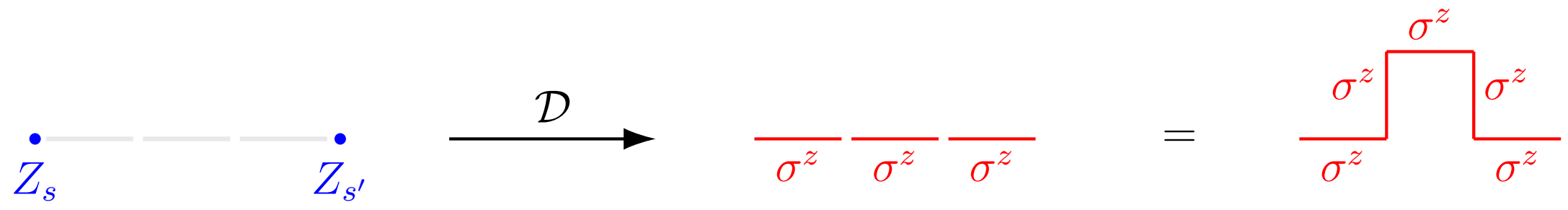

One way to overcome this issue is to impose a local operator constraint on every plaquette:
\ie\label{flat}
\prod_{\ell \in p} \sigma^z_\ell=1,~~~\forall ~p.
\fe
We work with the algebra ${\cal A}^\text{top}$ defined by first restricting to the subalgebra of local operators that commute with every plaquette operator $B_p = \prod_{\ell \in p} \sigma^z_\ell$, and then quotienting by the above constraints. The quotient amounts to the identification\footnote{Mathematically, we take the quotient by the closed ideal generated by $\prod_{\ell\in p}\sigma_\ell^z-1$ for all plaquettes $p$.}
\ie
    a \sim b \qquad \text{iff} \qquad a \left( \prod_{p \in R} \frac{1+B_p}{2} \right) = b \left( \prod_{p \in R} \frac{1+B_p}{2} \right) \,,
\fe
where the product is over the plaquettes in a finite rectangle $R$ containing the supports of both $a$ and $b$. The 1-form symmetry becomes topological in this algebra.  
We then define a locality-preserving map $\dd$ on ${\cal A}^\text{top}$ in infinite volume as follows. 
We first split ${\cal A}^\text{top}={\cal A}^\text{even}\oplus {\cal A}^\text{odd}$ into even and odd subspaces under the $\mathbb{Z}_2^{(0)}$ symmetry. 
This map obeys $\dd(aa') = \dd(a)\dd(a')$ if $a,a'\in \mathcal{A}^\text{even}$ and $\dd(b)=0$ if $b\in \mathcal{A}^\text{odd}$. 
For instance, $\dd(Z_s)=0$. 
This map is then completely determined by its action on the generators for ${\cal A}^\text{even}$:
\ie
&\dd (X_s )  = \prod_{\ell\ni s} \sigma_\ell^x,~~~~&&
\dd(Z_sZ_{s'}) = \prod_{\ell \in \gamma_{ss'}}\sigma^z_\ell ,\\
&\dd\left( \prod_{\ell \ni s} \sigma^x_\ell\right) = X_s,~~~~&&
\dd\left(\sigma^z_\ell\right)  = \prod_{s\in \ell}Z_s \,.
\fe
Here $\gamma_{ss'}$ is any path connecting $s$ and $s'$. Due to the constraints \eqref{flat}, the image of $Z_sZ_{s'}$ is independent of the choice of $\gamma_{ss'}$. See Figure \ref{fig:2+1d}. 
Note that we do not define the map on a single $\sigma^x_\ell$, since the latter violates the constraints \eqref{flat}.

Similar to the discussion for the Kramers--Wannier map in 1+1d in Section \ref{sec:LQC}, the locality-preserving condition for $\dd$ holds only for
regions with the topology of a disk. For instance, consider the operator $Z_sZ_{s'}$ in Figure \ref{fig:2+1d} as supported in two disconnected disks around the two points. The image $\dd(Z_sZ_{s'})$ is not contained in a thickening of these two disks. The condition also fails for an annulus: $\mathcal{D}$ maps the product of $\sigma^x_\ell$ along a loop on the dual lattice to a product of $X_s$ over the interior disk.

This non-invertible map obeys
\ie
\dd\circ\dd = E\,,
\fe
where $E$ is a conditional expectation projecting an operator onto its $\mathbb{Z}_2^{(0)}$ even component:
\ie
E(a) = \frac a2 
+\frac12\left(\prod_{s\in \text{supp}(a)}X_s\right)\,a\,
\left(\prod_{s\in \text{supp}(a)}X_s\right) \,.
\fe
We also have $\dd^\star = \dd$, where the dual is taken with respect to the tracial state of ${\cal A}^\text{top}$. This is to be contrasted with the algebra of the non-invertible operator $\op$ in finite volume \cite{Choi:2024rjm} where we have imposed the constraint \eqref{flat},
\ie
\op^2  =  \left(1+\prod_s X_s\right) \mathsf{C}_\eta,
\fe
where $\mathsf{C}_\eta =\frac{(1+\eta_x)(1+\eta_y)}{2}$ is the condensation operator \cite{Gaiotto:2019xmp, Roumpedakis:2022aik}, which projects to the $\mathbb{Z}_2^{(1)}$ invariant space. 
Here $\eta_i = \prod_{\ell\in \gamma_i} \sigma^z_\ell$ is the $\mathbb{Z}_2^{(1)}$ symmetry operator and  $\gamma_i$ is a non-contractible loop in the $i$-direction. The conditional expectation $E$ is the counterpart of the projector $1+\prod_s X_s$ in infinite volume. The condensation operator $\mathsf{C}_\eta$, on the other hand, becomes trivial in ${\cal A}^\text{top}$ in infinite volume.

\subsection{Wegner duality symmetry in 3+1d} \label{sec:wegner}

The most straightforward generalization of the 1+1d Kramers--Wannier symmetry is the non-invertible symmetry in 3+1d associated with the Wegner duality \cite{Wegner:1971app}. 
This non-invertible symmetry is related to the invariance under gauging a $\mathbb{Z}_2^{(1)}$ 1-form global symmetry. 
It is realized in 3+1d $\mathbb{Z}_2$ lattice gauge theory formulated in the Euclidean \cite{Koide:2021zxj, Koide:2023rqd,Choi:2021kmx} and Hamiltonian formalisms  \cite{Gorantla:2024ocs, Gorantla:2024ptu}. 
It was formulated as a QCA on the quotient algebra of $\mathbb{Z}_2^{(1)}$-symmetric operators in~\cite{Inamura:2026kco} (see also~\cite{Ma:2024ypm, Zhang:2026kjf}). This example is interesting because while the operator is non-invertible in finite volume, it extends to an (invertible) automorphism on the local operator algebra in infinite volume when the 1-form symmetry is topological.

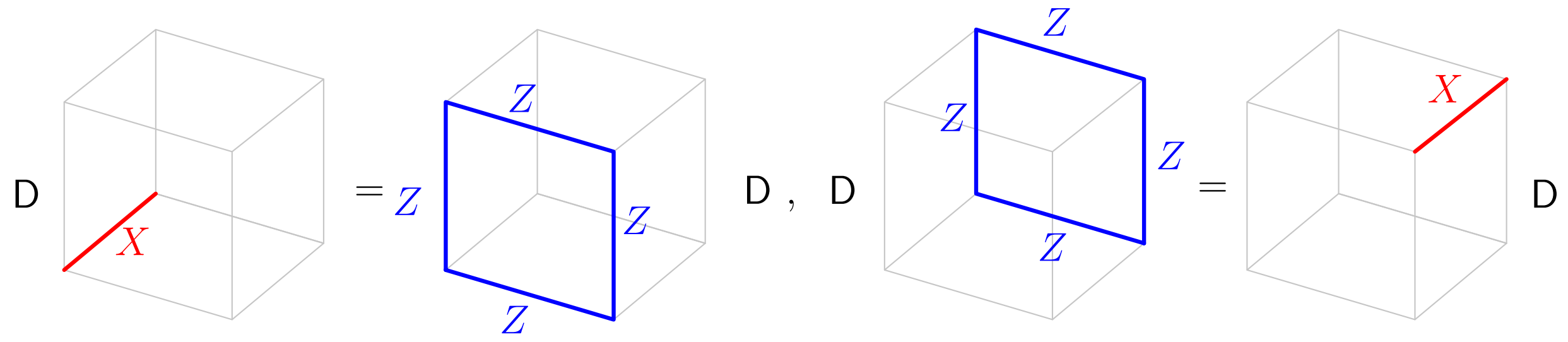
\begin{figure}

\begin{tikzpicture}[scale=0.4,line cap=round,line join=round]

\begin{scope}

\coordinate (A) at (0.0,0.0);   
\coordinate (B) at (4.4,-1.3);  
\coordinate (C) at (6.8,0.7);   
\coordinate (D) at (2.4,2.0);   

\coordinate (E) at (0.0,4.4);   
\coordinate (F) at (4.4,3.1);   
\coordinate (G) at (6.8,5.0);   
\coordinate (H) at (2.4,6.3);   

\draw[gray!45, thin] (A)--(B)--(C)--(D)--cycle;
\draw[gray!45, thin] (E)--(F)--(G)--(H)--cycle;
\draw[gray!45, thin] (A)--(E);
\draw[gray!45, thin] (B)--(F);
\draw[gray!45, thin] (C)--(G);
\draw[gray!45, thin] (D)--(H);


\draw[gray!45, thin] (D)--(C);

\draw[red, very thick] (A)--(D);

\node[ font=\itshape] at (-1,2) {$\op$};
\node[red, font=\itshape] at (1.8,0.75) {$X$};
\node[font=\itshape] at (8,2) {$=$};

\end{scope}

\begin{scope}[shift={(10,0)}]

\coordinate (A) at (0.0,0.0);   
\coordinate (B) at (4.4,-1.3);  
\coordinate (C) at (6.8,0.7);   
\coordinate (D) at (2.4,2.0);   

\coordinate (E) at (0.0,4.4);   
\coordinate (F) at (4.4,3.1);   
\coordinate (G) at (6.8,5.0);   
\coordinate (H) at (2.4,6.3);   

\draw[gray!45, thin] (A)--(B)--(C)--(D)--cycle;
\draw[gray!45, thin] (E)--(F)--(G)--(H)--cycle;
\draw[gray!45, thin] (A)--(E);
\draw[gray!45, thin] (B)--(F);
\draw[gray!45, thin] (C)--(G);
\draw[gray!45, thin] (D)--(H);


\draw[gray!45, thin] (D)--(C);

\draw[blue, very thick] (A)--(B)--(F)--(E)--(A);

\node[blue, font=\itshape] at (-1,1.8) {$Z$};
\node[blue, font=\itshape] at (1.8,-1.3) {$Z$};
\node[blue, font=\itshape] at (2,4.5) {$Z$};
\node[blue, font=\itshape] at (5,1.3) {$Z$};
\node[ font=\itshape] at (8.5,2) {$\op~,$};

\end{scope}

\begin{scope}[shift={(21.5,0)}]

\coordinate (A) at (0.0,0.0);   
\coordinate (B) at (4.4,-1.3);  
\coordinate (C) at (6.8,0.7);   
\coordinate (D) at (2.4,2.0);   

\coordinate (E) at (0.0,4.4);   
\coordinate (F) at (4.4,3.1);   
\coordinate (G) at (6.8,5.0);   
\coordinate (H) at (2.4,6.3);   

\draw[gray!45, thin] (A)--(B)--(C)--(D)--cycle;
\draw[gray!45, thin] (E)--(F)--(G)--(H)--cycle;
\draw[gray!45, thin] (A)--(E);
\draw[gray!45, thin] (B)--(F);
\draw[gray!45, thin] (C)--(G);
\draw[gray!45, thin] (D)--(H);


\draw[gray!45, thin] (D)--(C);

\draw[blue, very thick] (C)--(D)--(H)--(G)--(C);

\node[blue, font=\itshape] at (7.5,3) {$Z$};
\node[blue, font=\itshape] at (4.5,6.5) {$Z$};
\node[blue, font=\itshape] at (1.8,4) {$Z$};
\node[blue, font=\itshape] at (4.4,.6) {$Z$};
\node[ font=\itshape] at (-1.1,2.1) {$\op$};
\node[ font=\itshape] at (8.6,2.1) {$=$};

\end{scope}

\begin{scope}[shift={(31,0)}]

\coordinate (A) at (0.0,0.0);   
\coordinate (B) at (4.4,-1.3);  
\coordinate (C) at (6.8,0.7);   
\coordinate (D) at (2.4,2.0);   

\coordinate (E) at (0.0,4.4);   
\coordinate (F) at (4.4,3.1);   
\coordinate (G) at (6.8,5.0);   
\coordinate (H) at (2.4,6.3);   

\draw[gray!45, thin] (A)--(B)--(C)--(D)--cycle;
\draw[gray!45, thin] (E)--(F)--(G)--(H)--cycle;
\draw[gray!45, thin] (A)--(E);
\draw[gray!45, thin] (B)--(F);
\draw[gray!45, thin] (C)--(G);
\draw[gray!45, thin] (D)--(H);


\draw[gray!45, thin] (D)--(C);

\draw[red, very thick] (G)--(F);

\node[ font=\itshape] at (7.8,2) {$\op$};
\node[red, font=\itshape] at (5.2,4.75) {$X$};

\end{scope}

\end{tikzpicture}
\caption{The action of the non-invertible Wegner symmetry operator $\op$ in finite volume on the local operators \cite{Gorantla:2024ocs}.}\label{fig:3+1dDaction}
\end{figure}

We will follow the Hamiltonian lattice setup in \cite{Gorantla:2024ocs, Gorantla:2024ptu}. 
Let space be a cubic lattice with a qubit on every link $\ell$. 
In finite volume, the non-invertible operator $\op$ acts on the local operators as
\ie
\op X_\ell =\left(\prod_{\ell' \in\mathsf{t}(\ell)}Z_{\ell'}\right)\op,~~~
\op\left( \prod_{\ell\in p} Z_\ell \right)=  X_{\mathsf{t}(p)}\op,
\fe
where $\mathsf{t}$ is a half-translation, i.e., $\mathsf{t}(x,y,z) = (x+1/2,y+1/2,z+1/2)$. 
It maps a link $\ell$ centered at $(x+1/2,y,z)$ to a plaquette $\mathsf{t}(\ell)$ centered at $(x+1, y+1/2,z+1/2)$, etc. 
Similarly, it maps a plaquette $p$ centered at $(x+1/2,y+1/2,z)$ to a link $\mathsf{t}(p)$ centered at $(x+1,y+1,z+1/2)$. 
See Figure \ref{fig:3+1dDaction}. 
This operator is non-invertible; it annihilates every state that is charged under the $\mathbb{Z}_2^{(1)}$ 1-form global symmetry generated by $\eta(\widehat\Sigma)=\prod_{\ell \in \widehat\Sigma} X_\ell$, where $\widehat\Sigma$ is any closed two-surface on the dual lattice.

Similar to the discussion in Section \ref{sec:2+1d}, this operator in finite volume cannot be extended unambiguously to a map in infinite volume on the full algebra of local operators.  
One may attempt to define a map that takes a plaquette operator $B_p = \prod_{\ell\in p}Z_\ell$ to a single $X_{\mathsf{t}(p)}$ as in the right part of Figure \ref{fig:3+1dDaction}. 
However, this plaquette operator $B_p$ itself can be written as the product of five other plaquette operators in the same cube. Applying the proposed map to different decompositions of the same operator yields different images, so it is not well-defined.

To unambiguously extend this operator in finite volume to a map in infinite volume, we again need to work with the algebra ${\cal A}^\text{top}$ defined by taking the subalgebra of local operators that commute with the following constraint at every site $s$, and taking the quotient by it:
\ie\label{3+1dflat}
G_s = \prod_{\ell\ni s} X_\ell=1,~~~\forall ~s.
\fe
This can be interpreted as the Gauss-law constraint of the $\mathbb Z_2$ lattice gauge theory. Imposing the constraints \eqref{3+1dflat} renders the $\mathbb Z_2^{(1)}$ symmetry topological. The resulting algebra ${\cal A}^{\mathrm{top}}$ is generated by the local operators $X_\ell$ and the plaquette operators $\prod_{\ell\in p} Z_\ell$, and the map $\dd$ acts on them as
\ie
\dd(X_\ell) =\prod_{\ell' \in\mathsf{t}(\ell)}Z_{\ell'},~~~
\dd\left( \prod_{\ell\in p} Z_\ell \right)=  X_{\mathsf{t}(p)}.
\fe

\begin{figure}
\centering
 \begin{tikzpicture}[
    x={(1.05cm,-0.24cm)},
    y={(-0.72cm,-0.38cm)},
    z={(0cm,1.12cm)},
    line cap=round,
    line join=round
]

\begin{scope}[scale=0.6, transform shape]
    \foreach \k in {0,1,2,3,4} {
        \draw[black!25, line width=0.7pt] (0,0,\k) -- (4,0,\k);
        \draw[black!25, line width=0.7pt] (0,1,\k) -- (4,1,\k);

        \foreach \i in {0,1,2,3,4} {
            \draw[black!25, line width=0.7pt] (\i,0,\k) -- (\i,1,\k);
        }
    }

    \foreach \i in {0,1,2,3,4} {
        \draw[black!25, line width=0.7pt] (\i,0,0) -- (\i,0,4);
        \draw[black!25, line width=0.7pt] (\i,1,0) -- (\i,1,4);
    }

    \draw[blue, line width=2.8pt]
        (0,0,0) -- (4,0,0) -- (4,0,4) -- (0,0,4) -- cycle;

    \node[blue, font=\Large] at (2.15,-0.12,4.35) {$\gamma$};
\end{scope}

\begin{scope}[xshift=6.7cm, scale=0.7, transform shape]
    \foreach \k in {0,1,2,3,4} {
        \draw[black!25, line width=0.7pt] (0,0,\k) -- (4,0,\k);
        \draw[black!25, line width=0.7pt] (0,1,\k) -- (4,1,\k);

        \foreach \i in {0,1,2,3,4} {
            \draw[black!25, line width=0.7pt] (\i,0,\k) -- (\i,1,\k);
        }
    }

    \foreach \i in {0,1,2,3,4} {
        \draw[black!25, line width=0.7pt] (\i,0,0) -- (\i,0,4);
        \draw[black!25, line width=0.7pt] (\i,1,0) -- (\i,1,4);
    }

    \foreach \i in {1,2,3,4} {
        \foreach \k in {1,2,3,4} {
            \draw[red, line width=2.8pt]
                (\i,1,\k) -- (\i,0,\k);
        }
    }

      \draw[-{Latex[length=3mm,width=2mm]}, thick]
    (-4,0,0) -- (-2.5,-.9,0) node[midway, above] {$\dd$};
\end{scope}

\end{tikzpicture}
\caption{A loop $\gamma$ of Pauli-$Z$ operators (blue) is mapped to a square of Pauli-$X$ operators (red) under the Wegner duality map $\dd$ \cite{Gorantla:2024ocs}.}\label{fig:loop}
\end{figure}
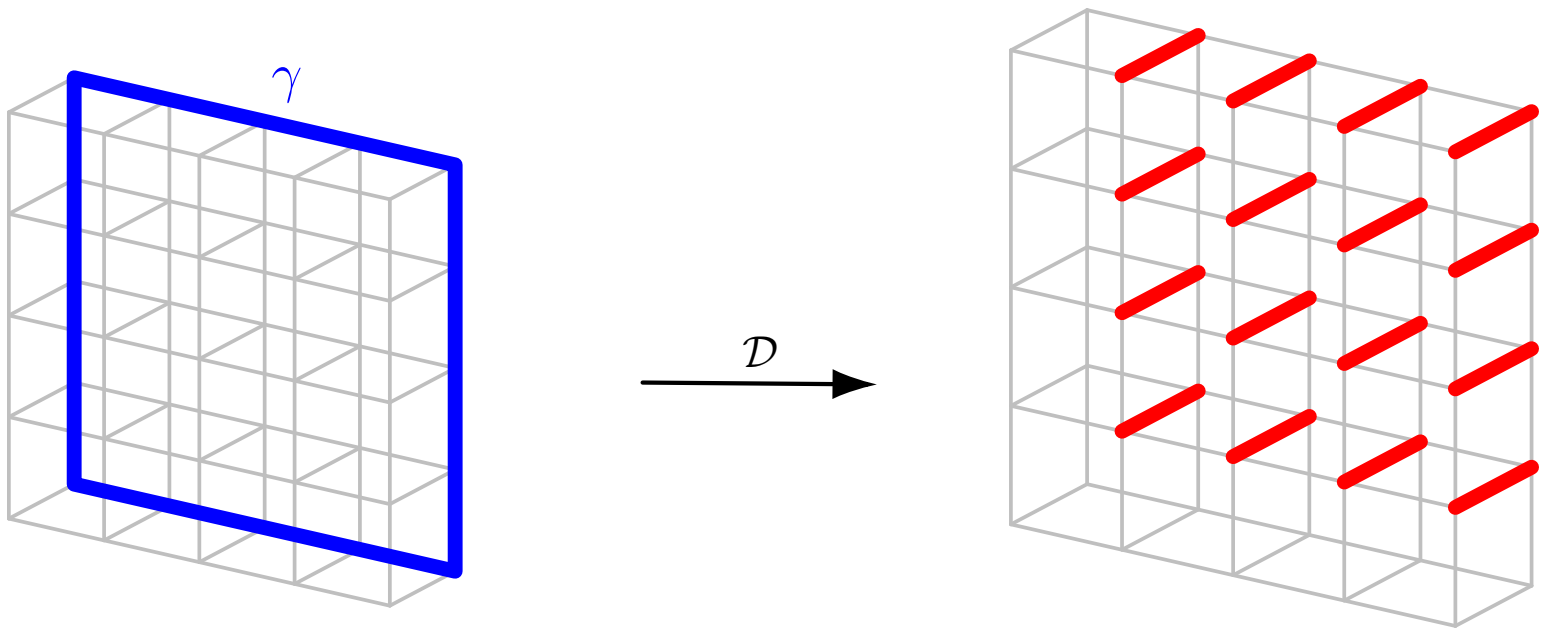

Similar to the discussion in Section \ref{sec:2+1d}, the locality-preserving condition for $\dd$ in 3+1d only holds for regions without non-contractible loops. More specifically, given an operator $O$ supported in a region $R$ with the topology of a 3-ball $B^3$, the image $\dd(O)$ is supported on a larger 3-ball $R^{+\ell}$ obtained by thickening $R$ to its finite neighborhood. However, if $R$ is not simply connected, then $\dd(O)$ need not be supported within a finite thickening of $R$. For example, take
\ie
O=\prod_{\ell\in\gamma} Z_\ell ,
\fe
where $\gamma$ is a closed loop. Then $\dd(O)=\prod_{p\in\Sigma}X_{\mathsf{t}(p)}$ is supported on a solid disk $\Sigma$ with $\partial \Sigma=\gamma$, which is not contained in a thickening of $\gamma$. See Figure \ref{fig:loop}. Here, locality refers to the existence of a representative of the operator so that it is supported in the specified region~\cite{Harlow:2025cqc}. This differs from the commutator-based support used in~\cite{Inamura:2026kco}, under which a membrane is supported on its boundary and the Wegner duality is a locality-preserving automorphism for arbitrary regions.

Interestingly, the map $\dd$ in infinite volume acts invertibly and defines an automorphism of ${\cal A}^\text{top}$. This is related to the fact that the non-invertible operator $\op$ in finite volume acts invertibly on all (gauge-invariant) local operators, and fails to be invertible only on extended line operators carrying nontrivial $\mathbb Z_2^{(1)}$ charge. Operators of the latter kind do not exist in the framework of the algebra of local operators in infinite volume. 
The map $\dd$ obeys
\ie
\dd\circ\dd = T_{1,1,1}\,,
\fe
where  $T_{1,1,1}$ is a lattice translation that shifts every operator by one lattice site in all three directions. 
Hence, the inverse of $\dd$ is $\dd^{-1} = \dd \,T_{1,1,1}^{-1}$.

This is to be contrasted with the algebra of the non-invertible operator $\op$ in finite volume \cite{Gorantla:2024ocs}, where we have imposed the constraint \eqref{3+1dflat},
\ie
\op^2 = T_{1,1,1} \mathsf{C}.
\fe
Here $\mathsf{C}=\frac 12 (1+\eta_{xy})(1+\eta_{yz})(1+\eta_{zx})$ is the condensation operator \cite{Choi:2021kmx,Kaidi:2021xfk,Choi:2022zal}, which projects to the $\mathbb{Z}_2^{(1)}$ invariant sector. Here, $\eta_{ij} = \prod_{\ell \in \widehat{\Sigma}_{ij}}X_\ell$ is the $\mathbb{Z}_2^{(1)}$ symmetry operator supported on the $ij$-plane. The condensation operator $\mathsf{C}$ in finite volume becomes trivial in infinite volume, and the non-invertible operator $\op$ becomes an invertible map $\dd$.

\subsection{Subsystem Kramers--Wannier symmetry in 2+1d} \label{sec:subsystem}

We now turn to an example of a non-invertible symmetry in 2+1d acting on a tensor-product Hilbert space. It is associated with a subsystem symmetry as introduced in \cite{Cao:2023doz, ParayilMana:2024txy}. The space is an infinite square lattice with a qubit at each site labeled $(i,j)$. The subsystem symmetry is generated by an automorphism $\eta_j^{(x)}$ for every $j$ and an automorphism $\eta_i^{(y)}$ for every $i$:
\ie
&\eta^{(x)}_j (a) =  
\left(\prod_{\substack{i\\ (i,j)\in \text{supp}(a) }}X_{i,j} \right) \,a \,
\left(\prod_{\substack{i\\ (i,j)\in \text{supp}(a) }} X_{i,j}\right),\\
&\eta^{(y)}_i (a) =  
\left(\prod_{\substack{j\\ (i,j)\in \text{supp}(a) }}X_{i,j} \right) \,a \,
\left(\prod_{\substack{j\\ (i,j)\in \text{supp}(a) }} X_{i,j}\right) \,.
\fe
As in the discussion of the 1+1d Kramers--Wannier map, the subsystem counterpart is a composition of an automorphism and a conditional expectation for the subsystem symmetry. 
It acts on the subsystem symmetric operators as
\ie
\dd(X_{i,j}) =Z_{i,j}Z_{i,j+1}Z_{i+1,j}Z_{i+1,j+1}\,,~~~~
\dd(Z_{i,j}Z_{i,j+1}Z_{i+1,j}Z_{i+1,j+1}) = X_{i+1,j+1}\,.
\fe
It annihilates every operator that is odd under some $\eta^{(x)}_j$ or $\eta^{(y)}_i$.

The quantum channel $\dd$ here satisfies the purity condition of Section \ref{sec:def}, by the same arguments as for the Kramers--Wannier map. The locality-preserving condition, however, holds only for certain regions such as rectangles. Unlike the examples above, the condition depends on the region's shape, not only on its topology. For instance, we can embed the support of $Z_{0,0}Z_{n,0}Z_{0,n}Z_{n,n}$ inside a U-shaped region that has the topology of a disk, while its image $\prod_{i,j=1}^{n}X_{i,j}$ fills the square. Relatedly, the corresponding non-invertible defect is not fully topological. As mentioned in \cite{Cao:2023doz}, the defect is invariant under translations in space and time, but not under arbitrary deformations. For these reasons, we do not consider this map a non-invertible QCA.

\section*{Acknowledgements}

We thank Arkya Chatterjee, Xie Chen, Yichul Choi, Dominic Else, Lukasz Fidkowski, Jeongwan Haah, Seishiro Ono, Salvatore Pace, Brandon Rayhaun, Sakura Sch\"afer-Nameki, Nikita Sopenko, Nat Tantivasadakarn, and Carolyn Zhang for helpful discussions. We are especially grateful to Corey Jones for participating in the initial stage of this project and for numerous insightful discussions. S.S. and S.H.S. were supported in part by the Simons Collaboration on Ultra-Quantum Matter, which is a grant from the Simons Foundation (651444, NS, SHS). S.S. also gratefully acknowledges support from the Ambrose Monell Foundation and an anonymous donor at the Institute for Advanced Study. 
S.H.S. was also supported in part by the U.S. Department of Energy, Office of Science, Office of High Energy
Physics under Contract Number DE-SC0012567, and in part by the Wade
Fund at the School of Science at MIT.

We used ChatGPT and Claude to help explore and check mathematical arguments and to edit the manuscript. In particular, GPT-5.6 Sol helped find the proof in Appendix \ref{app:symm.H} and the example in \eqref{non.example}. ChatGPT-6 and Claude Opus assisted with details of some derivations in Sections \ref{sec:onsite} and \ref{sec:kw.deformed} and Appendices \ref{app:lemma} and \ref{app:repG}.

\appendix

\section{Choi's theorem \label{app:choi}}

In this appendix, we review Choi's theorem \cite{choi1974schwarz} on the characterization of the multiplicative domain of UCP maps and generalize it to non-unital CP maps.

We begin with the Kadison--Schwarz inequality. Let $\f:\cA \to \cA$ be a 2-positive linear map. Applying $\f$ to the positive $2\times 2$ matrix
\ie
    A =\begin{pmatrix} a^\dagger  \\ 1 \end{pmatrix} \begin{pmatrix} a && 1  \end{pmatrix} = \begin{pmatrix} a^\dagger a && a^\dagger \\ a && 1 \end{pmatrix} \geq 0\,,
\fe
we find that $\f(A) \geq 0$. Thus $\sum_{i,j} v_i^\dagger \f(A_{ij}) v_j \geq 0$ for $v_1,v_2 \in \cA$. Setting $v_1 = 1$ and $v_2 = -c$, we find
\ie \label{generalized.ks.ineq}
    \f(a^\dagger a) \geq \f(a^\dagger) c + c^\dagger \f(a) - c^\dagger \f(1) c\,.
\fe
When $\f$ is unital, setting $c = \f(a)$, we recover the Kadison--Schwarz inequality. Below, we will show that if the inequality above is saturated for some $c \in \cA$, then $\f(xa)=\f(x)c$ for all $x \in \cA$.

If $c$ saturates \eqref{generalized.ks.ineq}, replacing $a$ by $a - \lambda x$ in the inequality, we find
\ie
    \lambda \left( \f(a^\dagger x) - c^\dagger \f(x) \right) + \lambda^* \left( \f(a^\dagger x) - c^\dagger \f(x) \right)^\dagger \leq |\lambda|^2 \f(x^\dagger x) \,.
\fe
Since the inequality holds for any $\lambda \in \bC$, by taking $|\lambda| \to 0$ and $\lambda/|\lambda| = \pm1,\pm i$ we find $\f(x^\dagger a) - \f(x^\dagger) c=0$. 

The above observation leads to a generalized notion of the multiplicative domain for non-unital CP maps. An operator $a \in \cA$ belongs to the multiplicative domain of $\f$, denoted by $\cA_\f$, if there exist $c,d \in \cA$ such that
\ie \label{generalized.ks}
    \f(a^\dagger a) &= \f(a^\dagger) c + c^\dagger \f(a) - c^\dagger \f(1) c \,, \\
    \f(a a^\dagger) &= \f(a) d^\dagger + d \f(a^\dagger) - d \f(1) d^\dagger \,.
\fe
These identities imply that
\ie
    \f(xa) = \f(x) c \qquad \text{ and } \qquad \f(ax) = d \f(x) \qquad \text{ for all } x \in \cA\,.
\fe
Since $\f(a)=\f(1)c=d\f(1)$, the operators $c$ and $d$ need not be unique unless $\f(1)$ is invertible.

\subsection{Multiplicative domain of \texorpdfstring{$\f = \f_1 + \f_2$}{phi = phi1 + phi2} \label{app:md.composite}}

Here, we review Choi's theorem on the characterization of the multiplicative domain of a UCP map $\f = \f_1 + \f_2$, where $\f_1$ and $\f_2$ are CP maps. We slightly generalize the result of \cite{choi1974schwarz} to the case where $\f_1$ and $\f_2$ are not proportional to UCP maps.

\begin{theorem*}[Choi '74] Given UCP maps $\f_1,\f_2 : \cA \to \cA$ and $\f = t \f_1 + (1-t) \f_2$ for $0 < t < 1$:
\ie \label{choi.theorem}
    \cA_\f = \cA_{\f_1} \cap \cA_{\f_2} \cap \big\{ a\in \cA \mid \f_1(a) = \f_2(a) = \f(a) \big\} \,.
\fe
\end{theorem*}
\begin{proof}
Using the Kadison--Schwarz inequality, we find
\ie
    \f(a^\dagger a) &=  t \f_1(a^\dagger a) + (1-t) \f_2(a^\dagger a) \\
        & \geq t \f_1(a^\dagger) \f_1(a) + (1-t) \f_2(a^\dagger) \f_2(a) \\
        &= \f(a^\dagger) \f(a) + t(1-t)\Big( \f_1(a) - \f_2(a) \Big)^\dagger \Big(\f_1(a) - \f_2(a) \Big) \\
        & \geq \f(a^\dagger) \f(a)\,.
\fe
If $a \in \cA_\f$, we have $\f(a^\dagger a) = \f(a^\dagger) \f(a)$. Thus all the inequalities above are saturated, and similarly for $a^\dagger$. Hence $a \in \cA_{\f_1}$, $a \in \cA_{\f_2}$, and $\f_1(a)=\f_2(a)$. Conversely, if $a \in \cA_{\f_1} \cap \cA_{\f_2}$ and $\f_1(a)=\f_2(a)$, it is straightforward to see $\f(a^\dagger a) = \f(a^\dagger) \f(a)$, meaning $a \in \cA_\f$. 
\end{proof}

Choi's theorem can be generalized as follows.
\begin{theorem*} Given a UCP map $\f = \f_1 + \f_2$ where $\f_1$ and $\f_2$ are CP maps:\footnote{This theorem also holds when $\f_1$ and $\f_2$ are 2-positive maps.}
\ie \label{gen.choi.th}
    \cA_\f = \Big\{a\in \cA \mid \f_i(ax) = \f(a)\f_i(x) \;\text{ and }\; \f_i(xa) = \f_i(x)\f(a) \;\text{ for all }\; x\in \cA \Big\} \,.
\fe
\end{theorem*}
\begin{proof}
Applying inequality \eqref{generalized.ks.ineq} for $\f_1$ and $\f_2$ with $c = \f(a)$, we find
\ie 
    \f(a^\dagger a) = \sum_{i=1,2} \f_i(a^\dagger a) &\geq \sum_{i=1,2}  \Big( \f_i(a^\dagger) \f(a) + \f(a)^\dagger \f_i(a) - \f(a)^\dagger \f_i(1) \f(a) \Big) \\ &= \f(a^\dagger) \f(a) \,.
\fe
For $a \in \cA_\f$ we have $\f(a^\dagger a) = \f(a^\dagger) \f(a)$ and hence all inequalities above are saturated. This implies that $a$ is inside the multiplicative domain of $\f_i$ with $c=d=\f(a)$ as defined in \eqref{generalized.ks}. Hence,
\ie
     \f_i(xa) = \f_i(x)\f(a) \quad \text{and} \quad \f_i(ax) = \f(a)\f_i(x) \quad \text{for all } x\in \cA\,.
\fe
The other direction is straightforward. Namely, given the identities above and setting $x=a^\dagger$, we find $\f(a^\dagger a) = \f(a^\dagger) \f(a)$ and $\f(a a^\dagger) = \f(a) \f(a^\dagger)$, therefore $a \in \cA_\f$.
\end{proof}

Another proof follows from the following inequalities
\ie \label{app:approximate.choi}
    \big( \f_i(xa)-\f_i(x)\f(a) \big)^\dagger \big( \f_i(xa)-\f_i(x)\f(a) \big) \leq \| x\|^2 \left(\f(a^\dagger a) -\f(a^\dagger)\f(a) \right) \,, \\
    \big( \f_i(ax)-\f(a)\f_i(x) \big)\big( \f_i(ax)-\f(a)\f_i(x) \big)^\dagger \leq \| x\|^2 \left(\f(a a^\dagger) -\f(a)\f(a^\dagger) \right) \,.
\fe
Indeed, for $a \in \cA_\f$ the right-hand sides vanish, and we recover $\f_i(xa) = \f_i(x)\f(a)$ and $\f_i(ax) = \f(a)\f_i(x)$ for all $x \in \cA$.

\begin{proof}
These inequalities follow from the Stinespring dilation of $\f_1$ and $\f_2$. Choose a faithful $*$-representation of $\cA$ on a Hilbert space $\cH$, so that we regard $\cA \subset B(\cH)$. For instance, one can take $\cH$ to be the GNS Hilbert space $\cH_\tr$ of the tracial state. Since a faithful representation of a C$^*$-algebra is isometric, an inequality between elements of $\cA$ holds in $\cA$ if and only if it holds in $B(\cH)$, so nothing is lost in this identification.

Viewing $\f_i$ as a CP map $\cA \to B(\cH)$, Stinespring's dilation theorem provides a Hilbert space $\cK_i$, a unital $*$-representation $\pi_i : \cA \to B(\cK_i)$, and a bounded operator $V_i : \cH \to \cK_i$ such that
\ie
    \f_i(y) = V_i^\dagger\, \pi_i(y)\, V_i \qquad \text{for all } y \in \cA \,.
\fe
In particular $V_i^\dagger V_i = \f_i(1)$, so the unitality of $\f$ gives $V_1^\dagger V_1 + V_2^\dagger V_2 = 1$. Hence $\|V_i\| \leq 1$, which also implies $V_i V_i^\dagger \leq 1$. Denoting $w_i = \pi_i(a) V_i - V_i \, \f(a)$, a direct computation gives
\ie
    \f_i(xa) - \f_i(x) \f(a) = V_i^\dagger \pi_i(x) \, w_i \,, \qquad \qquad \sum_{i=1,2} w_i^\dagger w_i = \f(a^\dagger a) - \f(a^\dagger) \f(a) \,,
\fe
where the second identity uses $\f(1)=1$. Therefore
\ie
    \left| \f_i(xa) - \f_i(x) \f(a) \right|^2 &= w_i^\dagger\, \pi_i(x^\dagger)\, V_i V_i^\dagger\, \pi_i(x)\, w_i \leq w_i^\dagger\, \pi_i(x^\dagger x)\, w_i \\& \leq \|x\|^2 \, w_i^\dagger w_i \leq \|x\|^2 \left( \f(a^\dagger a) - \f(a^\dagger) \f(a) \right) ,
\fe
where $|y|^2 = y^\dagger y$ and we used $V_i V_i^\dagger \leq 1$, then $x^\dagger x \leq \|x\|^2$ together with the positivity and unitality of $\pi_i$, and finally $w_j^\dagger w_j \geq 0$ for $j \neq i$. This proves the first inequality. The second one follows upon substituting $(x,a) \to (x^\dagger,a^\dagger)$ and using $\f_i(x^\dagger a^\dagger) - \f_i(x^\dagger)\f(a^\dagger) = \big( \f_i(ax) - \f(a) \f_i(x) \big)^\dagger$.
\end{proof}

\section{Trace-preserving UCP maps} \label{app:tp}

In this appendix, we state and prove some useful algebraic properties of UCP maps that are also trace-preserving. They involve the multiplicative domain of such maps.\footnote{See also \cite{rahaman2017multiplicative} for similar statements on the multiplicative domain of trace-preserving UCP maps in the finite-dimensional setting.}

\subsection{Contractivity \label{app:contractivity}}
The first property is contractivity in the $L^2$ norm, which is 
\ie
    \|\f(a)\|_2 \leq \|a\|_2 \qquad \text{where} \qquad \|x\|_2 =\sqrt{\tr(x^\dagger x)} \,.
\fe
Moreover, as we show below, the inequality is saturated if and only if $a$ is inside the multiplicative domain of $\f$. Namely,
\ie \label{l2.saturation}
    \|\f(a)\|_2 = \|a\|_2 \quad \Leftrightarrow \quad a \in \cA_\f \,.
\fe
In particular, this implies that $\f(a)=a$ only if $a \in \cA_\f$.

The contractivity essentially follows from the Kadison--Schwarz inequality \eqref{kd.inequality}. Namely,
\ie
    \| a \|_2^2 - \| \f(a) \|_2^2 = \tr\!\big( \f(a^\dagger a) - \f(a^\dagger) \f(a) \big) = \tr\!\big( \f(a a^\dagger) - \f(a) \f(a^\dagger) \big)\geq 0  \,,
\fe
where we have used the cyclicity of the trace and the trace-preserving property of $\f$ to write $\| a \|_2^2 = \tr( \f(a^\dagger a)) = \tr( \f(a a^\dagger))$. Moreover, since $\f(a^\dagger a) - \f(a^\dagger) \f(a)$ is a positive operator, this operator vanishes iff its trace vanishes. Hence, the RHS of the equation above vanishes iff $a$ and $a^\dagger$ saturate the Kadison--Schwarz inequality. Thus $a$ is inside the multiplicative domain of $\f$ iff $\f$ preserves the $L^2$-norm of $a$.

\subsection{Multiplicative domain of trace-preserving UCP maps} \label{app:md.tp}

Let $\f$ be a trace-preserving UCP map whose dual $\f^\star$ exists as a UCP map on $\cA$. We will show that the restriction of $\f$ to $\cA_\f$ is an injective $*$-homomorphism $\f|_{\cA_\f}:\cA_\f \to \cA_{\f^\star}$ whose inverse is given by $\f^\star$. In other words, a trace-preserving UCP map $\f$ acts invertibly on its multiplicative domain. Namely,
\ie \label{fixed.subalgebra}
     \cA_{\f} = \{ a\in \cA \mid \f^\star \circ \f (a) = a \} = \cA_{\f^\star\circ \f} = \f^\star(\cA_{\f^\star})\,.
\fe
\begin{proof}
    We first note that since $\f^\star$ is also a trace-preserving UCP map, the contractivity of Appendix \ref{app:contractivity} applies to $\f^\star$ and $\f^\star \circ \f$ as well as to $\f$.

    We begin by proving $\cA_{\f} = \{ a\in \cA \mid \f^\star \circ \f (a) = a \}$. Let $a \in \cA_\f$. Using the definition \eqref{def.dual} of the dual map, the multiplicativity of $\f$ on $\cA_\f$, and the trace-preserving property, we find
\ie
    \tr\big( x \, \f^\star \circ \f (a) \big) = \tr\big( \f(x) \f(a) \big) = \tr\big( \f(x a) \big) = \tr(x a) \qquad \text{for all } x \in \cA \,.
\fe
The faithfulness of the trace implies $\f^\star \circ \f(a)=a$. Conversely, given $\f^\star \circ \f(a)=a$, we find
\ie
    \tr(xa)=\tr\big(x \f^\star \circ \f(a)\big) = \tr(\f(x) \f(a))\,.
\fe
Setting $x=a^\dagger$ we conclude $\|a\|_2 = \| \f(a)\|_2$, and hence $a \in \cA_\f$ by the criterion \eqref{l2.saturation}. This proves $a \in \cA_\f \Leftrightarrow \f^\star \circ \f(a)=a$.

This set also coincides with $\cA_{\f^\star \circ \f}$. Indeed, a fixed point of the trace-preserving UCP map $\f^\star \circ \f$ lies in its multiplicative domain, as noted in Appendix \ref{app:contractivity}. Conversely, if $a \in \cA_{\f^\star \circ \f}$ then $\|\f^\star \f(a)\|_2 = \|a\|_2$ by \eqref{l2.saturation} and
\ie
    \|a\|_2 = \big\| \f^\star \f(a) \big\|_2 \leq \big\| \f(a) \big\|_2 \leq \|a\|_2 \,.
\fe
Hence $\|\f(a)\|_2 = \|a\|_2$, and $a \in \cA_\f$ by \eqref{l2.saturation}.

So far we have shown $\cA_{\f} = \{ a\in \cA \mid \f^\star \circ \f (a) = a \} = \cA_{\f^\star\circ \f}$ and it remains to show $\cA_{\f} = \f^\star(\cA_{\f^\star})$. By applying the argument above to the trace-preserving UCP map $\f^\star$, we find $\cA_{\f^\star} = \{b \in \cA \mid \f \circ \f^\star(b) = b\}$. For any $a \in \cA_\f$, let $b = \f(a)$. Then $\f(\f^\star(b)) = \f(\f^\star(\f(a))) = \f(a) = b$, which implies $\f(a) \in \cA_{\f^\star}$. Thus, $\f(\cA_\f) \subseteq \cA_{\f^\star}$ and by a symmetric argument, $\f^\star(\cA_{\f^\star}) \subseteq \cA_\f$. To show $\cA_\f \subseteq \f^\star(\cA_{\f^\star})$, note that for $a \in \cA_\f$, we have $a = \f^\star\circ \f(a) = \f^\star(b) \in \f^\star(\cA_{\f^\star})$ since $b \in \cA_{\f^\star}$.
\end{proof}

\subsection{Multiplicative domain of compositions}

Here we show that the multiplicative domain of the composition of two trace-preserving UCP maps $\f_1$ and $\f_2$ is given by (see also \cite{rahaman2017multiplicative})
\ie \label{md.composition}
    \cA_{\f_2 \f_1} = \Big\{ a \in \cA_{\f_1} \mid  \f_1(a) \in \cA_{\f_2} \Big\}\,.
\fe
To show this, we apply the Kadison--Schwarz inequality \eqref{kd.inequality} and investigate when it is saturated.

Applying the inequality for $\f_1$ and $\f_2$, we find
\ie
    \f_2 \f_1(a^\dagger a ) \geq  \f_2\!\left( \f_1(a^\dagger )\f_1( a )\right) \geq \f_2 \f_1\!\left(a^\dagger \right) \f_2 \f_1( a ) \,,
\fe
and a similar inequality for $a^\dagger$. Thus, if $a \in \cA_{\f_2 \f_1}$, the LHS and RHS above are equal, and thus both inequalities above must be saturated. The trace-preserving property of $\f_2$ implies that $\f_1(a^\dagger a ) =   \f_1(a^\dagger )\f_1( a )$ (and similarly for $a^\dagger$) hence $a \in \cA_{\f_1}$. Saturation of the second inequality implies $\f_1(a) \in \cA_{\f_2}$. Conversely, if $a \in \cA_{\f_1}$ and $\f_1(a) \in \cA_{\f_2}$, then $\f_2\f_1(xa)=\f_2(\f_1(x)\f_1(a)) = \f_2\f_1(x)\f_2\f_1(a)$ for all $x \in \cA$, and similarly for $ax$, so $a \in \cA_{\f_2 \f_1}$.

\section{Symmetric Hamiltonian in the thermodynamic limit \label{app:symm.H}}

In this appendix, we prove a lemma that is used in the main text to show that a quantum channel $\f_1 + \f_2$ is a symmetry of a Hamiltonian only if both $\f_1$ and $\f_2$ are symmetries.

\subsection*{Setup}
Consider a Hamiltonian $H$ with finite-range and uniformly bounded interactions as
\ie
    H=\sum_{j\in\mathbb Z}h_j\,, \qquad h_j=h_j^\dagger\in\cA_{[j,j+R]}\,, \qquad \|h_j\| \leq \lambda \,, \qquad \tr(h_j) = 0 \,,
\fe
where $R \in \bZ_{\geq 0}$ is the range and $\lambda > 0$ controls the strengths of interactions. For convenience, we assume that the interactions are traceless, $\tr(h_j)=0$.  Since $H$ is a formal infinite sum, by $\f(H)=H$ we mean that $\sum_j \tr\big( (\f(h_j)-h_j)\, b \big) = 0$ for all $b \in \cA_\text{loc}$, where only finitely many terms are non-zero. This holds, for instance, if $\sum_j \f(h_j)$ and $\sum_j h_j$ generate the same derivation of $\cA_\text{loc}$. Consider the local Hamiltonians
\ie \label{cutoff.H}
    H_n = \sum_{j \in \bZ} f_n(j) h_j\,, 
\fe
where the cutoff $f_n$ is a finitely supported real function $f_n : \bZ \to [0,1]$ such that
\ie \label{smooth.cutoff}
    \lim_{n \to \infty} f_n(j) = 1 \,, \qquad & \text{for every } j\in \bZ\,, \\
    \lim_{n \to \infty} \Delta(f_n) = 0\,, \qquad & \text{where } \Delta(f_n) = \sum_{j \in \bZ}|f_n(j+1)-f_n(j)|^2 \,.
\fe
Note that the summations over $j$ in \eqref{cutoff.H} and \eqref{smooth.cutoff} are actually finite sums since the function $f_n$ is chosen to have finite support. We can take $f_n=1$ on $[-n,n]$, and let it decrease linearly to zero over a region of width $n^2$ at each end, and set it to zero outside. Namely,
\ie
    f_{n}(j)=\begin{cases}1\,,  & |j|\le n\,,\\
    1-\frac{|j|-n}{n^{2}} \,, \qquad ~ & n<|j|<n+n^{2}\,,\\
    0\,,&|j|\ge n+n^{2} \,, \end{cases}
\fe
therefore $\Delta(f_n) \leq \frac{2}{n^2}$.

\begin{lemma} \label{lemma.symm.h}
    Let $H$ be a Hamiltonian with finite-range and uniformly bounded interactions, where $H_n$ is its cutoff version as above. Given $\f(H)=H$ and $\f^\star(H)=H$ for a locality-preserving and dualizable quantum channel $\f=\f_1 + \f_2$, where $\f_1$ and $\f_2$ are completely positive, we have
\ie \label{app:approximate.choi.theorem}
    \lim_{n \to \infty} \big \| \f_i(a H_n) - \f_i(a) H_n \big \|_2 = \lim_{n \to \infty} \big \| \f_i(H_n a) -  H_n \f_i(a) \big \|_2 = 0 \quad \text{for } a \in \cA \,.
\fe
\end{lemma}

\begin{proof}
To show \eqref{app:approximate.choi.theorem}, first note that
\ie
    \left\| \f_i(a H_n) - \f_i(a) H_n \right \|_2 &\leq  \left\| \f_i(a H_n) - \f_i(a) \f(H_n) \right \|_2 + \left \| \f_i(a) \right\|  \left\| \f(H_n) -H_n \right \|_2 \\
    &\leq \|a\| \sqrt{ \|H_n\|_2^2-\|\f(H_n)\|_2^2 } + \left \| \f_i(a) \right\|  \left\| \f(H_n) -H_n \right \|_2 \,,
\fe
where the first inequality follows from the triangle inequality and the fact that the $L^2$-norm is sub-multiplicative with respect to the operator norm, $\|ab\|_2  \leq \|a\| \|b\|_2$.\footnote{The sub-multiplicative inequality follows from $a^\dagger a \leq \|a\|^2 \, \mathbbm{1}$, which gives $\| ab \|_2^2 = \tr(b^\dagger a^\dagger a b) \leq \| a\|^2 \tr(b^\dagger b) = \|a\|^2 \|b\|^2_2$.} The second line follows from the inequality $|\f_i(aH_n)-\f_i(a)\f(H_n)|^2 \leq \| a\|^2 (\f(H_n^2) -\f(H_n)^2)$, which we proved at the end of Appendix \ref{app:md.composite} around \eqref{app:approximate.choi}. To establish \eqref{app:approximate.choi.theorem}, it remains to show
\ie \label{two.equalities}
    \lim_{n \to \infty} \Big(\|H_n\|_2^2-\|\f(H_n)\|_2^2 \Big) \stackrel{?}{=} 0 \qquad \text{and} \qquad \lim_{n \to \infty}   \left\| \f(H_n) -H_n \right \|_2 \stackrel{?}{=} 0\,.
\fe

We first establish the first equality in \eqref{two.equalities}. Denote
\ie
    G_{jk} = \tr(h_j h_k) - \tr(\f(h_j) \f(h_k))\,,
\fe
and note that the symmetric matrix $G$ satisfies three key conditions:
\begin{subequations}
\label{G.conditions}
\begin{align}
& G_{jk} = 0 \qquad \text{for } |j-k| > R + 2\ell \,, \label{G.range} \\
& \sum_{j} G_{jk} = \sum_{k} G_{jk} = 0 \,, \label{G.sum} \\
& |G_{jk}| \le 2\lambda^2 \,, \label{G.bound} 
\end{align}
\end{subequations}
where $R$ is the range of the interactions of the Hamiltonian and $\ell$ is the range of $\f$. The row/column sums of the matrix $G$ vanish since
\ie \label{G.2}
    \sum_j G_{jk} &= \tr(H h_k) - \tr(\f(H) \f(h_k)) \\ &= \tr(H h_k) - \tr(H \f(h_k))
    \\ &= \tr(H h_k) - \tr(\f^\star(H) h_k) = 0\,,
\fe
where we have used both $\f(H)=H$ and $\f^\star(H)=H$. Using \eqref{G.conditions}, we conclude
\ie
    \|H_n\|_2^2-\|\f(H_n)\|_2^2 &= \sum_{j,k} f_n(j) f_n(k) G_{jk} \\
    &= -\frac{1}{2} \sum_{j,k} \big( f_n(j) - f_n(k) \big)^2 G_{jk}\,,
\fe
where the summations are finite since $f_n$ is finitely supported and because of \eqref{G.range}. Finally, the uniform boundedness and finite-range condition imply that 
\ie
    \left| \sum_{j,k} \big( f_n(j) - f_n(k) \big)^2 G_{jk} \right| \leq C \Delta(f_n) \,,
\fe
where $C$ is a constant independent of $n$ that only depends on $R+2\ell$ and $\lambda$.\footnote{In particular, $C \leq 2\lambda^2 \frac{(R+2\ell)(R+2\ell+1)(2R+4\ell+1)}{3}$.} Since $\Delta(f_n) \leq \frac{2}{n^2}$, we conclude $\|H_n\|_2^2-\|\f(H_n)\|_2^2 \to0$ as $n \to \infty$. Showing the second equality in \eqref{two.equalities} is similar. Namely,
\ie
    & \left\| \f(H_n) -H_n \right \|_2^2 = \sum_{j,k} f_n(j) f_n(k) D_{jk} \,,\\
    & D_{jk} = \tr\Big( \f(h_j)\f(h_k) - \f(h_j)h_k - h_j\f(h_k) + h_jh_k \Big) \,,
\fe
where $D_{jk}/2$ satisfies the same three conditions in \eqref{G.conditions} that $G_{jk}$ satisfies. Therefore, $\left\| \f(H_n) -H_n \right \|_2^2 \leq 2C \Delta(f_n) \to 0$ as $n \to \infty$. The second limit in \eqref{app:approximate.choi.theorem} follows in the same way from the second inequality in \eqref{app:approximate.choi}. This concludes the proof.
\end{proof}

\section{More on anomalous on-site non-invertible symmetries \label{app:lemma}}

Here we prove the following lemma that is relevant for the anomalous on-site non-invertible symmetry in Section \ref{anomalous.onsite}.

\begin{lemma}\label{lem:finite-erasure}
Let $\mathcal{N} = \mathsf{H}\f_{\pi/2}$, where $\mathsf{H}$ is the on-site Hadamard conjugation and $\f_{\pi/2}$ is the cosine symmetry of Section \ref{sec:cosine}. For every interval $I$ of $L$ sites,
\ie
    \mathcal{N}^{2\lceil L/2\rceil}(a)
    = \tr(a)\mathbbm{1}
    \qquad \text{for } a\in\cA_I \,.
\fe
\end{lemma}

\begin{proof}
Label the sites $0,\ldots, L-1$ and encode a Pauli string by two Laurent polynomials in $t$ with coefficients in $\mathbb{F}_2$ (polynomials in $t$ and $t^{-1}$ with mod $2$ coefficients):
\ie
    P(x,z) = \prod_{j=0}^{L-1} X_j^{x_j}Z_j^{z_j} \,,
    \qquad
    x(t) = \sum_{j=0}^{L-1} x_jt^j \,,
    \qquad
    z(t) = \sum_{j=0}^{L-1} z_jt^j \,.
\fe
Both polynomials have degree less than $L$. The cosine formula
\eqref{cosine.ucp} (up to a sign that is irrelevant for the current discussion) gives
\ie \label{eq.update}
    \mathcal{N}\(P(x,z)\) =
    \begin{cases}
        0 \,, & (t+t^{-1})\nmid z \,, \\[2pt]
        \displaystyle
        \pm P\(z,\,x+\frac{z}{t+t^{-1}}\) \,,
        & (t+t^{-1})\mid z \,.
    \end{cases}
\fe
Indeed, $\mathcal{N}$ annihilates every Pauli string that is not invariant under both $\eta_\mathrm{e}$ and $\eta_\mathrm{o}$, which is precisely the condition $(t+t^{-1}) \nmid z$. The surviving strings (i.e., those that are not annihilated by $\mathcal{N}$) lie in the multiplicative domain and are products of $X_j$ and $Z_jZ_{j+2}$. Their images
\ie
    \mathcal{N}(X_j) = Z_j \,,
    \qquad
    \mathcal{N}(Z_jZ_{j+2}) = -X_jZ_{j+1}X_{j+2}
\fe
give the second line of \eqref{eq.update}.

Suppose a Pauli string $P(x^{(0)},z^{(0)})$ is not annihilated after $2m$ steps, and thus lies in the multiplicative domain of $\mathcal{N}^{2m}$. Let the polynomials $x^{(k)}$ and $z^{(k)}$ denote the associated Laurent polynomials after step $k$, ignoring the overall sign. Namely $\mathcal{N}(P(x^{(k)},z^{(k)}))=\pm P(x^{(k+1)},z^{(k+1)})$, where
\ie \label{recurrence}
    &x^{(k+1)}=z^{(k)} \,,
     &&0\leq k\leq 2m-1 \,, \\
    &z^{(k+1)} = x^{(k)} + \frac{z^{(k)}}{t+t^{-1}} \,,
    \qquad &&0\leq k\leq 2m-1 \,.
\fe
Noting that the coefficients are all modulo 2, we find
\ie
    &(t+t^{-1})(z^{(k-1)}+z^{(k+1)}) = z^{(k)} \,,
    \qquad &&1\leq k\leq 2m-1 \,, \\
    &(t+t^{-1})(x^{(k-1)}+x^{(k+1)}) = x^{(k)} \,,
    \qquad &&1\leq k\leq 2m-1 \,. \\
\fe
Thus, $z^{(k)}$ and $x^{(k)}$ are divisible by $t+t^{-1}$ for $1 \leq k \leq 2m-1$. By induction on $r$, we find
\ie
    (t+t^{-1})^r \mid z^{(k)}, x^{(k)}
    \qquad \text{for } r\leq k\leq 2m-r \,.
\fe
Namely, to show $(t+t^{-1})^{r+1} \mid z^{(k)}$ we need to show that $(t+t^{-1})^{r} \mid z^{(k-1)},z^{(k+1)}$. The latter holds when $r \leq k-1, k+1 \leq 2m-r$, which is equivalent to $r+1 \leq k \leq 2m-r-1$. In particular, $(1+t^2)^m\mid  z^{(m)}, x^{(m)}$ since $(t+t^{-1})^m=t^{-m}(1+t^2)^m$. Taking $m=\lceil L/2\rceil$, this divisor has degree $2m\geq L$. Since the channel preserves the interval $I$, $z^{(m)}$ and $x^{(m)}$ are still polynomials of degree less than $L$, so $z^{(m)}=x^{(m)}=0$. Since the recurrence relation \eqref{recurrence} is invertible with $z^{(k)} = x^{(k+1)}$ and $x^{(k)} = z^{(k+1)} + x^{(k+1)}/(t+t^{-1})$, this implies $x^{(0)}=z^{(0)}=0$, i.e., $P(x^{(0)},z^{(0)})$ is the identity. In other words, every non-trivial Pauli string inside an interval of length $L$ is annihilated by $2\lceil L/2\rceil$ steps. Expanding $a$ in the Pauli basis proves the claim.
\end{proof}

\section{Faithfulness and purity of the Rep(\texorpdfstring{$G$}{G}) symmetry \label{app:repG}}

In this appendix, we provide more details on the faithfulness and purity of the on-site non-invertible symmetries of Section \ref{sec:repg}. We first show that each representation of $\overline{G_0}$ defines a UCP map after finite blocking. We show these maps faithfully realize the normalized fusion rule of $\mathrm{Rep}(\overline{G_0})$, and that $\f_\rho$ is CP-pure precisely when $\rho|_{\overline{G_0}}$ is isotypical.

\subsection{Faithfulness \label{app:faithfulness}}

We start by showing that the UCP map $\f_{\pi}$ in \eqref{rep.g.0.maps}, associated with an arbitrary representation $\pi$ of $\overline{G_0}$, is well-defined after suitable blocking. As we will explain, this follows from the following lemma, which also enters the purity argument below.

\begin{lemma} \label{lemma1}
    Let $\sigma$ be a finite-dimensional unitary representation of $G$, and let $p_1,\cdots,p_r$ be the projections onto the isotypical components of $\sigma|_{G_0}$, associated with pairwise inequivalent irreducible representations $\pi_1,\cdots,\pi_r$ of $G_0$. Namely, $\sigma|_{G_0} = \pi_1^{\oplus m_1} \oplus \cdots \oplus \pi_r^{\oplus m_r}$.  
    There exists a positive integer $k$ such that $\sigma(U_{\bm{s}})$ commutes with every $p_i$ for all words of length $|\bm{s}|=k$.
\end{lemma}
\begin{proof}
Since $G_0$ is a normal subgroup of $G$, each $\sigma(u_s)$ must permute the projection operators $p_i$. The permutation is the same for every $s$ because $u_s u_t^{-1} \in G_0$. If $k$ is the order of this permutation, any word $\sigma(U_{\bm{s}})$ with $|\bm{s}|=k$ commutes with all these projections.
\end{proof}

The corollary that we will need is that after blocking every $k$ consecutive sites, we find
\ie \label{decomposition}
    \f_\sigma = \sum_{i=1}^r \frac{m_i d_{\pi_i}}{d_{\sigma}}\, \f_{\pi_i} \,, \qquad \f_{\pi_i}\!\left( \ket{\bm{s}} \! \bra{\bm{t}} \right) = \frac{\Tr\!\big(p_i\, \sigma(U_{\bm{s}} U_{\bm{t}}^{-1})\big)}{m_i d_{\pi_i}} \ket{\bm{s}} \! \bra{\bm{t}} = \frac{\chi_{\pi_i}(U_{\bm{s}} U_{\bm{t}}^{-1})}{d_{\pi_i}} \ket{\bm{s}} \! \bra{\bm{t}} \,,
\fe
where importantly each $\f_{\pi_i}$ is a well-defined UCP map. To see this, we block every $k$ consecutive sites into one to obtain the blocked MPO tensor
\ie \label{blocked.tensor}
    &\mathbb{U}_{jk,jk+1,\cdots, (j+1)k-1} = \sum_{\bm{s}} \sigma(U_{\bm{s}}) \otimes \ket{\bm{s}}\!\bra{\bm{s}}_{jk,jk+1,\cdots, (j+1)k-1} \,, \\
    &\bm{s} = (s_{jk}, s_{jk+1}, \cdots, s_{(j+1)k-1}) \,.
\fe
This MPO tensor only depends on words of length $k$, which commute with the projections $p_i$. Thus, the map in \eqref{decomposition} constructed from this blocked MPO tensor is well-defined when the interval is a union of blocks of size $k$ as above. We start at site $j=0$, and consider the blocks $\cdots,[-2k,-k),[-k,0) ,[0,k), [k,2k), \cdots$. Relatedly, the map $\f_{\pi_i}$ commutes with lattice translation by $k$ sites and need not commute with the translation by one site.

By the compactness of $\overline{G_0}$, every irreducible finite-dimensional representation $\pi$ of $\overline{G_0}$ appears in the restriction of $\sigma=\rho^{\otimes n} \otimes \overline{\rho}^{\otimes m}$ to $\overline{G_0}$ for some $m,n \in \bZ_{\geq0}$, which is a finite-dimensional representation of $G$.\footnote{More precisely, $\sigma|_{\overline{G_0}}$ is given by first extending $\sigma$ continuously to $\overline{G}$ and then restricting it to its subgroup $\overline{G_0}$. Also, note that since $G_0$ is dense in $\overline{G_0}$, the restriction of $\pi$ to $G_0$ is still irreducible.} Recall that $\rho$ is the defining representation of $G$ as discussed in Section \ref{sec:repg}. Thus, applying Lemma \ref{lemma1} to this $\sigma$, one of the $\pi_i$ is isomorphic to $\pi|_{G_0}$, and therefore $\f_\pi = \f_{\pi_i}$ is a well-defined UCP map after blocking every $k$ sites. When $\pi$ is reducible, the map $\f_\pi$ is a convex combination of the maps associated with its irreducible components and is also well-defined. Thus, we have established that $\f_{\pi}$ in \eqref{rep.g.0.maps} is well-defined after suitable blocking for any representation $\pi$ of $\overline{G_0}$.

Now we show faithfulness, that is, for any two representations $\pi_1,\pi_2$ of $\overline{G_0}$ 
\ie \label{faithful}
    \f_{\pi_1}=\f_{\pi_2} \quad \Leftrightarrow \quad \frac{\chi_{\pi_1}}{d_{\pi_1}}=\frac{\chi_{\pi_2}}{d_{\pi_2}} \text{ on } G_0 \quad \Leftrightarrow \quad \left(\pi_1\right)^{\oplus d_{\pi_2}} \simeq \left(\pi_2 \right)^{\oplus d_{\pi_1}} \,.
\fe
Thus, $\f_\pi$ depends only on the restriction of $\pi$ to $G_0$ up to an overall multiplicity, which cannot be detected since $\f_{\pi^{\oplus m}} = \f_\pi$. 

The second equivalence in \eqref{faithful} holds because a finite-dimensional unitary representation is determined by its character, and the character on $G_0$ determines the character on $\overline{G_0}$ by continuity. In the first equivalence, the direction $\Leftarrow$ is clear from \eqref{rep.g.0.maps}. The direction $\Rightarrow$ is less obvious, because the balanced words $U_{\bm{s}}U_{\bm{t}}^{-1}$ are only a subset of $G_0$. It follows from the following lemma.

\begin{lemma} \label{lemma2} For any choice of the unitary matrices $u_1,u_2,\cdots,u_N$, the balanced words span the same subspace of $\bC^{d\times d}$ as the group $G_0 \subseteq \mathrm{U}(d)$:
\ie
    \mathrm{span}_\bC \left\{ U_{\bm{s}} U_{\bm{t}}^{-1} \,:\, |\bm{s}| = |\bm{t}| \right\} = \mathrm{span}_\bC\left\{G_0 \right\} \,.
\fe
\end{lemma}

\begin{proof}
    Let $B$ denote the subspace on the left-hand side. $B \subseteq \mathrm{span}_\bC\{G_0\}$ since balanced words lie in $G_0$. To show the converse, first note that $u_s B u_s^{-1} \subseteq B$, and since $B$ is finite-dimensional, this inclusion is an equality. Therefore, $B$ is invariant under conjugation by all of $G$. Moreover, $B$ is closed under multiplication, since for $b \in B$ and a general element $\sum_{\bm{s},\bm{t}} \alpha_{\bm{s},\bm{t}} U_{\bm{s}} U_{\bm{t}}^{-1} \in B$ with $|\bm{s}|=|\bm{t}|$:
    \ie
        \left( \sum_{\bm{s},\bm{t}} \alpha_{\bm{s},\bm{t}} U_{\bm{s}} U_{\bm{t}}^{-1} \right) b =  \sum_{\bm{s},\bm{t}} \alpha_{\bm{s},\bm{t}} U_{\bm{s}} \left( U_{\bm{t}}^{-1} b U_{\bm{t}} \right) U_{\bm{t}}^{-1} \in B \,.
    \fe
Above, we have also used the fact that $U_{\bm{s}}BU_{\bm{t}}^{-1} \subseteq B$. Hence, $B$ is an algebra that is invariant under conjugation by $G$. It therefore also contains the
balanced words $U_{\bm{t}}^{-1} U_{\bm{s}} = U_{\bm{t}}^{-1} (U_{\bm{s}}U_{\bm{t}}^{-1}) U_{\bm{t}}$, and hence the group $G_0$ that they generate.
\end{proof}

Let $\sigma$ be a representation of $G$ whose restriction to $\overline{G_0}$ contains both $\pi_1$ and $\pi_2$, which exists as discussed above. Applying Lemma \ref{lemma2} to the matrices $\sigma(u_s)$, we find that for any $g \in G_0$, we have $\sigma(g) = \sum_{\bm{s},\bm{t}} \alpha_{\bm{s},\bm{t}}\, \sigma(U_{\bm{s}} U_{\bm{t}}^{-1})$ for some $\alpha_{\bm{s},\bm{t}} \in \bC$, and hence $\pi_i(g) = \sum_{\bm{s},\bm{t}} \alpha_{\bm{s},\bm{t}}\, \pi_i(U_{\bm{s}} U_{\bm{t}}^{-1})$ with the same coefficients for $i=1,2$. Thus $\f_{\pi_1}=\f_{\pi_2}$ implies the equality of normalized characters on the entire $G_0$, and therefore $(\pi_1)^{\oplus d_{\pi_2}} \simeq (\pi_2)^{\oplus d_{\pi_1}}$.

\subsection{Purity \label{app:purity}}

Here we show that $\f_\rho$ is pure if and only if $\rho|_{G_0} \simeq \pi^{\oplus m}$ for some irreducible representation $\pi$. Such a representation is called isotypical. Since $G_0$ is dense in $\overline{G_0}$, the two groups have the same commutants in $\bC^{d \times d}$, so $\rho|_{G_0}$ is isotypical if and only if $\rho|_{\overline{G_0}}$ is.

\textbf{CP-purity $\Rightarrow$ isotypicality:}
Assume $\rho|_{G_0}$ is not isotypical. Applying Lemma \ref{lemma1} and its corollary to $\sigma=\rho$, we can decompose $\f_\rho$ into $r \geq 2$ UCP maps $\f_{\pi_1}, \cdots, \f_{\pi_r}$ as in \eqref{decomposition}. They are distinct by \eqref{faithful}, hence the UCP map $\f_\rho$ is not CP-pure.

\textbf{Isotypicality $\Rightarrow$ CP-purity:} It remains to show that for an isotypical representation $\rho|_{G_0}$, the channel $\f_\rho$ is pure. We will first show that $\rho|_{G_0}$ can be taken to be irreducible without loss of generality, and then show that for irreducible $\rho|_{G_0}$, its corresponding Choi state is normal and hence $\f_\rho$ is CP-pure.

Let $\rho|_{G_0} \simeq \pi^{\oplus m}$ be an isotypical representation, where $\pi$ is an irreducible representation of $G_0$. This implies that $\mathrm{span}_\bC\{G_0\} = \bC^{d_\pi \times d_\pi} \otimes \mathbbm{1}_{m}$. Recall that $G$ is identified with its defining representation as unitary $d\times d$ matrices and hence $G_0 \subseteq G \subset \mathrm{U}(d)$. Since $G_0$ is a normal subgroup of $G$, conjugation by $u_s$ preserves $\mathrm{span}_\bC\{G_0\}$ and therefore $u_s = a_s \otimes b_s$. Moreover, $u_s u_t^{-1} \in G_0$ implies $b_s \propto b_t$. After absorbing the phase difference into $a_s$, we find $u_s = a_s \otimes b$ for a $b$ that is independent of $s$. The factor of $b$ does not appear in balanced words. Therefore, $\f_\rho$ does not change when $u_s$ is replaced by $a_s$, and we may assume that $\rho|_{G_0}$ is irreducible without loss of generality.

Assuming $\rho|_{G_0}$ is irreducible, we have $\mathrm{span}_\bC\{G_0\} = \bC^{d \times d}$. As for the cosine symmetry, the Choi state of $\f_\rho$ is the MPS with tensor $\mathbb{A}_j^\rho = \frac{1}{\sqrt{N}} \sum_s u_s \otimes \ket{ss}_j$, and it is pure if the transfer channel $\mathcal{E}(x)=\frac{1}{N}\sum_s u_s^{-1} x u_s$ is primitive. Let $\mathcal{E}(x)= \lambda x$ with $|\lambda|=1$. Since $\mathcal{E}$ is a trace-preserving UCP map and $\| \mathcal{E}(x) \|_2=\|x\|_2$, the operator $x$ belongs to the multiplicative domain of $\mathcal{E}$ by \eqref{l2.saturation}, and Choi's theorem (see Appendix \ref{app:md.composite}) gives $u_s^{-1} x u_s = \lambda x$ for every $s$. Thus $x$ commutes with the balanced words $U_{\bm{s}}U_{\bm{t}}^{-1}$ and $U_{\bm{t}}^{-1}U_{\bm{s}}$, and by Schur's lemma $x \propto \mathbbm{1}$. Thus $\mathcal{E}$ is primitive, $\f_\rho$ is Choi-pure and therefore CP-pure.

\section{Dipolar non-invertible symmetries} \label{app:dipolar}

Here we discuss the lattice dipolar Kramers--Wannier non-invertible symmetry in \cite{Cao:2024qjj, Pace:2024tgk, Kim:2025ttl} as a UCP map satisfying our three conditions. 
(Related dipolar non-invertible symmetries in the continuum were studied in \cite{Spieler:2024fby}.) 
It is defined on a 1d qunit chain whose local Hilbert space is $N$-dimensional, i.e., $\mathcal{H}_j \simeq \mathbb{C}^N$. 
Denote the clock and shift unitary operators on the $j$-th site as $Z_j$ and $X_j$, obeying
\begin{align}
Z_j X_{j'} = \omega^{\delta_{j,j'}} X_{j'}Z_j,~~~(X_j)^N=(Z_j)^N=1, 
\end{align}
where $\omega=e^{2\pi i /N}$.
The dipolar Kramers--Wannier symmetry is defined as the composition of three locality-preserving quantum channels:
\ie \label{dKW.map}
    &\dd =  \alpha_\text{dKW} \circ E_Q\circ E_D : \cA \to \cA_\dd \subset \cA \,.
\fe
Here, $E_Q, E_D$ are conditional expectations projecting an operator onto its component that is invariant under the $\mathbb{Z}_N^Q$ charge symmetry generated by $\prod_j X_j$ and under the $\mathbb{Z}_N^D$ dipole symmetry generated by $\prod_j X_j^j$, respectively:
\ie
    &E_Q(a)  
    = \frac{1}{N} \sum_{k=0}^{N-1} \left( \prod_{j \in \text{supp}(a)} X_j \right)^k a \left( \prod_{j \in \text{supp}(a)} X_j \right)^{-k} \,,\\
    &E_D(a)  
    = \frac{1}{N} \sum_{k=0}^{N-1} \left( \prod_{j \in \text{supp}(a)} X_j^j \right)^k a \left( \prod_{j \in \text{supp}(a)} X_j^j \right)^{-k} \,.
\fe
The map $\alpha_\text{dKW}: \cA_\dd \to \cA_\dd$ is an automorphism of the algebra of $\bZ_N^Q\times \mathbb{Z}_N^D$ invariant operators $\cA_\dd$. 
The latter is generated by the local operators $X_j$ and $ Z_{j-1}(Z_j^\dagger)^2 Z_{j+1}$, and $\alpha_\text{dKW}$ acts on them as
\ie
\alpha_\text{dKW}(X_j) = Z_{j-1}(Z_j^\dagger)^2 Z_{j+1}\,,~~~~
\alpha_\text{dKW}( Z_{j-1}(Z_j^\dagger)^2 Z_{j+1}) =X_j^\dagger\,.
\fe
For $N=2$, $\dd$ becomes the Rep(D$_8$) non-invertible symmetry on a qubit chain in \cite{Seifnashri:2024dsd, Seifnashri:2025fgd}.

Similar to the discussion for the Kramers--Wannier symmetry, the dipolar Kramers--Wannier map $\dd$ satisfies all three conditions. It is a locality-preserving UCP map because $\alpha_\text{dKW}$, $E_Q$, and $E_D$ are. By the same computation as for the Kramers--Wannier symmetry in Section \ref{sec:dual}, its dual is
\ie
\dd^\star =  \alpha_\text{dKW} ^{-1} \circ E_Q \circ E_D \,.
\fe
The inverse $\alpha_\text{dKW}^{-1}$ of $\alpha_\text{dKW}$ equals $ \alpha_\text{dKW}\circ C$, where $C$ is the charge conjugation automorphism of $\cA$ that acts as $C: X_j \to X_j^\dagger,~Z_j\to Z_j^\dagger$ and preserves $\cA_\dd$.
Hence, $\dd^\star = \dd \circ C$ is locality-preserving and $\dd$ is dualizable. In particular, $\dd\circ\dd = C \circ E_Q \circ E_D$, which involves charge conjugation rather than a lattice translation \cite{Cao:2024qjj}. Finally, the purity of $\dd$ follows from an argument similar to the one in Section \ref{sec:purity} for the Kramers--Wannier symmetry.

\section{Dualizable Sequential Matrix Product Operators} \label{app:dsMPO}

Here, we relate our perspective on generalized symmetries on the lattice to that of \cite{Tantivasadakarn:2025txn}. The authors of \cite{Tantivasadakarn:2025txn} identify generalized symmetries in 1+1d with sMPOs. They show that any simple (injective) sMPO $\mathsf{W}$  satisfies the fusion relation
\ie \label{annihilability}
    \mathsf{W}^\dagger \mathsf{W} = \mathbbm{1} + \cdots \,,
\fe
where the operators that appear in `$\cdots$' are all simple sMPOs.\footnote{Their simplicity (injectivity) assumption corresponds to our CP-purity and irreducibility of the defect.}
They refer to this property as annihilability, which is weaker than the dualizability of the sequential circuit that we impose. Dualizability leads to the fusion relation $\overline{\mathbb{W}} \otimes \mathbb{W} = \mathbbm{1} \oplus \cdots$ for the \emph{defects}, which is stronger than such a relation for the MPOs. The sMPO $\op_\theta$ of Section \ref{sec:kw.deformed} is simple and annihilable for every $\theta$ by \eqref{d.theta.fusion}, whereas the sequential circuit (the topological defect) associated with \eqref{U.theta} is dualizable only for $\theta \in \pi \bZ/4$.

The authors of \cite{Tantivasadakarn:2025txn} further show that sMPOs form a closed algebra under multiplication. Moreover, by normalizing simple sMPOs so that the identity appears with coefficient 1 in \eqref{annihilability}, they argue that all fusion coefficients must be non-negative integers. Namely, 
\ie 
    \mathsf{W}_\alpha \mathsf{W}_\beta = \sum_{\gamma} N_{\alpha \beta}^\gamma \mathsf{W}_\gamma \,, \qquad N_{\alpha \beta}^\gamma \in \bZ_{\geq 0}\,, \qquad  N_{\alpha \bar{\alpha}}^1 = 1\,,
\fe
where $\mathsf{W}_{\bar \alpha} = \mathsf{W}_\alpha^\dagger$ and $\mathsf{W}_1 = \mathbbm{1}$. 
What was implicitly assumed in their argument is the existence of a dual sequential circuit, without which the fusion coefficients are non-integral in general. For instance, the sMPOs $\op_\theta$ and $X_1 \cdots X_L$ are both simple and normalized as above, and yet the fusion coefficient $\sin(2\theta)^L$ in \eqref{d.theta.fusion} is not an integer for $\theta \notin \pi \bZ/4$. Such non-integral fusion coefficients vanish exponentially as $L \to \infty$. Indeed, the symmetry associated with $\op_\theta$ becomes invertible in the thermodynamic limit as it generates an automorphism of the quasi-local algebra $\cA$.

To summarize, the fusion coefficients involving simple sMPOs are non-negative integers only up to exponentially small terms in the system size. For a \emph{generic} choice of sequential circuit, the corresponding sMPO becomes a unitary operator in the thermodynamic limit with the fusion relation $\mathsf{W}^\dagger \mathsf{W} \sim \mathbbm{1}$ up to exponentially suppressed terms. Thus, generic sequential circuits lead to invertible symmetries that are not locality-preserving in the thermodynamic limit. To obtain exact non-negative integer fusion coefficients, we must work with dualizable sMPOs, i.e., MPOs that admit a dualizable sequential circuit presentation. Translationally invariant dualizable MPOs are studied in~\cite{Inamura:2026hif} under the name \emph{topological injective MPOs}.

\bibliographystyle{JHEP}
\bibliography{refs}

\providecommand{\href}[2]{#2}\begingroup\raggedright\begin{thebibliography}{100}

\bibitem{Gaiotto:2014kfa}
D.~Gaiotto, A.~Kapustin, N.~Seiberg, and B.~Willett, {\it {Generalized Global Symmetries}},  {\em JHEP} {\bf 02} (2015) 172, [\href{http://arxiv.org/abs/1412.5148}{{\tt arXiv:1412.5148}}].

\bibitem{McGreevy:2022oyu}
J.~McGreevy, {\it {Generalized Symmetries in Condensed Matter}},  {\em Ann. Rev. Condensed Matter Phys.} {\bf 14} (2023) 57--82, [\href{http://arxiv.org/abs/2204.03045}{{\tt arXiv:2204.03045}}].

\bibitem{Cordova:2022ruw}
C.~Cordova, T.~T. Dumitrescu, K.~Intriligator, and S.-H. Shao, {\it {Snowmass White Paper: Generalized Symmetries in Quantum Field Theory and Beyond}},  in {\em {Snowmass 2021}}, 5, 2022.
\newblock \href{http://arxiv.org/abs/2205.09545}{{\tt arXiv:2205.09545}}.

\bibitem{Schafer-Nameki:2023jdn}
S.~Schafer-Nameki, {\it {ICTP lectures on (non-)invertible generalized symmetries}},  {\em Phys. Rept.} {\bf 1063} (2024) 1--55, [\href{http://arxiv.org/abs/2305.18296}{{\tt arXiv:2305.18296}}].

\bibitem{Brennan:2023mmt}
T.~D. Brennan and S.~Hong, {\it {Introduction to Generalized Global Symmetries in QFT and Particle Physics}},  \href{http://arxiv.org/abs/2306.00912}{{\tt arXiv:2306.00912}}.

\bibitem{Shao:2023gho}
S.-H. Shao, {\it {What's Done Cannot Be Undone: TASI Lectures on Non-Invertible Symmetries}},  in {\em {Theoretical Advanced Study Institute in Elementary Particle Physics 2023}: {Aspects of Symmetry}}, 8, 2023.
\newblock \href{http://arxiv.org/abs/2308.00747}{{\tt arXiv:2308.00747}}.

\bibitem{shao2025noninvertible}
S.-H. Shao, {\it Noninvertible symmetries: What’s done cannot be undone},  {\em Physics Today} {\bf 78} (2025), no.~7 30--35.

\bibitem{Kapustin:2005py}
A.~Kapustin, {\it {Wilson-'t Hooft operators in four-dimensional gauge theories and S-duality}},  {\em Phys. Rev. D} {\bf 74} (2006) 025005, [\href{http://arxiv.org/abs/hep-th/0501015}{{\tt hep-th/0501015}}].

\bibitem{Seifnashri:2023dpa}
S.~Seifnashri, {\it {Lieb-Schultz-Mattis anomalies as obstructions to gauging (non-on-site) symmetries}},  {\em SciPost Phys.} {\bf 16} (2024), no.~4 098, [\href{http://arxiv.org/abs/2308.05151}{{\tt arXiv:2308.05151}}].

\bibitem{Schumacher:2004hix}
B.~Schumacher and R.~F. Werner, {\it {Reversible quantum cellular automata}},  \href{http://arxiv.org/abs/quant-ph/0405174}{{\tt quant-ph/0405174}}.

\bibitem{Farrelly:2019zds}
T.~Farrelly, {\it {A review of Quantum Cellular Automata}},  {\em Quantum} {\bf 4} (2020) 368, [\href{http://arxiv.org/abs/1904.13318}{{\tt arXiv:1904.13318}}].

\bibitem{Bultinck:2015bot}
N.~Bultinck, M.~Mari{\"e}n, D.~J. Williamson, M.~B. {\c{S}}ahino{\u{g}}lu, J.~Haegeman, and F.~Verstraete, {\it {Anyons and matrix product operator algebras}},  {\em Annals Phys.} {\bf 378} (2017) 183--233, [\href{http://arxiv.org/abs/1511.08090}{{\tt arXiv:1511.08090}}].

\bibitem{Vanhove:2018wlb}
R.~Vanhove, M.~Bal, D.~J. Williamson, N.~Bultinck, J.~Haegeman, and F.~Verstraete, {\it {Mapping topological to conformal field theories through strange correlators}},  {\em Phys. Rev. Lett.} {\bf 121} (2018), no.~17 177203, [\href{http://arxiv.org/abs/1801.05959}{{\tt arXiv:1801.05959}}].

\bibitem{Inamura:2021szw}
K.~Inamura, {\it {On lattice models of gapped phases with fusion category symmetries}},  {\em JHEP} {\bf 03} (2022) 036, [\href{http://arxiv.org/abs/2110.12882}{{\tt arXiv:2110.12882}}].

\bibitem{Lootens:2021tet}
L.~Lootens, C.~Delcamp, G.~Ortiz, and F.~Verstraete, {\it {Dualities in One-Dimensional Quantum Lattice Models: Symmetric Hamiltonians and Matrix Product Operator Intertwiners}},  {\em PRX Quantum} {\bf 4} (2023), no.~2 020357, [\href{http://arxiv.org/abs/2112.09091}{{\tt arXiv:2112.09091}}].

\bibitem{Molnar:2022nmh}
A.~Molnar, A.~R. de~Alarc{\'o}n, J.~Garre-Rubio, N.~Schuch, J.~I. Cirac, and D.~P{\'e}rez-Garc{\'\i}a, {\it {Matrix Product Operator Algebras I: Representations of Weak Hopf Algebras and Projected Entangled Pair States}},  {\em Commun. Math. Phys.} {\bf 407} (2026), no.~10 203, [\href{http://arxiv.org/abs/2204.05940}{{\tt arXiv:2204.05940}}].

\bibitem{Lootens:2022avn}
L.~Lootens, C.~Delcamp, and F.~Verstraete, {\it {Dualities in One-Dimensional Quantum Lattice Models: Topological Sectors}},  {\em PRX Quantum} {\bf 5} (2024), no.~1 010338, [\href{http://arxiv.org/abs/2211.03777}{{\tt arXiv:2211.03777}}].

\bibitem{Garre-Rubio:2022uum}
J.~Garre-Rubio, L.~Lootens, and A.~Moln{\'a}r, {\it {Classifying phases protected by matrix product operator symmetries using matrix product states}},  {\em Quantum} {\bf 7} (2023) 927, [\href{http://arxiv.org/abs/2203.12563}{{\tt arXiv:2203.12563}}].

\bibitem{Lootens:2023wnl}
L.~Lootens, C.~Delcamp, D.~Williamson, and F.~Verstraete, {\it {Low-Depth Unitary Quantum Circuits for Dualities in One-Dimensional Quantum Lattice Models}},  {\em Phys. Rev. Lett.} {\bf 134} (2025), no.~13 130403, [\href{http://arxiv.org/abs/2311.01439}{{\tt arXiv:2311.01439}}].

\bibitem{Fechisin:2023odt}
C.~Fechisin, N.~Tantivasadakarn, and V.~V. Albert, {\it {Noninvertible Symmetry-Protected Topological Order in a Group-Based Cluster State}},  {\em Phys. Rev. X} {\bf 15} (2025), no.~1 011058, [\href{http://arxiv.org/abs/2312.09272}{{\tt arXiv:2312.09272}}].

\bibitem{Seiberg:2024gek}
N.~Seiberg, S.~Seifnashri, and S.-H. Shao, {\it {Non-invertible symmetries and LSM-type constraints on a tensor product Hilbert space}},  {\em SciPost Phys.} {\bf 16} (2024) 154, [\href{http://arxiv.org/abs/2401.12281}{{\tt arXiv:2401.12281}}].

\bibitem{Chatterjee:2024ych}
A.~Chatterjee, {\"O}.~M. Aksoy, and X.-G. Wen, {\it {Quantum phases and transitions in spin chains with non-invertible symmetries}},  {\em SciPost Phys.} {\bf 17} (2024), no.~4 115, [\href{http://arxiv.org/abs/2405.05331}{{\tt arXiv:2405.05331}}].

\bibitem{Lootens:2024gfp}
L.~Lootens, C.~Delcamp, and F.~Verstraete, {\it {Entanglement and the density matrix renormalization group in the generalized Landau paradigm}},  {\em Nature Phys.} {\bf 21} (2025), no.~10 1657--1663, [\href{http://arxiv.org/abs/2408.06334}{{\tt arXiv:2408.06334}}].

\bibitem{Inamura:2024jke}
K.~Inamura and S.~Ohyama, {\it {1+1d SPT phases with fusion category symmetry: interface modes and non-abelian Thouless pump}},  \href{http://arxiv.org/abs/2408.15960}{{\tt arXiv:2408.15960}}.

\bibitem{Pace:2024acq}
S.~D. Pace, H.~T. Lam, and {\"O}.~M. Aksoy, {\it {(SPT-)LSM theorems from projective non-invertible symmetries}},  {\em SciPost Phys.} {\bf 18} (2025), no.~1 028, [\href{http://arxiv.org/abs/2409.18113}{{\tt arXiv:2409.18113}}].

\bibitem{Zhang:2020pco}
H.-C. Zhang and G.~Sierra, {\it {Kramers-Wannier self-duality and non-invertible translation symmetry in quantum chains: a wave-function perspective}},  {\em JHEP} {\bf 05} (2025) 157, [\href{http://arxiv.org/abs/2410.06727}{{\tt arXiv:2410.06727}}].

\bibitem{Jones:2024lws}
C.~Jones, K.~Schatz, and D.~J. Williamson, {\it {Quantum Cellular Automata and Categorical Dualities of Spin Chains}},  {\em Commun. Math. Phys.} {\bf 407} (2026), no.~4 66, [\href{http://arxiv.org/abs/2410.08884}{{\tt arXiv:2410.08884}}].

\bibitem{Pace:2024oys}
S.~D. Pace, A.~Chatterjee, and S.-H. Shao, {\it {Lattice T-duality from non-invertible symmetries in quantum spin chains}},  {\em SciPost Phys.} {\bf 18} (2025), no.~4 121, [\href{http://arxiv.org/abs/2412.18606}{{\tt arXiv:2412.18606}}].

\bibitem{Meng:2024nxx}
C.~Meng, X.~Yang, T.~Lan, and Z.~Gu, {\it {Non-invertible SPTs: an on-site realization of (1+1)d anomaly-free fusion category symmetry}},  \href{http://arxiv.org/abs/2412.20546}{{\tt arXiv:2412.20546}}.

\bibitem{Vancraeynest-DeCuiper:2025fpn}
B.~Vancraeynest-De~Cuiper, J.~Garre-Rubio, F.~Verstraete, K.~Vervoort, D.~J. Williamson, and L.~Lootens, {\it {From gauging to duality in one-dimensional quantum lattice models}},  {\em SciPost Phys.} {\bf 20} (2026), no.~4 113, [\href{http://arxiv.org/abs/2509.22051}{{\tt arXiv:2509.22051}}].

\bibitem{Lu:2026rhb}
D.-C. Lu, A.~Chatterjee, and N.~Tantivasadakarn, {\it {Generalized Kramers-Wannier Self-Duality in Hopf-Ising Models}},  \href{http://arxiv.org/abs/2602.10183}{{\tt arXiv:2602.10183}}.

\bibitem{Inamura:2026hif}
K.~Inamura, {\it {Remarks on non-invertible symmetries on a tensor product Hilbert space in 1+1 dimensions}},  {\em SciPost Phys.} {\bf 21} (2026), no.~4 088, [\href{http://arxiv.org/abs/2602.12053}{{\tt arXiv:2602.12053}}].

\bibitem{Wen:2026ncw}
R.~Wen, K.~Inamura, and S.~Schafer-Nameki, {\it {Non-Invertible Symmetries on Tensor-Product Hilbert Spaces and Quantum Cellular Automata}},  \href{http://arxiv.org/abs/2605.15194}{{\tt arXiv:2605.15194}}.

\bibitem{Vernier:2026kxf}
E.~Vernier, Y.~Miao, and M.~Yamazaki, {\it {Lattice non-invertible symmetry from non-commuting transfer matrices}},  \href{http://arxiv.org/abs/2606.25660}{{\tt arXiv:2606.25660}}.

\bibitem{Bosgoed:2026ayl}
D.~Bosgoed, F.~Verstraete, and L.~Lootens, {\it {Dualities in one-dimensional quantum lattice models: Lie groups, Schur-Weyl duality and symmetry breaking}},  \href{http://arxiv.org/abs/2611.08293}{{\tt arXiv:2611.08293}}.

\bibitem{Tantivasadakarn:2025txn}
N.~Tantivasadakarn, X.~Liu, and X.~Chen, {\it {Sequential circuits as generalized symmetry on the lattice}},  {\em Phys. Rev. B} {\bf 114} (2026), no.~22 225101, [\href{http://arxiv.org/abs/2507.22394}{{\tt arXiv:2507.22394}}].

\bibitem{Bhardwaj:2017xup}
L.~Bhardwaj and Y.~Tachikawa, {\it {On finite symmetries and their gauging in two dimensions}},  {\em JHEP} {\bf 03} (2018) 189, [\href{http://arxiv.org/abs/1704.02330}{{\tt arXiv:1704.02330}}].

\bibitem{Chang:2018iay}
C.-M. Chang, Y.-H. Lin, S.-H. Shao, Y.~Wang, and X.~Yin, {\it {Topological Defect Lines and Renormalization Group Flows in Two Dimensions}},  {\em JHEP} {\bf 01} (2019) 026, [\href{http://arxiv.org/abs/1802.04445}{{\tt arXiv:1802.04445}}].

\bibitem{Lin:2022dhv}
Y.-H. Lin, M.~Okada, S.~Seifnashri, and Y.~Tachikawa, {\it {Asymptotic density of states in 2d CFTs with non-invertible symmetries}},  {\em JHEP} {\bf 03} (2023) 094, [\href{http://arxiv.org/abs/2208.05495}{{\tt arXiv:2208.05495}}].

\bibitem{Okada:2024qmk}
M.~Okada and Y.~Tachikawa, {\it {Noninvertible Symmetries Act Locally by Quantum Operations}},  {\em Phys. Rev. Lett.} {\bf 133} (2024), no.~19 191602, [\href{http://arxiv.org/abs/2403.20062}{{\tt arXiv:2403.20062}}].

\bibitem{Bischoff:2016jmy}
M.~Bischoff, {\it {Generalized Orbifold Construction for Conformal Nets}},  {\em Rev. Math. Phys.} {\bf 29} (2016), no.~01 1750002, [\href{http://arxiv.org/abs/1608.00253}{{\tt arXiv:1608.00253}}].

\bibitem{Bischoff:2016rpu}
M.~Bischoff and K.-H. Rehren, {\it {The hypergroupoid of boundary conditions for local quantum observables.}},  {\em Adv. Stud. Pure Math.} {\bf 80} (2019) 32--42, [\href{http://arxiv.org/abs/1612.02972}{{\tt arXiv:1612.02972}}].

\bibitem{Bischoff:2020iac}
M.~Bischoff, S.~Del~Vecchio, and L.~Giorgetti, {\it {Compact hypergroups from discrete subfactors}},  {\em J. Funct. Anal.} {\bf 281} (2021) 109004, [\href{http://arxiv.org/abs/2007.12384}{{\tt arXiv:2007.12384}}].

\bibitem{Bischoff:2021snn}
M.~Bischoff, S.~Del~Vecchio, and L.~Giorgetti, {\it {Galois Correspondence and Fourier Analysis on Local Discrete Subfactors}},  {\em Annales Henri Poincare} {\bf 23} (2022), no.~8 2979--3020, [\href{http://arxiv.org/abs/2107.09345}{{\tt arXiv:2107.09345}}].

\bibitem{Bischoff:2022fxf}
M.~Bischoff, S.~Del~Vecchio, and L.~Giorgetti, {\it {Quantum operations on conformal nets}},  {\em Rev. Math. Phys.} {\bf 35} (2023), no.~04 2350007, [\href{http://arxiv.org/abs/2204.14105}{{\tt arXiv:2204.14105}}].

\bibitem{Evans:2025msy}
D.~E. Evans and C.~Jones, {\it {An operator algebraic approach to fusion category symmetry on the lattice}},  \href{http://arxiv.org/abs/2507.05185}{{\tt arXiv:2507.05185}}.

\bibitem{Jones:2026dcb}
C.~Jones and X.~Yang, {\it {On the structure of categorical duality operators}},  \href{http://arxiv.org/abs/2603.09949}{{\tt arXiv:2603.09949}}.

\bibitem{richter1996ergodicity}
S.~Richter and R.~F. Werner, {\it {Ergodicity of quantum cellular automata}},  {\em J. Stat. Phys.} {\bf 82} (1996) 963--998, [\href{http://arxiv.org/abs/cond-mat/9504001}{{\tt cond-mat/9504001}}].

\bibitem{Seifnashri:QCAStructure}
S.~Seifnashri, ``{Non-Invertible Quantum Cellular Automata II: Defect Reconstruction and Index Theory}.'' In preparation.

\bibitem{Hauru:2015abi}
M.~Hauru, G.~Evenbly, W.~W. Ho, D.~Gaiotto, and G.~Vidal, {\it {Topological conformal defects with tensor networks}},  {\em Phys. Rev. B} {\bf 94} (2016), no.~11 115125, [\href{http://arxiv.org/abs/1512.03846}{{\tt arXiv:1512.03846}}].

\bibitem{Aasen:2016dop}
D.~Aasen, R.~S.~K. Mong, and P.~Fendley, {\it {Topological Defects on the Lattice I: The Ising model}},  {\em J. Phys. A} {\bf 49} (2016), no.~35 354001, [\href{http://arxiv.org/abs/1601.07185}{{\tt arXiv:1601.07185}}].

\bibitem{Aasen:2020jwb}
D.~Aasen, P.~Fendley, and R.~S.~K. Mong, {\it {Topological Defects on the Lattice: Dualities and Degeneracies}},  \href{http://arxiv.org/abs/2008.08598}{{\tt arXiv:2008.08598}}.

\bibitem{Cheng:2022sgb}
M.~Cheng and N.~Seiberg, {\it {Lieb-Schultz-Mattis, Luttinger, and 't Hooft - anomaly matching in lattice systems}},  {\em SciPost Phys.} {\bf 15} (2023), no.~2 051, [\href{http://arxiv.org/abs/2211.12543}{{\tt arXiv:2211.12543}}].

\bibitem{Sinha:2023hum}
M.~Sinha, F.~Yan, L.~Grans-Samuelsson, A.~Roy, and H.~Saleur, {\it {Lattice realizations of topological defects in the critical (1+1)-d three-state Potts model}},  {\em JHEP} {\bf 07} (2024) 225, [\href{http://arxiv.org/abs/2310.19703}{{\tt arXiv:2310.19703}}].

\bibitem{Sinha:2025jhh}
M.~Sinha, T.~S. Tavares, A.~Roy, and H.~Saleur, {\it {Integrability and lattice discretizations of all topological defect lines in minimal CFTs}},  {\em Nucl. Phys. B} {\bf 1023} (2026) 117308, [\href{http://arxiv.org/abs/2509.04257}{{\tt arXiv:2509.04257}}].

\bibitem{Chen:2023qst}
X.~Chen, A.~Dua, M.~Hermele, D.~T. Stephen, N.~Tantivasadakarn, R.~Vanhove, and J.-Y. Zhao, {\it {Sequential quantum circuits as maps between gapped phases}},  {\em Phys. Rev. B} {\bf 109} (2024), no.~7 075116, [\href{http://arxiv.org/abs/2307.01267}{{\tt arXiv:2307.01267}}].

\bibitem{Vanhove:2024lmj}
R.~Vanhove, V.~Ravindran, D.~T. Stephen, X.-G. Wen, and X.~Chen, {\it {Duality via sequential quantum circuit in the topological holography formalism}},  {\em Phys. Rev. B} {\bf 112} (2025), no.~3 035173, [\href{http://arxiv.org/abs/2409.06647}{{\tt arXiv:2409.06647}}].

\bibitem{Sato:2026rpj}
S.~Sato, Y.~Hidaka, and R.~Kobayashi, {\it {Invariants of Sequential Circuits and Generalized Non-Abelian Statistics}},  \href{http://arxiv.org/abs/2606.11527}{{\tt arXiv:2606.11527}}.

\bibitem{Seifnashri:2025fgd}
S.~Seifnashri, S.-H. Shao, and X.~Yang, {\it {Gauging non-invertible symmetries on the lattice}},  {\em SciPost Phys.} {\bf 19} (2025), no.~2 063, [\href{http://arxiv.org/abs/2503.02925}{{\tt arXiv:2503.02925}}].

\bibitem{Freedman:2019ncn}
M.~Freedman and M.~B. Hastings, {\it {Classification of Quantum Cellular Automata}},  {\em Commun. Math. Phys.} {\bf 376} (2020) 1171--1222, [\href{http://arxiv.org/abs/1902.10285}{{\tt arXiv:1902.10285}}].

\bibitem{Gross:2011yvb}
D.~Gross, V.~Nesme, H.~Vogts, and R.~F. Werner, {\it {Index Theory of One Dimensional Quantum Walks and Cellular Automata}},  {\em Commun. Math. Phys.} {\bf 310} (2012), no.~2 419--454, [\href{http://arxiv.org/abs/0910.3675}{{\tt arXiv:0910.3675}}].

\bibitem{Haah:2022abh}
J.~Haah, {\it {Invertible Subalgebras}},  {\em Commun. Math. Phys.} {\bf 403} (2023), no.~2 661--698, [\href{http://arxiv.org/abs/2211.02086}{{\tt arXiv:2211.02086}}].

\bibitem{Jones:2026gmd}
C.~Jones, N.~Sopenko, and R.~Thorngren, {\it {Quantum cellular automata and invertible phases of matter}},  \href{http://arxiv.org/abs/2608.26456}{{\tt arXiv:2608.26456}}.

\bibitem{Jones:2023imy}
C.~Jones and J.~Lim, {\it {An Index for Quantum Cellular Automata on Fusion Spin Chains}},  {\em Annales Henri Poincare} {\bf 25} (2024), no.~10 4399--4422, [\href{http://arxiv.org/abs/2309.10961}{{\tt arXiv:2309.10961}}].

\bibitem{Ma:2024ypm}
R.~Ma, Y.~Li, and M.~Cheng, {\it {Quantum Cellular Automata on Symmetric Subalgebras}},  {\em Quantum} {\bf 10} (2026) 2123, [\href{http://arxiv.org/abs/2411.19280}{{\tt arXiv:2411.19280}}].

\bibitem{bratteli2012operator}
O.~Bratteli and D.~W. Robinson, {\em Operator Algebras and Quantum Statistical Mechanics 1: {$C^*$}- and {$W^*$}-Algebras, Symmetry Groups, Decomposition of States}.
\newblock Springer Science \& Business Media, 2012.

\bibitem{Naaijkens:2013vqa}
P.~Naaijkens, {\it {Quantum spin systems on infinite lattices}},  \href{http://arxiv.org/abs/1311.2717}{{\tt arXiv:1311.2717}}.

\bibitem{Tasaki:2022gka}
H.~Tasaki, {\it {The Lieb-Schultz-Mattis theorem. A topological point of view.}},  \href{http://arxiv.org/abs/2202.06243}{{\tt arXiv:2202.06243}}.

\bibitem{choi1974schwarz}
M.-D. Choi, {\it A {S}chwarz inequality for positive linear maps on {$C^*$}-algebras},  {\em Illinois Journal of Mathematics} {\bf 18} (1974), no.~4 565--574.

\bibitem{paschke1972completely}
W.~Paschke, {\it Completely positive maps on {$U^*$}-algebras},  {\em Proceedings of the American Mathematical Society} {\bf 34} (1972), no.~2 412--416.

\bibitem{Paulsen_2003}
V.~Paulsen, {\em Completely Bounded Maps and Operator Algebras}.
\newblock Cambridge Studies in Advanced Mathematics. Cambridge University Press, 2003.

\bibitem{rahaman2017multiplicative}
M.~Rahaman, {\it Multiplicative properties of quantum channels},  {\em Journal of Physics A: Mathematical and Theoretical} {\bf 50} (2017), no.~34 345302.

\bibitem{Piroli:2020jqb}
L.~Piroli and J.~{\,}. Cirac, {\it {Quantum Cellular Automata, Tensor Networks, and Area Laws}},  {\em Phys. Rev. Lett.} {\bf 125} (2020), no.~19 190402, [\href{http://arxiv.org/abs/2007.15371}{{\tt arXiv:2007.15371}}].

\bibitem{Bartsch:2026wqq}
T.~Bartsch, Y.~Gai, and S.~Schafer-Nameki, {\it {Beyond Wigner: Non-Invertible Symmetries Preserve Probabilities}},  \href{http://arxiv.org/abs/2602.07110}{{\tt arXiv:2602.07110}}.

\bibitem{Nill:1995nk}
F.~Nill and K.~Szlach{\'a}nyi, {\it {Quantum chains of Hopf algebras with quantum double cosymmetry}},  {\em Commun. Math. Phys.} {\bf 187} (1997) 159--200, [\href{http://arxiv.org/abs/hep-th/9509100}{{\tt hep-th/9509100}}].

\bibitem{Jones:2023ptg}
C.~Jones, {\it {DHR bimodules of quasi-local algebras and symmetric quantum cellular automata}},  \href{http://arxiv.org/abs/2304.00068}{{\tt arXiv:2304.00068}}.

\bibitem{Cao:2025qnc}
W.~Cao, Y.~Miao, and M.~Yamazaki, {\it {Global symmetries of quantum lattice models under non-invertible dualities}},  {\em SciPost Phys. Core} {\bf 8} (2025) 070, [\href{http://arxiv.org/abs/2501.12514}{{\tt arXiv:2501.12514}}].

\bibitem{Chang:2020imq}
C.-M. Chang and Y.-H. Lin, {\it {Lorentzian dynamics and factorization beyond rationality}},  {\em JHEP} {\bf 10} (2021) 125, [\href{http://arxiv.org/abs/2012.01429}{{\tt arXiv:2012.01429}}].

\bibitem{Thorngren:2021yso}
R.~Thorngren and Y.~Wang, {\it {Fusion category symmetry. Part II. Categoriosities at c = 1 and beyond}},  {\em JHEP} {\bf 07} (2024) 051, [\href{http://arxiv.org/abs/2106.12577}{{\tt arXiv:2106.12577}}].

\bibitem{Fannes:1990ur}
M.~Fannes, B.~Nachtergaele, and R.~F. Werner, {\it {Finitely correlated states on quantum spin chains}},  {\em Commun. Math. Phys.} {\bf 144} (1992) 443--490.

\bibitem{Perez-Garcia:2006nqo}
D.~Perez-Garcia, F.~Verstraete, M.~M. Wolf, and J.~I. Cirac, {\it {Matrix product state representations}},  {\em Quant. Inf. Comput.} {\bf 7} (2007), no.~5-6 401--430, [\href{http://arxiv.org/abs/quant-ph/0608197}{{\tt quant-ph/0608197}}].

\bibitem{Cirac:2020obd}
J.~I. Cirac, D.~Perez-Garcia, N.~Schuch, and F.~Verstraete, {\it {Matrix product states and projected entangled pair states: Concepts, symmetries, theorems}},  {\em Rev. Mod. Phys.} {\bf 93} (2021), no.~4 045003, [\href{http://arxiv.org/abs/2011.12127}{{\tt arXiv:2011.12127}}].

\bibitem{watatani1990index}
Y.~Watatani, {\em Index for $C^*$-Subalgebras}, vol.~424.
\newblock American Mathematical Soc., 1990.

\bibitem{Seifnashri:2024dsd}
S.~Seifnashri and S.-H. Shao, {\it {Cluster State as a Noninvertible Symmetry-Protected Topological Phase}},  {\em Phys. Rev. Lett.} {\bf 133} (2024), no.~11 116601, [\href{http://arxiv.org/abs/2404.01369}{{\tt arXiv:2404.01369}}].

\bibitem{Xie:2025ccn}
W.~Xie, S.~Ono, and H.~C. Po, {\it {Unitary network: Tensor network unitaries with local unitarity}},  \href{http://arxiv.org/abs/2508.16890}{{\tt arXiv:2508.16890}}.

\bibitem{etingof2015tensor}
P.~Etingof, S.~Gelaki, D.~Nikshych, and V.~Ostrik, {\em Tensor Categories}, vol.~205 of {\em Mathematical Surveys and Monographs}.
\newblock American Mathematical Society, Providence, RI, 2015.

\bibitem{epstein1971almost}
D.~B. Epstein, {\it Almost all subgroups of a {L}ie group are free},  {\em Journal of Algebra} {\bf 19} (1971), no.~2 261--262.

\bibitem{Zadnik:2020xcd}
L.~Zadnik and M.~Fagotti, {\it {The Folded Spin-1/2 XXZ Model: I. Diagonalisation, Jamming, and Ground State Properties}},  {\em SciPost Phys. Core} {\bf 4} (2021) 010, [\href{http://arxiv.org/abs/2009.04995}{{\tt arXiv:2009.04995}}].

\bibitem{Pozsgay:2021dyy}
B.~Pozsgay, T.~Gombor, A.~Hutsalyuk, Y.~Jiang, L.~Pristy{\'a}k, and E.~Vernier, {\it {Integrable spin chain with Hilbert space fragmentation and solvable real-time dynamics}},  {\em Phys. Rev. E} {\bf 104} (2021), no.~4 044106, [\href{http://arxiv.org/abs/2105.02252}{{\tt arXiv:2105.02252}}].

\bibitem{Borsi:2023nyt}
M.~Borsi, L.~Pristy{\'a}k, and B.~Pozsgay, {\it {Matrix Product Symmetries and Breakdown of Thermalization from Hard Rod Deformations}},  {\em Phys. Rev. Lett.} {\bf 131} (2023), no.~3 037101, [\href{http://arxiv.org/abs/2302.07219}{{\tt arXiv:2302.07219}}].

\bibitem{alcaraz1999exactly}
F.~C. Alcaraz and R.~Z. Bariev, {\it {An Exactly Solvable Constrained XXZ Chain}},  \href{http://arxiv.org/abs/cond-mat/9904042}{{\tt cond-mat/9904042}}.

\bibitem{Verresen:2019jvy}
R.~Verresen, A.~Vishwanath, and F.~Pollmann, {\it {Stable Luttinger liquids and emergent $U(1)$ symmetry in constrained quantum chains}},  \href{http://arxiv.org/abs/1903.09179}{{\tt arXiv:1903.09179}}.

\bibitem{Kretschmann:2005mya}
D.~Kretschmann and R.~F. Werner, {\it {Quantum channels with memory}},  {\em Phys. Rev. A} {\bf 72} (2005), no.~6 062323, [\href{http://arxiv.org/abs/quant-ph/0502106}{{\tt quant-ph/0502106}}].

\bibitem{Ranard:2020jkb}
D.~Ranard, M.~Walter, and F.~Witteveen, {\it {A Converse to Lieb{\textendash}Robinson Bounds in One Dimension Using Index Theory}},  {\em Annales Henri Poincare} {\bf 23} (2022), no.~11 3905--3979, [\href{http://arxiv.org/abs/2012.00741}{{\tt arXiv:2012.00741}}].

\bibitem{Ranard:2026hqu}
D.~Ranard, M.~Walter, and F.~Witteveen, {\it {Approximate QCAs in one dimension using approximate algebras}},  \href{http://arxiv.org/abs/2603.08702}{{\tt arXiv:2603.08702}}.

\bibitem{Seiberg:2023cdc}
N.~Seiberg and S.-H. Shao, {\it {Majorana chain and Ising model - (non-invertible) translations, anomalies, and emanant symmetries}},  {\em SciPost Phys.} {\bf 16} (2024), no.~3 064, [\href{http://arxiv.org/abs/2307.02534}{{\tt arXiv:2307.02534}}].

\bibitem{Cao:2023doz}
W.~Cao, L.~Li, M.~Yamazaki, and Y.~Zheng, {\it {Subsystem non-invertible symmetry operators and defects}},  {\em SciPost Phys.} {\bf 15} (2023), no.~4 155, [\href{http://arxiv.org/abs/2304.09886}{{\tt arXiv:2304.09886}}].

\bibitem{Choi:2024rjm}
Y.~Choi, Y.~Sanghavi, S.-H. Shao, and Y.~Zheng, {\it {Non-invertible and higher-form symmetries in 2+1d lattice gauge theories}},  {\em SciPost Phys.} {\bf 18} (2025), no.~1 008, [\href{http://arxiv.org/abs/2405.13105}{{\tt arXiv:2405.13105}}].

\bibitem{Seiberg:2019vrp}
N.~Seiberg, {\it {Field Theories With a Vector Global Symmetry}},  {\em SciPost Phys.} {\bf 8} (2020), no.~4 050, [\href{http://arxiv.org/abs/1909.10544}{{\tt arXiv:1909.10544}}].

\bibitem{Qi:2020jrf}
M.~Qi, L.~Radzihovsky, and M.~Hermele, {\it {Fracton phases via exotic higher-form symmetry-breaking}},  {\em Annals Phys.} {\bf 424} (2021) 168360, [\href{http://arxiv.org/abs/2010.02254}{{\tt arXiv:2010.02254}}].

\bibitem{Oh:2023bnk}
Y.-T. Oh, S.~D. Pace, J.~H. Han, Y.~You, and H.-Y. Lee, {\it {Aspects of ZN rank-2 gauge theory in (2+1) dimensions: Construction schemes, holonomies, and sublattice one-form symmetries}},  {\em Phys. Rev. B} {\bf 107} (2023), no.~15 155151, [\href{http://arxiv.org/abs/2301.04706}{{\tt arXiv:2301.04706}}].

\bibitem{Gorantla:2024ocs}
P.~Gorantla, S.-H. Shao, and N.~Tantivasadakarn, {\it {Tensor Networks for Noninvertible Symmetries in 3+1D and Beyond}},  {\em Phys. Rev. X} {\bf 15} (2025), no.~4 041006, [\href{http://arxiv.org/abs/2406.12978}{{\tt arXiv:2406.12978}}].

\bibitem{Liu:2026tcl}
R.~Liu, P.~M. Tam, H.~T. Lam, and L.~Zou, {\it {When does a lattice higher-form symmetry flow to a topological higher-form symmetry at low energies?}},  \href{http://arxiv.org/abs/2601.20935}{{\tt arXiv:2601.20935}}.

\bibitem{Brito:2026eqi}
L.~Z. Brito and S.~D. Pace, {\it {Lattice 2-group symmetries: operators, defects, and gauging}},  \href{http://arxiv.org/abs/2609.10682}{{\tt arXiv:2609.10682}}.

\bibitem{Gaiotto:2019xmp}
D.~Gaiotto and T.~Johnson-Freyd, {\it {Condensations in higher categories}},  \href{http://arxiv.org/abs/1905.09566}{{\tt arXiv:1905.09566}}.

\bibitem{Roumpedakis:2022aik}
K.~Roumpedakis, S.~Seifnashri, and S.-H. Shao, {\it {Higher Gauging and Non-invertible Condensation Defects}},  {\em Commun. Math. Phys.} {\bf 401} (2023), no.~3 3043--3107, [\href{http://arxiv.org/abs/2204.02407}{{\tt arXiv:2204.02407}}].

\bibitem{Wegner:1971app}
F.~J. Wegner, {\it {Duality in Generalized Ising Models and Phase Transitions Without Local Order Parameters}},  {\em J. Math. Phys.} {\bf 12} (1971) 2259--2272.

\bibitem{Koide:2021zxj}
M.~Koide, Y.~Nagoya, and S.~Yamaguchi, {\it {Non-invertible topological defects in 4-dimensional $\mathbb {Z}_2$ pure lattice gauge theory}},  {\em PTEP} {\bf 2022} (2022), no.~1 013B03, [\href{http://arxiv.org/abs/2109.05992}{{\tt arXiv:2109.05992}}].

\bibitem{Koide:2023rqd}
M.~Koide, Y.~Nagoya, and S.~Yamaguchi, {\it {Noninvertible symmetries and boundaries in four dimensions}},  {\em Phys. Rev. D} {\bf 108} (2023), no.~6 065009, [\href{http://arxiv.org/abs/2304.01550}{{\tt arXiv:2304.01550}}].

\bibitem{Choi:2021kmx}
Y.~Choi, C.~Cordova, P.-S. Hsin, H.~T. Lam, and S.-H. Shao, {\it {Noninvertible duality defects in 3+1 dimensions}},  {\em Phys. Rev. D} {\bf 105} (2022), no.~12 125016 \text{(Editors' Suggestion)}, [\href{http://arxiv.org/abs/2111.01139}{{\tt arXiv:2111.01139}}].

\bibitem{Gorantla:2024ptu}
P.~Gorantla and T.-C. Huang, {\it {Duality-preserving deformation of 3+1d lattice Z2 gauge theory with exact gapped ground states}},  {\em Phys. Rev. B} {\bf 111} (2025), no.~24 245110, [\href{http://arxiv.org/abs/2409.10612}{{\tt arXiv:2409.10612}}].

\bibitem{Inamura:2026kco}
K.~Inamura, O.~Wojdel, L.~Fidkowski, and S.~Schafer-Nameki, {\it {Non-Invertible Symmetries Mixing with Witt Non-Trivial Quantum Cellular Automata}},  \href{http://arxiv.org/abs/2607.21698}{{\tt arXiv:2607.21698}}.

\bibitem{Zhang:2026kjf}
C.~Zhang and P.-S. Hsin, {\it {Quantum Cellular Automata from Kramers-Wannier Dualities and Modular Relations}},  \href{http://arxiv.org/abs/2607.21728}{{\tt arXiv:2607.21728}}.

\bibitem{Harlow:2025cqc}
D.~Harlow, S.-H. Shao, J.~Sorce, and M.~Srivastava, {\it {Disjoint additivity and local quantum physics}},  {\em JHEP} {\bf 06} (2026) 025, [\href{http://arxiv.org/abs/2509.03589}{{\tt arXiv:2509.03589}}].

\bibitem{Kaidi:2021xfk}
J.~Kaidi, K.~Ohmori, and Y.~Zheng, {\it {Kramers-Wannier-like Duality Defects in (3+1)D Gauge Theories}},  {\em Phys. Rev. Lett.} {\bf 128} (2022), no.~11 111601, [\href{http://arxiv.org/abs/2111.01141}{{\tt arXiv:2111.01141}}].

\bibitem{Choi:2022zal}
Y.~Choi, C.~Cordova, P.-S. Hsin, H.~T. Lam, and S.-H. Shao, {\it {Non-invertible Condensation, Duality, and Triality Defects in 3+1 Dimensions}},  {\em Commun. Math. Phys.} {\bf 402} (2023), no.~1 489--542, [\href{http://arxiv.org/abs/2204.09025}{{\tt arXiv:2204.09025}}].

\bibitem{ParayilMana:2024txy}
A.~Parayil~Mana, Y.~Li, H.~Sukeno, and T.-C. Wei, {\it {Kennedy-Tasaki transformation and noninvertible symmetry in lattice models beyond one dimension}},  {\em Phys. Rev. B} {\bf 109} (2024), no.~24 245129, [\href{http://arxiv.org/abs/2402.09520}{{\tt arXiv:2402.09520}}].

\bibitem{Cao:2024qjj}
W.~Cao, L.~Li, and M.~Yamazaki, {\it {Generating lattice non-invertible symmetries}},  {\em SciPost Phys.} {\bf 17} (2024), no.~4 104, [\href{http://arxiv.org/abs/2406.05454}{{\tt arXiv:2406.05454}}].

\bibitem{Pace:2024tgk}
S.~D. Pace, G.~Delfino, H.~T. Lam, and {\"O}.~M. Aksoy, {\it {Gauging modulated symmetries: Kramers-Wannier dualities and non-invertible reflections}},  {\em SciPost Phys.} {\bf 18} (2025), no.~1 021, [\href{http://arxiv.org/abs/2406.12962}{{\tt arXiv:2406.12962}}].

\bibitem{Kim:2025ttl}
J.~Kim, Y.~You, and J.~H. Han, {\it {Noninvertible symmetry and topological holography for modulated SPT in one dimension}},  {\em SciPost Phys.} {\bf 19} (2025), no.~4 110, [\href{http://arxiv.org/abs/2507.02324}{{\tt arXiv:2507.02324}}].

\bibitem{Spieler:2024fby}
R.~C. Spieler, {\it {Non-invertible duality interfaces in field theories with exotic symmetries}},  {\em JHEP} {\bf 06} (2024) 042, [\href{http://arxiv.org/abs/2402.14944}{{\tt arXiv:2402.14944}}].

\end{thebibliography}\endgroup

\end{document}